\documentclass[12pt,preprint]{aastex}

\newcommand{\masyr}{\mbox{mas\,yr$^{-1}$}}
\newcommand{\uv}{\mbox{$u$-$v$}}
\newcommand{\GPB}{\mbox{{\em GP-B}}}

\newcommand{\kmsMpc}{\mbox{km\,s$^{-1}$\,Mpc$^{-1}$}}

\slugcomment{Accepted to the Astrophysical Journal Supplement Series}

\shorttitle{VLBI for {\em GP-B}. II.}
\shortauthors{Ransom et al.}

\begin{document}

\title{VLBI for {\em Gravity Probe B}. II. Monitoring of the Structure
of the Reference Sources 3C~454.3, B2250+194, and B2252+172}

\author{R. R. Ransom\altaffilmark{1,2}, N. Bartel\altaffilmark{1},
M. F. Bietenholz\altaffilmark{1,3}, D. E. Lebach\altaffilmark{4}
J. I. Lederman\altaffilmark{5}, P. Luca\altaffilmark{1},
M. I. Ratner\altaffilmark{4} and I. I. Shapiro\altaffilmark{4}}

\altaffiltext{1}{York University, Department of Physics and Astronomy,
4700 Keele Street, Toronto, ON, M3J~1P3, Canada}
\altaffiltext{2}{Now at Okanagan College, 583 Duncan Avenue West,
Penticton, B.C., V2A 2K8, Canada and also at the National Research
Council of Canada, Herzberg Institute of Astrophysics, Dominion Radio
Astrophysical Observatory, P.O.\@ Box 248, Penticton, B.C., V2A 6K3,
Canada}
\altaffiltext{3}{Now also at Hartebeesthoek Radio Astronomy
Observatory, P.O.\@ Box 443, Krugersdorp 1740, South Africa}
\altaffiltext{4}{Harvard-Smithsonian Center for Astrophysics, 60 Garden
Street, Cambridge, MA 02138}
\altaffiltext{5}{York University, Centre for Research in Earth and Space
Sciences, 4700 Keele Street, Toronto, ON M3J~1P3, Canada}


\notetoeditor{Here's where we can insert notes to the editor which
do not show up in the typeset text.}

\begin{abstract}

We used 8.4~GHz VLBI images obtained at up to 35 epochs between 1997
and 2005 to examine the radio structures of the main reference source,
3C~454.3, and two secondary reference sources, B2250+194 and
B2252+172, for the guide star for the NASA/Stanford relativity mission
{\em Gravity Probe B} (\GPB\@).  For one epoch in 2004 May, we also
obtained images at 5.0 and 15.4~GHz.  The 35 8.4~GHz images for quasar
3C~454.3 confirm a complex, evolving, core-jet structure.  We
identified at each epoch a component, C1, near the easternmost edge of
the core region.  Simulations of the core region showed that C1 is
located, on average, $0.18 \pm 0.06$~mas west of the unresolved
``core'' identified in 43~GHz images.  We also identified in 3C~454.3
at 8.4~GHz several additional components which moved away from C1 with
proper motions ranging in magnitude between 0.9\,$c$ and 5\,$c$.  The
detailed motions of the components exhibit two distinct bends in the
jet axis located $\sim$3 and $\sim$5.5~mas west of C1.  The spectra
between 5.0 and 15.4~GHz for the ``moving'' components are steeper
than that for C1.  The 8.4~GHz images of B2250+194 and B2252+172, in
contrast to those of 3C~454.3, reveal compact structures.  The
spectrum between 5.0 and 15.4~GHz for B2250+194 is inverted while that
for B2252+172 is flat.

Based on its position near the easternmost edge of the 8.4~GHz radio
structure, close spatial association with the 43~GHz core, and
relatively flat spectrum, we believe 3C~454.3 component C1 to be the
best choice for the ultimate reference point for the \GPB\ guide
star.  The compact structures and inverted to flat spectra of
B2250+194 and B2252+172 make these objects valuable secondary
reference sources.

\end{abstract}

\section{Introduction \label{intro}}

{\em Gravity Probe B} (\GPB\@) is the spaceborne relativity experiment
developed by NASA and Stanford university to test two predictions of
general relativity (GR). The experiment used four superconducting
gyroscopes, contained in a low-altitude, polar orbiting spacecraft, to
measure the geodetic effect and the much smaller frame-dragging
effect. According to GR, each of these effects induces precessions of
the gyroscopes. The predicted precession is 6.6 arcsec\,yr$^{-1}$\@
for the geodetic effect and 39 milliarcseconds\,yr$^{-1}$ (\masyr) for
the frame-dragging effect. \GPB\ was designed to measure the
precessions with a standard error $<$0.5 \masyr\ relative to distant
inertial space.  This space may be closely linked to the extragalactic
International Celestial Reference Frame (ICRF), which is based on
radio very-long-baseline interferometry (VLBI) measurements
\citep{Ma+1998,Fey+2004}. Because of technical limitations, the
spacecraft could not measure the precessions relative to such a frame
directly, but rather only to an optically bright star.  The proper
motion of this chosen ``guide star,'' \objectname[]{IM~Pegasi}
(=HR~8703; HD~216489; FK5~3829), was independently determined using
VLBI measurements made relative to a fiducial point in the quasar
\objectname[]{3C~454.3} (=4C~15.76; QSO J2253+1608).  Two other radio
sources, \objectname[]{B2250+194} and \objectname[]{B2252+172}, are
used as secondary reference sources to set a limit on the stability of
the fiducial point.  The radio structures of these three reference
sources and their possible temporal changes are the subject of this
paper.  It is one of a series of seven papers reporting on the
astronomical support for \GPB\@.  For an overview of the paper series
see Paper~I \citep{GPB-I}.

The quasar 3C~454.3 \citep[$z = 0.859$;][]{Schmidt1968} is an
optically violent variable, and one of the brightest extragalactic
radio sources \citep*[see, e.g.,][]{StickelMK1994}.  Its
milliarcsecond-scale structure has been extensively studied with VLBI.
At radio frequencies between 1.7~GHz and 15~GHz, the structure of
3C~454.3 consists of a bright, relatively compact (unresolved) core
region and an extended one-sided jet which is directed west-southwest
near the core region and bends to the northwest $\sim$5~mas from the
core region (\citealt{Pauliny-Toth+1987}; \citealt{Bondi+1996};
\citealt{CawthorneG1996}; \citealt{Kellermann+1998};
\citealt{Pauliny-Toth1998}; \citealt{Chen+2005};
\citealt{Lister+2009}).  The $\sim$5~yr VLBI observing program at
10.7~GHz of \citet{Pauliny-Toth+1987} traced the motions of four
compact components which emerged from the core region and moved at
superluminal apparent speeds in the range 5--9\,$c$ along the curved
path of the jet.\footnote{Throughout this paper we assume a Hubble
constant $H_0 = 70\ \kmsMpc$ and a Friedmann-Robertson-Walker
cosmology with $\Omega_M = 0.27$ and $\Omega_{\Lambda} = 0.73$.}  The
results at 5 and 8.4~GHz by \citet{Pauliny-Toth1998} give for one
particular component a mean motion between 1983 and 1991 of
$\sim$17\,$c$.  At frequencies of 22 and 43~GHz, the core region
itself is resolved into two principal components oriented
approximately east-west and separated by $\sim$0.6~mas
(\citealt*{KemballDP1996}; \citealt{Marscher1998};
\citealt*{GomezMA1999}; \citealt{Jorstad+2001b}; \citealt{Chen+2005};
\citealt{Jorstad+2005}).  Moreover, observations at 43~GHz reveal that
new components are occasionally ejected from the easternmost
core-region component and move toward the westernmost core-region
component with superluminal speeds anywhere in the range 3--13\,$c$
\citep{Jorstad+2001b,Jorstad+2005}.  The easternmost component is very
compact at 43~GHz ($<$0.1~mas full-width-at-half-maximum: FWHM) and is
considered at that frequency to be the ``core,'' though its
stationarity with respect to a celestial reference frame (e.g., the
ICRF) has not been determined.  The stationarity of a spatially
related component at 8.4~GHz has recently been determined relative to
an updated celestial reference frame (CRF), defined by $\sim$4000
compact extragalactic radio sources (see Paper~III;
\citealt{GPB-III}).

The radio structure of quasar B2250+194 \citep[$z =
0.284$;][]{Snellen+2002} is not nearly so well studied as that of
3C~454.3.  In fact, only a handful of VLBI images of B2250+194 have
been produced by others, and mostly as part of the United States Naval
Observatory's astrometry program and the Very Long Baseline Array's
(VLBA's) calibrator survey \citep{Beasley+2002}.  These
images show that the source is only partially resolved at both 2.3
and 8.4~GHz on intercontinental baselines.

The radio source B2252+172 is catalogued in the National Radio
Astronomy Observatory Very Large Array Sky Survey
\citep[NVSS;][]{Condon+1998}, but has not previously been studied with
VLBI.  A spectrum for its optical counterpart has not yet been
obtained, so its distance is not yet known.

In \S~\ref{select}, we briefly summarize the process of selecting the
extragalactic reference sources for our astrometry of the \GPB\ guide
star.  In \S~\ref{obs}, we describe our 35 sessions of VLBI
observations, made between 1997 January and 2005 July.  In
\S~\ref{reduc}, we describe our data reduction and analysis
procedures.  In \S~\ref{3c454}, we show the 35 images made for quasar
3C~454.3 at 8.4~GHz, use model fitting to investigate the relative
motions of the observed 8.4~GHz components, and present images made
from VLBI observations at 5.0 and 15.4~GHz obtained on 2004 May~18.
In \S~\ref{2250}, we show the 35 images made for quasar B2250+194 at
8.4~GHz, study its radio structure via model fitting, and present
images made from VLBI observations at 5.0 and 15.4~GHz on 2004 May~18.
In \S~\ref{2252}, we show the 12 images made for radio source
B2252+172 at 8.4~GHz (which we observed starting only in 2002
November), investigate its radio structure via model fitting, and show
the images made from the 5.0 and 15.4~GHz VLBI observations on 2004
May~18.  We discuss our results in \S~\ref{discuss} and give our
conclusions in \S~\ref{conclus}.

\section{The Selection of Extragalactic Reference Sources for the \GPB\ Guide Star \label{select}}

The principal astrometric requirements for our VLBI reference sources
are that they be (1) radio bright and compact, (2) in close proximity
on the sky to IM~Peg, and (3) extragalactic in origin, with a core, or
other structural reference point, stationary on the sky to a very high
degree.  These requirements are consistent with those of targeted
astrometric VLBI experiments employing phase referencing \citep[see,
e.g.,][]{Lestrade+1990}.  To summarize briefly, phase referencing
requires that the reference source be detected in scans $\sim$1~minute
or less in length, so that the interferometric phase delay, which
changes with time, often very rapidly, can be successfully
interpolated to scans of the target source.  For this reason, the
phase reference source must have a correlated flux density
$\gtrsim$100~mJy, i.e., a flux density of at least this magnitude
arising from a sufficiently compact emission region.  Similarly, since
the phase delay can change significantly with pointing direction, due
mainly to radio propagation through the troposphere and ionosphere,
the reference source should be located within $\sim$1\arcdeg\ of the
target source.  Finally, if the morphology of the reference source is
variable between observing sessions, then a point must be identified
within the source image at every session that may confidently be
assumed to be a (nearly) stationary feature of the source over the
time spanned by our observing program.

\begin{figure}
\plotone{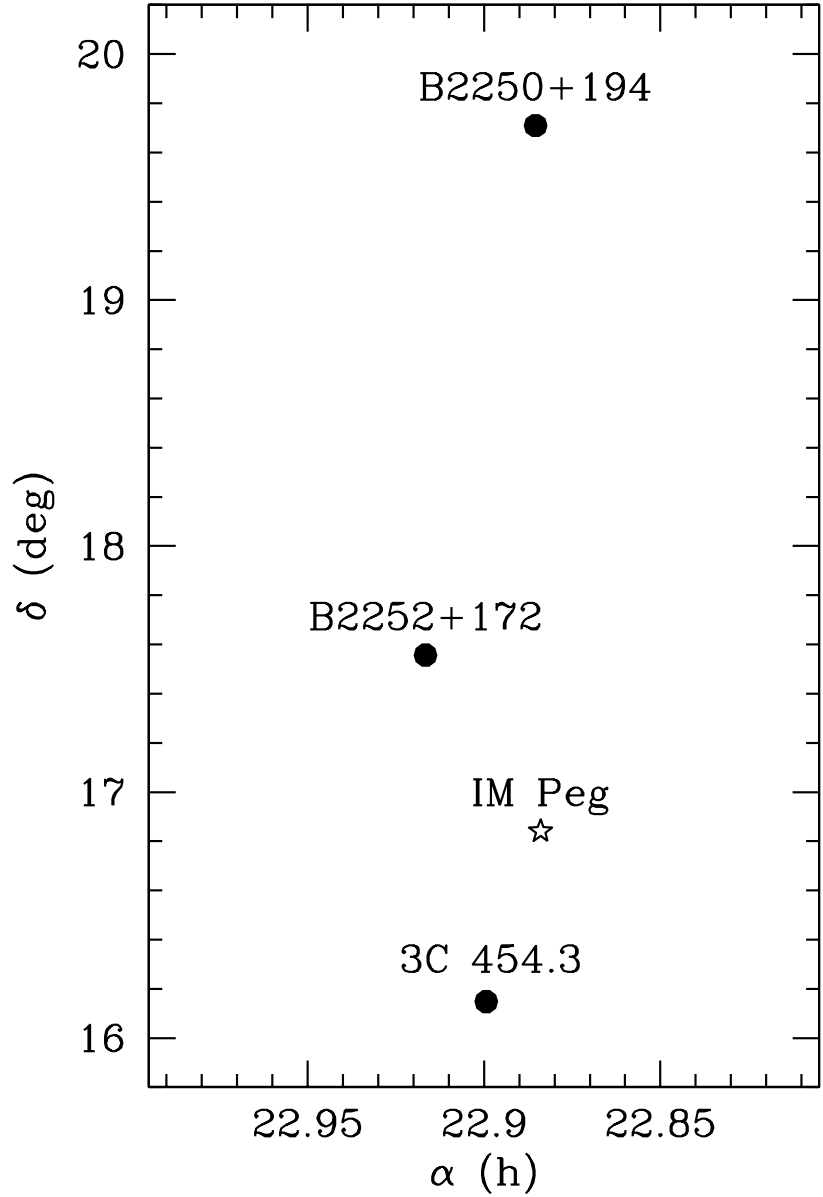} 
\figcaption{Positions (J2000) on the sky of the guide star, IM~Peg,
and the three reference sources.  The linear scales are the same for
$\alpha$ and $\delta$.  Here and in all images, north is up and east
to the left.
\label{foursourcepos}}
\end{figure}

The quasar 3C~454.3 is the only radio source which has both a
correlated flux density $>$100~mJy and a relatively compact,
milliarcsecond-scale structure within 1\arcdeg\ of IM~Peg (see
Figure~\ref{foursourcepos}).  Furthermore, 3C~454.3 was used as the
only reference source for IM~Peg in the phase-referenced mapping VLBI
observations made for the Hipparcos frame-tie project from 1991 to
1994 \citep[see][]{Lestrade+1999}.  Our continued observation of
3C~454.3 therefore provides a long time baseline for the guide-star
proper-motion measurement.  On the other hand, 3C~454.3 is a classic
superluminal source with, in general, moving jet components.  Without
the benefit of a detailed and evolving model of its radio structure,
the position accuracy of 3C~454.3 is limited \citep[in the ICRF
of][]{Ma+1998} to $\sim$1.5~mas, which is not quite sufficient to
yield proper motions to within the $\sim$0.1 \masyr\ standard error
desired for \GPB.  However, since our astrometric VLBI observations
were designed so that we could produce high quality images of our
reference sources, we can identify a point in 3C~454.3 which is
stationary on the sky to a much higher degree.  All things considered,
3C~454.3 was the best choice to be our phase reference and main
astrometric reference source.

The quasar B2250+194 is the next closest radio source on the sky to
IM~Peg which has a flux density $>$100~mJy and is compact on
milliarcsecond scales.  Indeed, B2250+194 has a more compact structure
and displays more limited structural changes than 3C~454.3, as
evidenced by the accuracy to which its position is measured in the
ICRF \citep[$\sim$0.3~mas; see][]{Fey+2004}.  However, at a separation
on the sky from IM~Peg of 2.9\arcdeg\ (see
Figure~\ref{foursourcepos}), its use as the primary astrometric
reference could have introduced large errors in the stellar
proper-motion measurement \citep*[see, e.g.,][]{PradelCL2006}.
Consequently, on the basis of its compact structure and ``reasonably''
small sky separation from IM~Peg, we selected B2250+194 to be a
secondary reference source.

The radio source B2252+172 is very compact and favorably located
within $\sim$2\arcdeg\ of, and approximately along the same
north-south axis as, 3C~454.3, B2250+194, and IM~Peg (see
Figure~\ref{foursourcepos}).  B2252+172 is relatively weak
($\sim$20~mJy at 8.4~GHz), and was discovered only recently by the
NVSS.  In light of its low centimetric spectral index, we assume that
B2252+172 is also extragalactic, even though its $R = 24$ optical
counterpart is so faint that no spectrum has yet been obtained for it.
We selected B2252+172 to serve as an additional reference source and
astrometric check source, since its north-south alignment with
3C~454.3, B2250+194, and IM~Peg helps to better model
propagation-medium effects.

\section{VLBI Observations \label{obs}}

We carried out 35 sessions of 8.4~GHz ($\lambda = 3.6$~cm) VLBI
observations of the \GPB\ guide star IM~Peg between 1997 January and
2005 July using a global VLBI array of between 12 and 16 telescopes
(see below).  The considerations governing the scheduling of the
observing sessions are outlined in Paper~V \citep{GPB-V}.  For each
session, we interleaved observations (``scans'') of IM~Peg with those
of our references sources so that we could employ the
phase-referencing technique
\citep[e.g.,][]{Shapiro+1979,Bartel+1986,Lestrade+1990,BeasleyC1995}
to determine an accurate astrometric position for IM~Peg.  The phase
reference source for all sessions was 3C~454.3.  B2250+194 was also
observed during all sessions.  B2252+172 was included for the final 12
sessions, starting in 2002 November.  We give the sky coordinates for
each reference source in Paper~III.

Our global VLBI array for each set of observations included the ten
25-m VLBA telescopes and, for increased sensitivity to the relatively
weak IM~Peg (flux density 0.2--80~mJy; see Paper~V), two or more of
the following large-aperture telescopes: the phased Very Large Array
(VLA), the 100-m Effelsberg telescope, the 43-m Green Bank telescope,
the 46-m Algonquin Radio Observatory telescope, and the three 70-m
NASA-JPL Deep Space Network (DSN) telescopes.  The total observing
time after data editing and calibration for each session was 11--15
hr.  During each session, we switched between IM~Peg and each of the
reference sources with a cycle time of between 5.5 and 7.3 minutes, in
which IM~Peg was observed for a minimum of 120 seconds, 3C~454.3 and
B2250+194 for approximately 60 seconds each, and B2252+172 (when
included) for approximately 75 seconds.  The data were recorded with
either the VLBA or Mark III system with sampling rates of 112, 128, or
256 $\rm{Mbits}\,\rm{s}^{-1}$, and were correlated with the NRAO VLBA
processor in Socorro, New Mexico.  At most epochs, both right-hand and
left-hand circular polarizations were recorded.  The specific
characteristics of each set of observations are given in
Table~\ref{antab}.

For our session on 2004 May~18, we observed IM~Peg and the three
reference sources at 5.0~GHz ($\lambda = 6.0$~cm) and 15.4~GHz
($\lambda = 2.0$~cm) in addition to using our standard frequency of
8.4~GHz.  We observed for $\sim$2~hr at 8.4~GHz at the start of the
observations, alternated $\sim$40~minute observing cycles at each of
the three frequencies for the next $\sim$8~hr, observed at 8.4~GHz
again for an extended period of $\sim$4~hr, and returned to the
$\sim$40~minute cycle of three frequencies for the remaining
$\sim$2~hr.

We used the VLA as a stand-alone interferometer (in addition to using
it as an element of our VLBI array) for most of our sessions in order
to monitor the variability of the total flux density and total
circular polarization of IM~Peg over the course of the 11--15 hr
observing sessions.  To set the flux-density scale for each set of VLA
observations, we used one or both of the flux-density calibrator
sources 3C~286 and 3C~48, with flux densities defined on the scale of
\citet{Baars+1977}.  We assume a standard error of $\sim$5\% in the
VLA-determined flux densities at each session.  The range of values
for the total flux density of IM~Peg for each session, as well as the
radio light curves in total intensity and total fractional circular
polarization for a selected sample of sessions, are presented in
Paper~VII \citep{GPB-VII}.  The flux densities\footnote{B2250+194 and
B2252+172 are unresolved by the VLA in all configurations, and we take
as the total flux density of each source the peak flux density in the
VLA interferometric image.  In contrast, 3C~454.3 has an
arcsecond-scale jet which is resolved by the VLA in the A, B, and C
configurations.  For observations made in these more extended
configurations, we take as the flux density of 3C~454.3 the peak flux
density of the core component in the VLA interferometric image.  In
the D configuration, we take as the flux density of 3C~454.3 the peak
flux density in the VLA interferometric image, which includes
contributions from both the variable core component and the
$\sim$0.2~mJy arcsecond-scale jet.} of reference sources 3C~454.3,
B2250+194, and B2252+172 for each session are given in
Table~\ref{fluxdensities}.

\begin{deluxetable}{l@{~}l@{~}l@{~~~~}l@{~~~}l@{~~~}l@{~~~}l@{~~~}l@{~~~}l@{~~~}l@{~~~}l@{~~~}l@{~~~}l@{~~~}l@{~~~}l@{~~~}l@{~~~}l@{~~~}l@{~~~}l@{~~~}l c c c c}
\tabletypesize{\scriptsize}
\rotate
\tablecaption{8.4~GHz VLBI Observations of the \GPB\ Guide Star and Reference Sources \label{antab}}
\tablewidth{0pt}
\tablehead{
\multicolumn{3}{c}{Start Date} &
\multicolumn{17}{c}{Telescope\tablenotemark{a}} &
\colhead{Total} &
\colhead{Cycle} &
\colhead{Recording} &
\colhead{Circ.} \\
\multicolumn{3}{c}{} &
\multicolumn{17}{c}{~} &
\colhead{Time\tablenotemark{b}} &
\colhead{Time} &
\colhead{Mode\tablenotemark{c}} &
\colhead{Pol.\tablenotemark{d}} \\
\multicolumn{3}{c}{} & Br & Fd & Hn & Kp & La & Mk & Nl & Ov & Pt & Sc & Y  & Aq & Eb & Gb & Go & Ro & Ti & \colhead{(hr)} & \colhead{(min)} & &
}
\startdata
1997 & Jan & 16 & X  & X  & X  & X  & X  & X  & X  & X  & X  & X  & X  & X  & X  &    & X  & X  &    & 14.2 & 7.1  & III-A   & R \\
1997 & Jan & 18 & X  & X  & X  & X  & X  & X  & X  & X  & X  & X  & X  & X  & X  & X  & X  & X  &    & 14.2 & 7.1  & III-A   & R \\
1997 & Nov & 29 & X  & X  & X  & X  & X  & X  & X  & X  & X  & X  & X  &    &    &    & X  & X  &    & 11.7 & 5.5  & 128-4-1 & R,L \\
1997 & Dec & 21 & X  & X  & X  & X  & X  & X  & X  & X  & X  & X  & X  &    &    &    & X  & X  &    & 14.8 & 5.5  & 128-4-1 & R,L \\
1997 & Dec & 27 & X  & X  & X  & X  & X  & X  & X  & X  & X  & X  & X  &    &    &    & X  & X  &    & 12.5 & 5.5  & 128-4-1 & R,L \\
1998 & Mar & 01 & X  & X  & X  & X  & X  & X  & X  & X  & X  & X  & X  &    &    &    & X  & X  & X  & 14.9 & 5.5  & 128-4-1 & R,L \\
1998 & Jul & 12 & X  & X  & X  & X  & X  & X  & X  & X  & X  & X  & X  &    &    &    & X  & X  & X  & 15.1 & 5.5  & 128-4-1 & R,L \\
1998 & Aug & 08 & X  & X  & X  & X  & X  & X  & X  & X  & X  & X  & X  &    & X  &    & X  & X  & X  & 15.1 & 5.5  & 128-4-1 & R,L \\
1998 & Sep & 16 & X  & X  & X  & X  & X  & X  & X  & X  & X  & X  & X  &    & X  &    & X  & X  & X  & 15.1 & 5.5  & 128-4-1 & R,L \\
1999 & Mar & 13 & X  & X  & X  & X  & X  & X  & X  & X  & X  & X  & X  &    & X  &    & X  & X  &    & 15.1 & 5.5  & 128-4-1 & R,L \\
1999 & May & 15 & X  & X  & X  & X  & X  & X  & X  & X  & X  & X  & X  &    & X  &    & X  & X  & X  & 15.1 & 5.5  & 128-4-1 & R,L \\
1999 & Sep & 18 & X  & X  & X  & X  & X  & X  & X  & X  & X  & X  & X  &    & X  &    & X  &    & X  & 15.2 & 5.5  & 128-4-1 & R,L \\
1999 & Dec & 09 & X  & X  & X  & X  & X  & X  & X  & X  & X  & X  & X  &    & X  &    & X  & X  &    & 14.5 & 5.5  & 128-4-1 & R,L \\
2000 & May & 15 & X  & X  & X  & X  &    & X  & X  & X  & X  & X  & X  &    & X  &    &    & X  & X  & 14.6 & 5.5  & 128-4-1 & R,L \\
2000 & Aug & 07 & X  & X  & X  & X  & X  & X  & X  & X  & X  & X  & X  &    & X  &    & X  & X  &    & 15.1 & 5.5  & 128-4-1 & R,L \\
2000 & Nov & 05 & X  & X  & X  & X  & X  & X  & X  & X  & X  & X  & X  &    & X  &    & X  & X  & X  & 15.1 & 5.5  & 128-4-1 & R,L \\
2000 & Nov & 06 & X  & X  & X  & X  & X  & X  & X  & X  & X  & X  & X  &    &    &    & X  &    &    & 13.3 & 5.5  & 128-4-1 & R,L \\
2001 & Mar & 31 & X  & X  &    & X  & X  & X  & X  & X  & X  & X  & X  &    & X  &    & X  & X  & X  & 15.2 & 5.5  & 128-4-1 & R,L \\
2001 & Jun & 29 & X  & X  & X  & X  & X  & X  & X  & X  & X  & X  & X  &    &    &    & X  & X  & X  & 13.9 & 5.5  & 128-4-1 & R,L \\
2001 & Oct & 19 & X  & X  & X  & X  & X  & X  & X  & X  & X  & X  & X  &    & X  &    & X  & X  & X  & 14.9 & 5.5  & 128-4-1 & R,L \\
2001 & Dec & 21 & X  & X  & X  & X  & X  & X  & X  & X  & X  & X  &    &    & X  &    & X  &    & X  & 15.1 & 5.5  & 128-4-1 & R,L \\
2002 & Apr & 14 & X  & X  & X  & X  & X  & X  & X  &    & X  &    & X  &    & X  &    & X  & X  & X  & 13.9 & 5.5  & 128-4-1 & R,L \\
2002 & Jul & 14 & X  & X  & X  & X  & X  & X  & X  & X  & X  & X  &    &    & X  &    & X  & X  & X  & 14.8 & 5.5  & 128-4-1 & R,L \\
2002 & Nov & 20 & X  & X  & X  & X  & X  & X  & X  & X  & X  & X  &    &    &    &    & X  & X  &    & 10.8 & 7.3  & 128-4-1 & R,L \\
2003 & Jan & 26 & X  & X  & X  & X  & X  & X  & X  & X  & X  & X  & X  &    & X  &    & X  & X  &    & 14.2 & 7.3  & 128-4-1 & R,L \\
2003 & May & 18 & X  & X  & X  & X  & X  & X  & X  & X  & X  & X  & X  &    & X  &    & X  & X  & X  & 15.2 & 7.3  & 128-4-1 & R,L \\
2003 & Sep & 08 & X  & X  & X  & X  & X  & X  & X  & X  & X  & X  & X  &    & X  &    & X  & X  & X  & 15.1 & 7.3  & 128-4-1 & R,L \\
2003 & Dec & 05 & X  & X  & X  & X  & X  & X  & X  & X  & X  & X  &    &    & X  &    & X  & X  & X  & 15.2 & 11.0\tablenotemark{e} & 128-4-1 & R,L \\
2004 & Mar & 06 & X  & X  & X  & X  & X  & X  & X  & X  & X  & X  & X  &    & X  &    & X  & X  & X  & 15.2 & 11.0 & 128-4-1 & R,L \\
2004 & May & 18 & X  & X  & X  & X  & X  & X  & X  & X  &    & X  & X  &    & X  &    & X  & X  & X  & 15.2 & 11.0 & 128-4-1 & R,L \\
2004 & Jun & 26 & X  & X  & X  & X  & X  & X  & X  & X  & X  & X  & X  &    &    &    & X  & X  & X  & 15.2 & 11.0 & 128-4-1 & R,L \\
2004 & Dec & 11 & X  & X  & X  & X  & X  & X  & X  & X  & X  & X  & X  &    & X  &    &    & X  & X  & 15.1 & 11.0 & 128-4-1 & R,L \\
2005 & Jan & 15 & X  & X  & X  & X  & X  & X  & X  & X  & X  & X  & X  &    & X  &    & X  & X  & X  & 13.4 & 11.0 & 256-4-2 & R,L \\
2005 & May & 28 & X  &    & X  & X  & X  & X  & X  & X  & X  & X  & X  &    & X  &    & X  & X  & X  & 13.3 & 11.0 & 256-4-2 & R,L \\
2005 & Jul & 16 & X  & X  & X  & X  & X  & X  & X  & X  & X  & X  & X  &    & X  &    & X  & X  & X  & 13.3 & 11.0 & 256-4-2 & R,L \\
\enddata
\tablenotetext{a}{
  Br = 25\,m, NRAO, Brewster, WA, USA;
  Fd = 25\,m, NRAO, Fort Davis, TX, USA;
  Hn = 25\,m, NRAO, Hancock, NH, USA;
  Kp = 25\,m, NRAO, Kitt Peak, AZ, USA;
  La = 25\,m, NRAO, Los Alamos, NM, USA;
  Mk = 25\,m, NRAO, Mauna Kea, HI, USA;
  Nl = 25\,m, NRAO, North Liberty, IA, USA;
  Ov = 25\,m, NRAO, Owens Valley, CA, USA;
  Pt = 25\,m, NRAO, Pie Town, NM, USA;
  Sc = 25\,m, NRAO, St. Croix, Virgin Islands, USA;
  Y = phased VLA equivalent diameter 130\,m, NRAO, near Socorro, NM, USA;
  Aq = 46\,m, ISTS (now CRESTech/York U.), Algonquin Park, Ontario, Canada;
  Eb = 100\,m, MPIfR, Effelsberg, Germany;
  Gb = 43\,m, NRAO, Green Bank, WV, USA;
  Go = 70\,m, NASA-JPL, Goldstone, CA, USA;
  Ro = 70\,m, NASA-JPL, Robledo, Spain;
  Ti = 70\,m, NASA-JPL, Tidbinbilla, Australia.
  }
\tablenotetext{b}{Session duration (including telescope slew time) after data editing.}
\tablenotetext{c}{Recording mode: III-A = Mk III mode A, 56~MHz
recorded; 128-4-1 = VLBA format, 128 Mbps recorded in 4 baseband
channels with 1-bit sampling; 256-4-2 = VLBA format, 256 Mbps recorded
in 4 baseband channels with 2-bit sampling.}
\tablenotetext{d}{The sense of circular polarization recorded: R =
right and L = left circular polarization (IEEE convention).}
\tablenotetext{e}{For the last eight sessions, we employed an
11.0-minute major cycle consisting of two minor cycles, one 6.3
minutes in length and including all four sources (3C~454.3, IM~Peg,
B2250+194, and B2252+172) and one 4.7 minutes in length and including
all sources except 2252+172.}
\end{deluxetable}

\begin{deluxetable}{l@{~}l@{~}l@{\,}r@{\hspace{70pt}}r@{\,}r@{\,}r@{}l@{\hspace{70pt}}c@{\hspace{70pt}}c}
\tabletypesize{\scriptsize}
\tablecaption{Flux Densities\tablenotemark{a} of 3C~454.3, B2250+194, and B2252+172 \label{fluxdensities}}
\tablewidth{0pt}
\tablehead{
  \multicolumn{3}{c}{Start Date} &
  \colhead{} &
  \multicolumn{3}{c}{3C~454.3\tablenotemark{b}} &
  \colhead{} &
  \colhead{B2250+194} &
  \colhead{B2252+172} \\
  \multicolumn{3}{c}{} &
  \colhead{} &
  \multicolumn{3}{c}{(Jy)} &
  \colhead{} &
  \colhead{(mJy)} &
  \colhead{(mJy)}
}
\startdata
1997 & Jan & 16 &                      & 12.16 & $\pm$ & 0.65 &     & 438\,$\pm$\,22 & \\
1997 & Jan & 18 &                      & 12.06 & $\pm$ & 0.64 &     & 430\,$\pm$\,21 & \\
1997 & Nov & 29 &                      & 11.31 & $\pm$ & 0.57 & (D) & 469\,$\pm$\,24 & \\
1997 & Dec & 21 &                      & 10.54 & $\pm$ & 0.54 & (D) & 451\,$\pm$\,23 & \\
1997 & Dec & 27 &                      & 10.01 & $\pm$ & 0.51 & (D) & 432\,$\pm$\,22 & \\
1998 & Mar & 01 &                      & 10.84 & $\pm$ & 0.56 &     & 488\,$\pm$\,24 & \\
1998 & Jul & 12 &                      & 11.80 & $\pm$ & 0.60 &     & 447\,$\pm$\,22 & \\
1998 & Aug & 08 &                      & 11.49 & $\pm$ & 0.61 &     & 417\,$\pm$\,24 & \\
1998 & Sep & 16 &                      & 11.46 & $\pm$ & 0.61 &     & 455\,$\pm$\,23 & \\
1999 & Mar & 13 &                      & 11.16 & $\pm$ & 0.58 & (D) & 432\,$\pm$\,23 & \\
1999 & May & 15 &                      & 11.09 & $\pm$ & 0.57 & (D) & 455\,$\pm$\,23 & \\
1999 & Sep & 18 &                      & 12.05 & $\pm$ & 0.74 &     & 418\,$\pm$\,21 & \\
1999 & Dec & 09 &                      & 10.74 & $\pm$ & 0.55 &     & 468\,$\pm$\,23 & \\
2000 & May & 15 &                      &  9.86 & $\pm$ & 0.52 &     & 433\,$\pm$\,22 & \\
2000 & Aug & 07 &                      &  9.13 & $\pm$ & 0.47 & (D) & 420\,$\pm$\,21 & \\
2000 & Nov & 05 &                      &  9.14 & $\pm$ & 0.46 &     & 436\,$\pm$\,22 & \\
2000 & Nov & 06 &                      &  9.12 & $\pm$ & 0.46 &     & 441\,$\pm$\,22 & \\
2001 & Mar & 31 &                      & 11.17 & $\pm$ & 0.59 &     & 443\,$\pm$\,24 & \\
2001 & Jun & 29 &                      & 10.92 & $\pm$ & 0.56 &     & 395\,$\pm$\,20 & \\
2001 & Oct & 19 &                      &  9.70 & $\pm$ & 0.49 & (D) & 380\,$\pm$\,19 & \\
2001 & Dec & 21 & (*)                  &  9.25 & $\pm$ & 0.93 &     & 406\,$\pm$\,41 & \\
2002 & Apr & 14 &                      &  9.30 & $\pm$ & 0.47 &     & 359\,$\pm$\,18 & \\
2002 & Jul & 14 & (*)                  &  9.32 & $\pm$ & 0.93 &     & 354\,$\pm$\,35 & \\
2002 & Nov & 20 & (*)                  &  9.47 & $\pm$ & 0.95 &     & 396\,$\pm$\,40 & 18.0\,$\pm$\,1.9 \\
2003 & Jan & 26 &                      & 10.80 & $\pm$ & 0.54 &     & 421\,$\pm$\,21 & 25.0\,$\pm$\,1.3 \\
2003 & May & 18 &                      & 10.42 & $\pm$ & 0.54 &     & 428\,$\pm$\,22 & 27.0\,$\pm$\,1.4 \\
2003 & Sep & 08 &                      &  9.66 & $\pm$ & 0.48 &     & 480\,$\pm$\,24 & 21.6\,$\pm$\,1.1 \\
2003 & Dec & 05 & (*)                  &  9.31 & $\pm$ & 0.93 &     & 488\,$\pm$\,49 & 19.1\,$\pm$\,1.9 \\
2004 & Mar & 06 &                      &  8.99 & $\pm$ & 0.45 &     & 512\,$\pm$\,26 & 17.2\,$\pm$\,0.9 \\
2004 & May & 18 &                      &  8.58 & $\pm$ & 0.43 &     & 533\,$\pm$\,27 & 16.0\,$\pm$\,0.8 \\
2004 & May & 18 & (C)\tablenotemark{c} & 10.45 & $\pm$ & 0.53 &     & 381\,$\pm$\,19 & 17.3\,$\pm$\,0.9 \\
2004 & May & 18 & (U)\tablenotemark{c} &  5.94 & $\pm$ & 0.42 &     & 561\,$\pm$\,39 & 11.5\,$\pm$\,0.8 \\
2004 & Jun & 26 &                      &  8.59 & $\pm$ & 0.43 & (D) & 525\,$\pm$\,26 & 15.9\,$\pm$\,0.8 \\
2004 & Dec & 11 &                      &  9.36 & $\pm$ & 0.47 &     & 507\,$\pm$\,26 & 18.3\,$\pm$\,0.9 \\
2005 & Jan & 15 &                      &  9.58 & $\pm$ & 0.48 &     & 531\,$\pm$\,27 & 17.5\,$\pm$\,0.9 \\
2005 & May & 28 &                      &  9.41 & $\pm$ & 0.47 &     & 498\,$\pm$\,25 & 23.0\,$\pm$\,1.2 \\
2005 & Jul & 16 &                      &  9.56 & $\pm$ & 0.48 &     & 519\,$\pm$\,25 & 26.5\,$\pm$\,1.3 \\
\enddata
\tablenotetext{a}{For all but four epochs (marked with ``*''), the
tabulated value for the flux density is that estimated with the VLA
Stokes-I data.  For the remaining epochs, the (Stokes-I) flux density
was estimated from the VLBI data.  The standard error is the
root-sum-square (rss) of the statistical standard error found in
estimating the flux density and a 5\% (10\% for VLBI) systematic error
allowed for calibration of the flux-density scale (see
\S~\ref{reduc}).}
\tablenotetext{b}{Epochs marked with a ``D'' indicate that the VLA was
in its D configuration.  The nominal value for the flux density of
3C~454.3 for these epochs includes a $\sim$0.2~mJy contribution from
the arcsecond-scale jet.}
\tablenotetext{c}{C indicates 5.0~GHz; U indicates 15.4~GHz; all other
observations were at 8.4~GHz.}
\end{deluxetable}

\section{VLBI Data Reduction and Analysis \label{reduc}}

We carried out the reduction and analysis of the VLBI data for the
three reference sources with NRAO's Astronomical Image Processing
System (AIPS).  We performed the preliminary steps in the data
reduction for each session, i.e., initial amplitude calibration (using
system-temperature measurements and antenna gain curves),
intermediate-frequency-channel bandpass and phase-offset correction,
and data editing, in the usual manner \citep[e.g.,][]{Walker1999}.  We
describe the remaining data reduction and analysis steps in some
detail below.

For each of the stronger sources, 3C~454.3 and B2250+194, we fringe
fit and then iteratively self-calibrated
\citep[e.g.,][]{CornwellF1999} and imaged the data.  For fringe
fitting, we found a model consisting of a single point source to be
adequate for determining the phase-delay rates and group delays for
all telescopes (relative to a reference telescope) for both 3C~454.3
and B2250+194.  For phase self-calibration, we found a starting model
consisting of a single point source to be adequate for B2250+194 at
all epochs, but inadequate for 3C~454.3 at most epochs.\footnote{We
judged the adequacy of the starting model by noting the number of
iterations of phase self-calibration required to achieve convergence
in the phase gains.  At some epochs, a single-point-source starting
model for 3C~454.3 would require several tens of iterations for
convergence, whereas a two-point-source starting model would always
require less than ten.} For 3C~454.3, we used, instead, a starting
model consisting of two point sources, which more accurately reflects
the internal structure of the bright core region of the source (see
\S~\ref{3c454}).  (We estimated the flux densities and relative
positions of the two point sources by first fitting the model to the
post-fringe-fit \uv\ data.)  After self-calibrating in phase, we
self-calibrated the data for each of 3C~454.3 and B2250+194 in
amplitude.  We stopped the iterative process when we achieved
convergence in both the phase and amplitude self-calibrations.  All
images of 3C~454.3 and B2250+194 were generated using the CLEAN
deconvolution algorithm \citep[see][]{Clark1980} and the robust
data-weighting scheme \citep{Briggs1995}.

For our weakest reference source, B2252+172, our fringe fitting failed
for each session to find a reliable rate and delay solution (with
signal-to-noise ratio, SNR $\geq$ 5) on 10--50\% of scans for at least
one, and usually several, telescopes (relative to a reference
telescope).  Instead we used the final image of 3C~454.3 to estimate
the phase component of the complex antenna gains at the position of
B2252+172 as a function of time (see \S~\ref{2252phaseref}), and the
images of both 3C~454.3 and B2250+194 to estimate the amplitude
component as a function of time.  (Note that, in contrast to the phase
gains, the amplitude gains are largely unaffected by inaccuracies in
the propagation-medium model.)  For each telescope, the amplitude
gains for sources 3C~454.3 and B2250+194 were identical to within 2\%,
so we smoothed them together over intervals of 30 minutes to 1~hr
before using them to fine-tune the amplitude calibration for
B2252+172.

\subsection{Phase-Referenced Imaging of B2252+172 \label{2252phaseref}}

At each session, the phase components of the complex antenna gains
were derived for every scan of 3C~454.3, with a cadence of 5.5--7.3
minutes, and interpolated to the effective time of the B2252+172
scans.  We found that phase connection between successive scans of
3C~454.3 was not possible in some instances for time ranges in which
at least one telescope on a baseline was viewing the source at an
elevation $\lesssim$20\arcdeg.  Any time range of data for which phase
connection on a given baseline was problematic was flagged but not
deleted.  We Fourier-inverted the unflagged B2252+172 data and used
the CLEAN deconvolution algorithm to produce a phase-referenced image
of the source.  Using the phase-referenced image as a starting model,
we then phase self-calibrated the full set of B2252+172 data at each
session (including the data previously flagged) over time intervals
from scan lengths to $\sim$20 minutes, depending on the flux density
of the source.  The quality of the image (judged by the ratio of peak
brightness to root-mean-square, ``rms,'' background and the ratio of
peak to minimum brightness) produced after a few iterations of phase
self-calibration was, for each session, significantly higher
($\gtrsim$50\% improvement in each ratio) than that of the strictly
phase-referenced image.  We also self-calibrated the B2252+172 data at
each session in amplitude, but found little improvement ($\lesssim$5\%
improvement in each ratio) in the resulting image over the image
produced in the phase-only self-calibration process.  The final images
for B2252+172 were produced with data that were phase self-calibrated
only.  As for 3C~454.3 and B2250+194, all images of B2252+172 were
generated using the CLEAN deconvolution algorithm and the robust
data-weighting scheme.

\subsection{Model Fitting \label{fitting}}

To derive quantitative results for each of the three reference
sources, we also fit simple brightness-distribution models to
either the image plane data using the AIPS program JMFIT, or the \uv\
plane data using the AIPS program OMFIT.  Both programs utilize
weighted least-squares algorithms.  In the case of the \uv\ plane
model fits, we first averaged the data within each individual scan.
We describe the model used in each fit, and the process of estimating
the standard errors in the fit parameters, source by source, in
\S~\ref{3c454}, \S~\ref{2250}, and \S~\ref{2252}.

\subsection{Standard Error in the VLBI Amplitude Calibration \label{VLBIamp}}

The mean ratios of VLBI-determined to VLA-determined flux densities
are $0.94 \pm 0.04$ for 3C~454.3 (see Table~\ref{3Cimstat}) and $0.97
\pm 0.04$ for B2250+194 (see Table~\ref{2250imstat}).  However, the
ratios are systematically lower (and more variable) for epochs before
2003 January, compared to those after, likely due to improved accuracy
and implementation of the system-temperature measurements and/or gain
curves for the large-aperture telescopes.  We estimate that our VLBI
array ``resolves out'' only $\sim$4\% of the total (VLA-determined)
flux density for 3C~454.3 and $\sim$3\% of the total flux density for
B2250+194.  In order to simplify our quantitative analysis, we adopt,
unless otherwise stated, a conservative standard error of 10\% for all
VLBI-determined flux densities.

\section{Quasar 3C~454.3: The Main Reference Source \label{3c454}}

\subsection{VLBI Images at 8.4~GHz}

\begin{figure}
\plotone{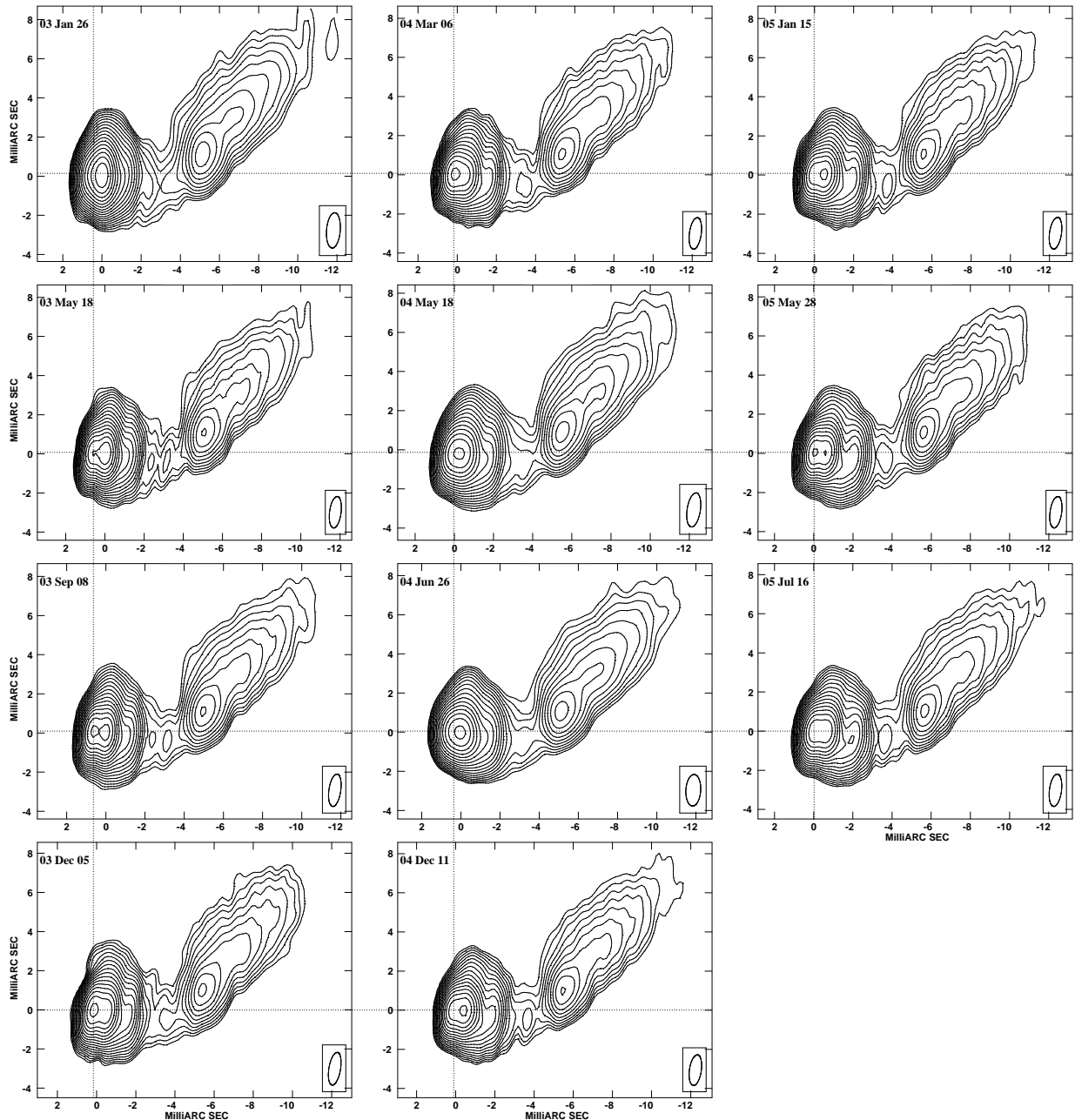}
\figcaption{8.4~GHz VLBI images of 3C~454.3 for each of our 35
observing sessions between 1997 January and 2005 July.  Contour levels
differ by factors of $\sqrt{2}$, starting in each image at
$10\,\rm{mJy}\,\rm{beam}^{-1}$.  Image characteristics are summarized
in Table~\ref{3Cimstat}.  The restoring beam is indicated in the
bottom-right-hand corner of each image.  The images are aligned, as
indicated by the dotted lines, on the center of the easternmost
component, C1 (see \S~\ref{coreregion}).
\label{3Cimages}}
\end{figure}

We present in Figure~\ref{3Cimages} the 8.4~GHz images of 3C~454.3
generated from each of the 35 sessions of VLBI observations made between
1997 January and 2005 July.  Each image is convolved with a Gaussian
(CLEAN) restoring beam of size and shape determined, as usual, from a
fit to the main lobe of the interferometer beam for that set of
observations.  The median size of the minor axis of the restoring beam
is 0.67~mas (FWHM), and the median ratio of major to minor axis is
2.5.  The image characteristics are summarized in
Table~\ref{3Cimstat}.  The images in the figure are aligned on the
center of the easternmost component of the radio structure observed at
each epoch (see \S~\ref{implanefits}).  The coordinates in each image
are relative to the phase center,\footnote{The phase center (i.e.,
position [0,0] in the image) refers to the assumed source position
used at the correlator.  During the phase self-calibration process,
the offset between the true and assumed positions, for a
flux-density-averaged point in the radio structure, is removed from
the \uv\ data.} which is generally offset from the brightness peak by
$<$0.02~mas in each of $\alpha$ and $\delta$.

\begin{figure}
\plotone{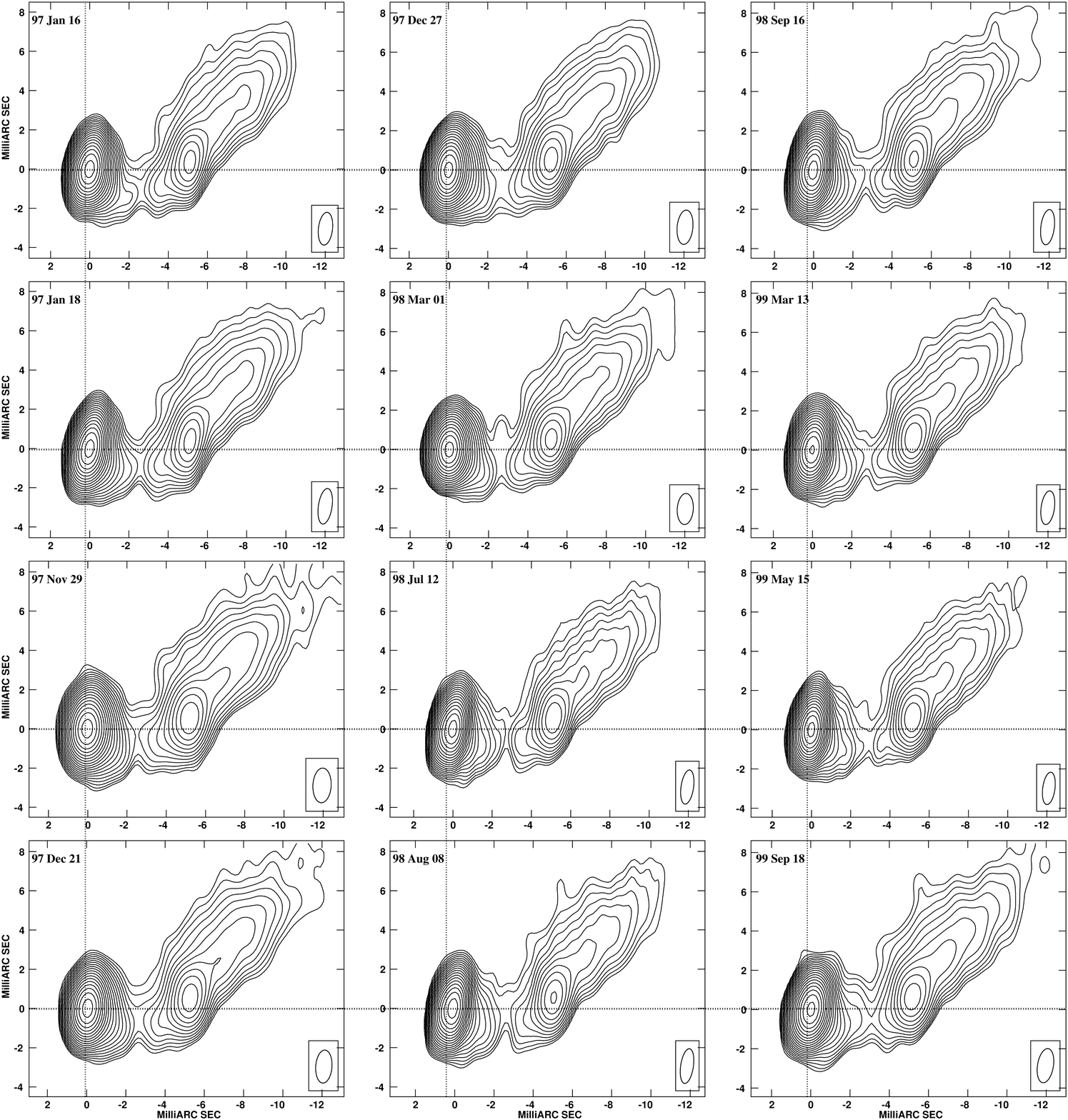}
\end{figure}
\begin{figure}
\plotone{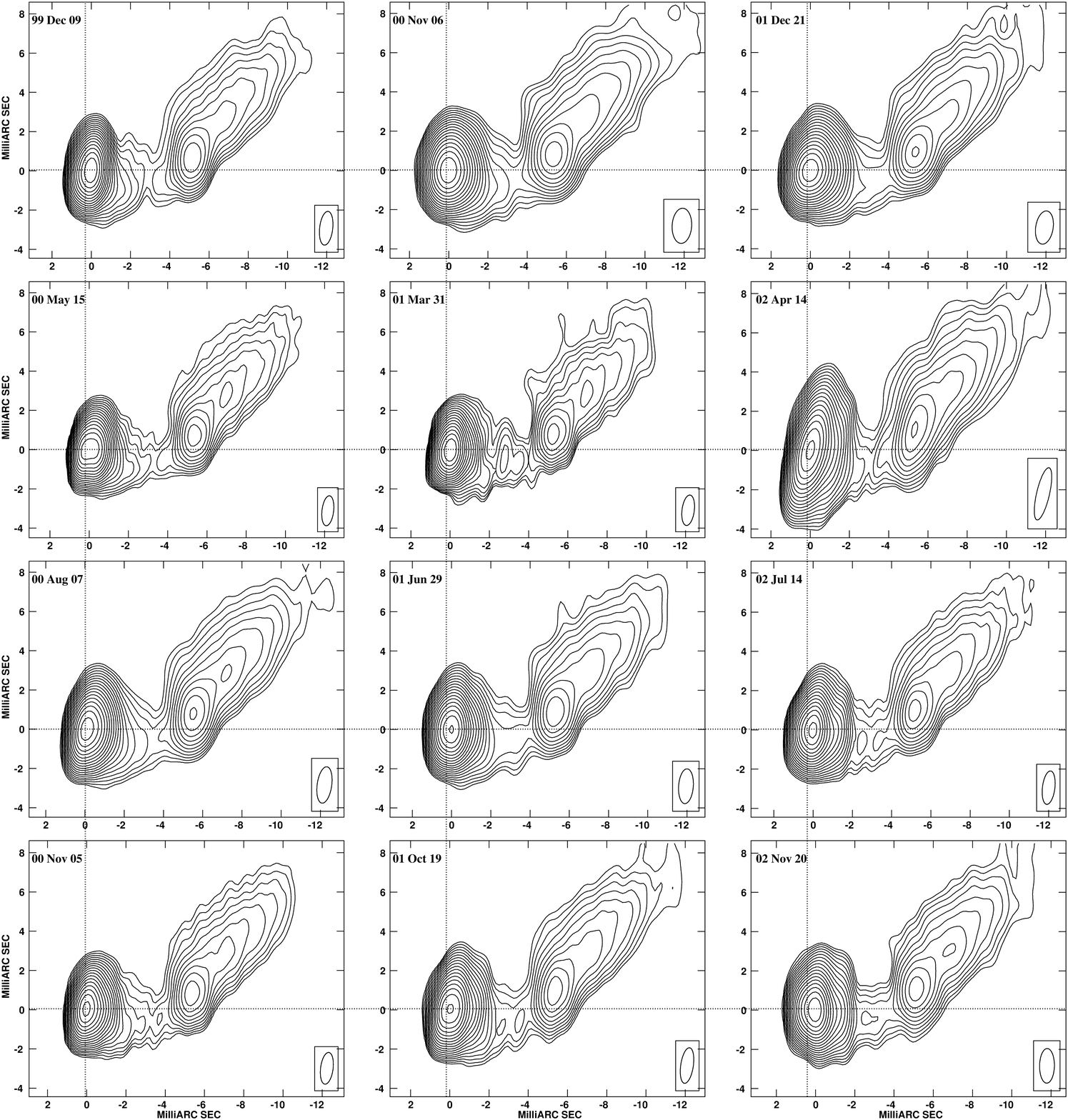}
\end{figure}

\begin{deluxetable}{l@{~}l@{~}l@{\,}r c c c c c c c c}
\tabletypesize{\scriptsize}
\tablecaption{3C~454.3 Image Characteristics \label{3Cimstat}}
\tablewidth{0pt}
\tablehead{
  \multicolumn{3}{c}{Start Date} &
  \colhead{} &
  \colhead{Peak} &
  \colhead{Min.} &
  \colhead{rms} &
  \colhead{$\Theta_{\rm{maj}}$} &
  \colhead{$\Theta_{\rm{min}}$} &
  \colhead{p.a.} &
  \colhead{CLEAN} &
  \colhead{$[\frac{\rm{CLEAN}}{\rm{VLA}}]$} \\
  \multicolumn{3}{c}{} &
  \colhead{} &
  \colhead{($\rm{Jy}\,\Omega_b^{-1}$)} &
  \colhead{($\rm{mJy}\,\Omega_b^{-1}$)} &
  \colhead{($\rm{mJy}\,\Omega_b^{-1}$)} &
  \colhead{(mas)} &
  \colhead{(mas)} &
  \colhead{($\arcdeg$)} &
  \colhead{(Jy)} &
  \colhead{} \\
  \multicolumn{3}{c}{} &
  \colhead{} &
  \colhead{[1]} &
  \colhead{[2]} &
  \colhead{[3]} &
  \colhead{[4]} &
  \colhead{[5]} &
  \colhead{[6]} &
  \colhead{[7]} &
  \colhead{[8]}
}
\startdata
1997 & Jan & 16 &                      & 6.14 & \phn$-4.87$ & 0.76    & 1.68 & 0.72 & \phn$-6.43$ & 10.70    & 0.88 \\
1997 & Jan & 18 &                      & 6.05 & \phn$-4.39$ & 0.66    & 1.83 & 0.70 & \phn$-8.43$ & 10.61    & 0.88 \\
1997 & Nov & 29 &                      & 6.20 & \phn$-5.87$ & 0.80    & 1.77 & 0.93 & \phn$-2.93$ & \phn9.72 & 0.86 \\
1997 & Dec & 21 &                      & 6.23 & \phn$-4.33$ & 0.58    & 1.68 & 0.83 & \phn$-4.08$ & 10.15    & 0.96 \\
1997 & Dec & 27 &                      & 5.91 & \phn$-3.88$ & 0.72    & 1.75 & 0.80 & \phn$-5.40$ & \phn9.54 & 0.95 \\
1998 & Mar & 01 &                      & 6.00 & \phn$-5.05$ & 0.56    & 1.59 & 0.82 & \phn$-3.80$ & \phn9.58 & 0.88 \\
1998 & Jul & 12 &                      & 6.10 & \phn$-5.93$ & 0.71    & 1.75 & 0.64 & \phn$-8.19$ & 10.48    & 0.89 \\
1998 & Aug & 08 &                      & 6.27 & \phn$-5.79$ & 0.74    & 1.78 & 0.65 & \phn$-8.36$ & 11.04    & 0.96 \\
1998 & Sep & 16 &                      & 6.18 & \phn$-5.77$ & 0.73    & 1.80 & 0.69 & \phn$-7.05$ & 10.69    & 0.93 \\
1999 & Mar & 13 &                      & 5.40 & \phn$-5.91$ & 0.59    & 1.73 & 0.68 & \phn$-7.90$ & \phn9.62 & 0.86 \\
1999 & May & 15 &                      & 5.90 & \phn$-4.39$ & 0.55    & 1.66 & 0.62 & \phn$-7.61$ & 10.41    & 0.94 \\
1999 & Sep & 18 &                      & 8.16 & \phn$-7.71$ & 1.02    & 1.76 & 0.81 & \phn$-9.28$ & 12.32    & 1.02 \\
1999 & Dec & 09 &                      & 5.04 & \phn$-5.18$ & 0.52    & 1.73 & 0.67 & \phn$-6.46$ & \phn9.46 & 0.88 \\
2000 & May & 15 &                      & 3.37 & \phn$-5.47$ & 0.77    & 1.54 & 0.59 & \phn$-6.66$ & \phn9.30 & 0.94 \\
2000 & Aug & 07 &                      & 4.60 & \phn$-3.09$ & 0.38    & 1.87 & 0.74 & \phn$-7.95$ & \phn9.31 & 1.02 \\
2000 & Nov & 05 &                      & 4.14 & \phn$-4.92$ & 0.71    & 1.63 & 0.64 & \phn$-7.20$ & \phn8.77 & 0.96 \\
2000 & Nov & 06 &                      & 4.98 & \phn$-5.18$ & 0.69    & 1.79 & 0.99 & \phn$-3.38$ & \phn8.84 & 0.97 \\
2001 & Mar & 31 &                      & 5.13 & \phn$-8.92$ & 0.90    & 1.56 & 0.63 & \phn$-7.18$ & \phn9.65 & 0.86 \\
2001 & Jun & 29 &                      & 5.32 & \phn$-5.10$ & 0.54    & 1.78 & 0.75 & \phn$-4.19$ & 10.08    & 0.92 \\
2001 & Oct & 19 &                      & 3.78 & \phn$-5.93$ & 0.88    & 1.83 & 0.64 & \phn$-8.35$ & \phn9.35 & 0.96 \\
2001 & Dec & 21 &                      & 4.78 & \phn$-3.53$ & 0.51    & 1.71 & 0.89 & \phn$-8.94$ & \phn9.25 &  --- \\
2002 & Apr & 14 &                      & 3.97 & \phn$-7.74$ & 0.89    & 2.73 & 0.66 & $-13.04$    & \phn8.67 & 0.93 \\
2002 & Jul & 14 &                      & 4.18 & \phn$-3.88$ & 0.41    & 1.71 & 0.63 & \phn$-6.91$ & \phn9.32 &  --- \\
2002 & Nov & 20 &                      & 4.90 & \phn$-7.05$ & 0.70    & 1.76 & 0.75 & \phn$-0.90$ & \phn9.47 &  --- \\
2003 & Jan & 26 &                      & 4.93 & \phn$-5.02$ & 0.70    & 1.81 & 0.70 & \phn$-4.59$ & 10.35    & 0.96 \\
2003 & May & 18 &                      & 3.54 & \phn$-5.07$ & 0.53    & 1.62 & 0.59 & \phn$-7.27$ & \phn9.80 & 0.94 \\
2003 & Sep & 08 &                      & 3.02 & \phn$-4.32$ & 0.63    & 1.67 & 0.60 & \phn$-8.15$ & \phn9.34 & 0.97 \\
2003 & Dec & 05 &                      & 2.90 & \phn$-7.79$ & 0.98    & 1.69 & 0.59 & \phn$-9.52$ & \phn9.31 &  --- \\
2004 & Mar & 06 &                      & 2.85 & \phn$-7.48$ & 0.97    & 1.62 & 0.60 & \phn$-7.87$ & \phn8.97 & 1.00 \\
2004 & May & 18 &                      & 2.73 & \phn$-4.25$ & 0.83    & 1.76 & 0.68 & \phn$-8.64$ & \phn8.41 & 0.98 \\
2004 & May & 18 & (C)\tablenotemark{a} & 3.93 & $-10.6\phn$ & 1.81    & 3.03 & 0.99 & \phn$-9.99$ & \phn9.73 & 0.93 \\
2004 & May & 18 & (U)\tablenotemark{a} & 1.13 & $-14.0\phn$ & 2.24    & 0.98 & 0.33 & \phn$-8.58$ & \phn5.82 & 0.98 \\
2004 & Jun & 26 &                      & 2.85 & \phn$-3.63$ & 0.45    & 1.61 & 0.76 & \phn$-3.30$ & \phn8.03 & 0.94 \\
2004 & Dec & 11 &                      & 2.78 & \phn$-3.02$ & 0.34    & 1.58 & 0.61 & \phn$-8.16$ & \phn8.94 & 0.96 \\
2005 & Jan & 15 &                      & 2.76 & \phn$-2.77$ & 0.41    & 1.64 & 0.61 & \phn$-8.22$ & \phn9.00 & 0.94 \\
2005 & May & 28 &                      & 2.68 & \phn$-4.94$ & 0.73    & 1.62 & 0.59 & \phn$-7.80$ & \phn9.17 & 0.98 \\
2005 & Jul & 16 &                      & 2.50 & \phn$-1.89$ & 0.23    & 1.66 & 0.62 & \phn$-7.13$ & \phn9.07 & 0.95 \\
\enddata
\tablenotetext{a}{C indicates 5.0~GHz; U indicates 15.4~GHz; all other
observations were at 8.4~GHz.}  \tablecomments{[1,2] Positive and
negative extrema ($\Omega_b$ $\equiv$ CLEAN restoring beam area); [3]
rms background level (``far'' away from source); [4--6] CLEAN
restoring beam major axis (FWHM), minor axis (FWHM), and position
angle (east of north); [7] Total CLEAN flux density; [8] Ratio of
VLBI-determined (column 7) to VLA-determined
(Table~\ref{fluxdensities}) flux density for all epochs at which the
VLA observed.  The estimated standard error in the VLBI-determined
flux densities is 10\%.}
\end{deluxetable}

Each of our VLBI images of 3C~454.3 exhibits a radio structure
generally consistent with that observed previously for this source at
frequencies between 1.7 and 15~GHz, namely a bright, relatively
compact core region and a bent milliarcsecond-scale jet.
Nevertheless, a number of structural changes are apparent over the
$\sim$8.5~yr period of our observing program.  The most significant
changes take place in the core and inner-jet regions.  The core
region, which we roughly define as the easternmost $\sim$1~mas of the
source, is visibly extended mainly east-west, ranging in appearance
from a single elongated emission region to a marginally resolved
double-peaked emission region.  We model the core region at each of
our 35 epochs with two compact components separated by 0.45--0.7~mas
(see \S~\ref{coreregion}).  The two-component, or indeed
multi-component, structure of the core and inner-jet regions of
3C~454.3 was first observed by \citet{Pauliny-Toth+1987} during their
$\sim$5~yr VLBI program at 10.7~GHz.  Their images show that the two
easternmost components have a relatively constant 0.5--0.6~mas
separation, and that others move approximately westward away from them
with superluminal apparent velocities.  Our images likewise show that
new components occasionally emerge to the west-southwest of the core
region and move approximately westward toward the extended jet.  We
discuss the motions in the inner-jet region in
\S~\ref{innerjetregion}.  The brightness peak of the jet, located
$\sim$5.5~mas west of the easternmost core-region component, also
appears to move with respect to the core region, but at a lower
apparent velocity, and northward rather than westward.  Beyond this
point, the axis of the jet bends sharply to the northwest.  We discuss
the motion of the jet peak, and extended jet as a whole beyond the
peak, in \S~\ref{outerjetregion}.

To assign a systematic error to quantities estimated in our
image-plane analysis in the next section, we also produced for each
set of observations an image (not shown) of 3C~454.3 using a common
restoring beam.  We chose a beam with major axis 2.00~mas (FWHM),
minor axis 0.80~mas (FWHM), and position angle (p.a.) 0\arcdeg\ (east
of north) to encompass the majority of the CLEAN restoring beams in
the nominal-resolution images (see Table~\ref{3Cimstat}), and to
reflect the median ratio of major to minor axis of the CLEAN restoring
beams.  For 28 epochs, the common restoring beam fully encompassed the
CLEAN restoring beam, so we simply convolved the nominal-resolution
image with the larger common restoring beam.  For five of the seven
remaining epochs, we first generated an image with a CLEAN restoring
beam slightly smaller than the common restoring beam, by increasing
the relative weights on the longer baselines during imaging, and then
convolved the new image with the common restoring beam.  For two
epochs, namely 2000 November 6 and 2002 April 14, no choice of data
weights produced a beam smaller than the common restoring beam, but we
still forced the CLEAN restoring beam to have the dimensions and
orientation of the common restoring beam (resulting in a mild
super-resolution).

\subsection{Image-Plane Model Fitting at 8.4~GHz \label{implanefits}}

\begin{figure}
\plotone{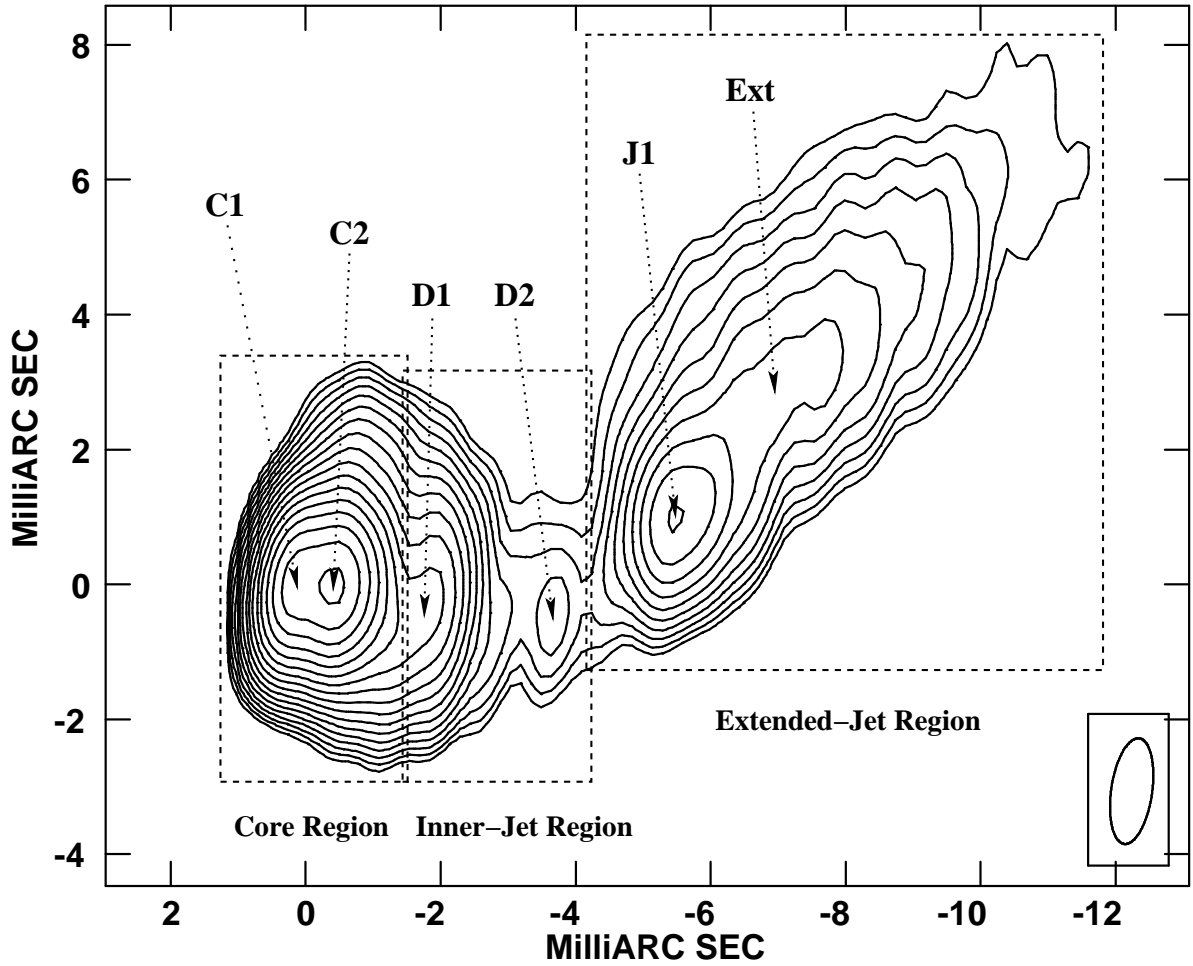} 
\figcaption{A VLBI image of 3C~454.3 (2004 December 11) showing the
approximate locations of the model components (labeled) and fitting
windows (dashed rectangles).  The restoring beam is at the lower
right.
\label{windowandcompid}}
\end{figure}

We performed a detailed analysis of the changing 8.4~GHz radio
structure of 3C~454.3 via model fitting in the image plane.  We used
for this analysis the nominal-resolution images shown in
Figure~\ref{3Cimages}.  Fitting in the image plane facilitates
dissection of the complex radio structure over its full brightness
range.  In the \uv\ plane (see \S~\ref{uvplanefits}), the bright
core-region components dominate the low-surface-brightness jet
components.  In the image plane, we can isolate structure and
investigate in turn the core and jet regions.  On the other hand,
concerns about systematic errors in the image plane analysis such as
from (CLEAN) deconvolution errors and the differing resolution at each
epoch have to be investigated and quantified as best as possible.  In
the remainder of this section, we discuss the image-plane-based model
fitting of the core, inner-jet, and extended-jet regions of 3C~454.3,
using windows allocated approximately as shown in
Figure~\ref{windowandcompid}.  For ease of reference, we also indicate
in Figure~\ref{windowandcompid} the approximate positions of the model
components discussed below.

\subsubsection{Core Region \label{coreregion}}

Several of our images of 3C~454.3, in particular those after 2003
January, show that the easternmost $\sim$1~mas of the 8.4~GHz radio
structure is marginally resolved into two components oriented
approximately east-west.  We fit to the core region at each epoch a
model consisting of two point sources, and, for comparison, a model
consisting of a single elliptical Gaussian.  The two-component model
fit the image data for the core region better (yielding $30\%$ to
$70\%$ lower values of the postfit rms residuals) at all epochs,
including those in which the core region appears only slightly
elongated in Figure~\ref{3Cimages}.  The results of the two-component
fit are given in Table~\ref{3Ccorecomp}.  We refer to the easternmost
component as C1 and the westernmost component as C2.  The tabulated
standard errors in the flux densities and positions of C1 and C2 at
each epoch include contributions, added in quadrature, from the
statistical standard error of the fit and from the systematic errors
associated with deconvolution (see \S~\ref{deconv}) and the ``choice''
of image resolution.  The last error is taken at each epoch to be the
difference between the parameter estimate from the fit to the
nominal-resolution image (Figure~\ref{3Cimages}) and the corresponding
parameter estimate from the fit to the common-resolution image.  At
every epoch, the systematic errors are $\sim$10--20 times larger than
the statistical standard errors.  Unless otherwise noted, the quoted
errors for all model parameters are standard errors, which include the
systematic-error estimate.

\begin{deluxetable}{l@{~}l@{~}l@{~~}c@{}c@{}c@{}c@{}c@{}c@{}c@{}c@{}c@{}c@{}c@{}c}
\tabletypesize{\scriptsize}
\rotate
\tablecaption{3C~454.3 Image-Plane 8.4~GHz Model Parameters: Core Region \label{3Ccorecomp}}
\tablewidth{0pt}
\tablehead{
  \multicolumn{3}{c}{Start Date} &
  \multicolumn{6}{c}{----------------------Component C1----------------------} &
  \multicolumn{6}{c}{----------------------Component C2----------------------} \\
  \multicolumn{3}{c}{} &
  \colhead{$S_{8.4}$} &
  \colhead{$\sigma_{S}$} &
  \colhead{$\Delta$$\alpha$} &
  \colhead{$\sigma_{\alpha}$} &
  \colhead{$\Delta$$\delta$} &
  \colhead{$\sigma_{\delta}$} &
  \colhead{$S_{8.4}$} &
  \colhead{$\sigma_{S}$} &
  \colhead{$\Delta$$\alpha$} &
  \colhead{$\sigma_{\alpha}$} &
  \colhead{$\Delta$$\delta$} &
  \colhead{$\sigma_{\delta}$} \\
  \multicolumn{3}{c}{} &
  \colhead{(Jy)} &
  \colhead{(Jy)} &
  \colhead{(mas)} &
  \colhead{(mas)} &
  \colhead{(mas)} &
  \colhead{(mas)} &
  \colhead{(Jy)} &
  \colhead{(Jy)} &
  \colhead{(mas)} &
  \colhead{(mas)} &
  \colhead{(mas)} &
  \colhead{(mas)} \\
  \multicolumn{3}{c}{} &
  \colhead{[1]} &
  \colhead{[2]} &
  \colhead{[3]} &
  \colhead{[4]} &
  \colhead{[5]} &
  \colhead{[6]} &
  \colhead{[7]} &
  \colhead{[8]} &
  \colhead{[9]} &
  \colhead{[10]} &
  \colhead{[11]} &
  \colhead{[12]}
}
\startdata
1997 & Jan & 16 & 3.39 & 0.21 & \phm{$-$}0.243 & 0.019 & $-0.044$       & 0.013 & 3.41 & 0.18 & $-0.224$ & 0.029 & \phm{$-$}0.013 & 0.011 \\
1997 & Jan & 18 & 3.41 & 0.25 & \phm{$-$}0.239 & 0.026 & $-0.047$       & 0.021 & 3.32 & 0.19 & $-0.225$ & 0.039 & \phm{$-$}0.008 & 0.014 \\
1997 & Nov & 29 & 5.24 & 0.34 & \phm{$-$}0.141 & 0.017 & $-0.004$       & 0.010 & 2.15 & 0.16 & $-0.378$ & 0.078 & \phm{$-$}0.046 & 0.011 \\
1997 & Dec & 21 & 5.04 & 0.56 & \phm{$-$}0.082 & 0.041 & \phm{$-$}0.004 & 0.017 & 2.61 & 0.39 & $-0.386$ & 0.110 & \phm{$-$}0.037 & 0.012 \\
1997 & Dec & 27 & 4.79 & 0.44 & \phm{$-$}0.126 & 0.029 & \phm{$-$}0.000 & 0.016 & 2.45 & 0.29 & $-0.339$ & 0.092 & \phm{$-$}0.003 & 0.018 \\
1998 & Mar & 01 & 4.19 & 0.49 & \phm{$-$}0.172 & 0.035 & \phm{$-$}0.000 & 0.010 & 3.07 & 0.36 & $-0.253$ & 0.083 & $-0.007$       & 0.010 \\
1998 & Jul & 12 & 3.04 & 0.16 & \phm{$-$}0.353 & 0.016 & \phm{$-$}0.000 & 0.012 & 5.24 & 0.26 & $-0.117$ & 0.017 & $-0.008$       & 0.010 \\
1998 & Aug & 08 & 3.30 & 0.18 & \phm{$-$}0.445 & 0.015 & \phm{$-$}0.002 & 0.010 & 5.44 & 0.27 & $-0.034$ & 0.019 & $-0.014$       & 0.010 \\
1998 & Sep & 16 & 3.39 & 0.18 & \phm{$-$}0.341 & 0.015 & \phm{$-$}0.006 & 0.016 & 4.97 & 0.25 & $-0.133$ & 0.017 & $-0.022$       & 0.010 \\
1999 & Mar & 13 & 3.38 & 0.20 & \phm{$-$}0.285 & 0.016 & \phm{$-$}0.036 & 0.010 & 4.03 & 0.20 & $-0.178$ & 0.028 & $-0.011$       & 0.010 \\
1999 & May & 15 & 3.72 & 0.22 & \phm{$-$}0.229 & 0.016 & \phm{$-$}0.042 & 0.010 & 4.17 & 0.21 & $-0.174$ & 0.027 & $-0.032$       & 0.010 \\
1999 & Sep & 18 & 4.91 & 0.43 & \phm{$-$}0.194 & 0.019 & \phm{$-$}0.057 & 0.012 & 4.97 & 0.34 & $-0.218$ & 0.047 & $-0.036$       & 0.016 \\
1999 & Dec & 09 & 3.03 & 0.16 & \phm{$-$}0.321 & 0.015 & \phm{$-$}0.048 & 0.011 & 4.06 & 0.21 & $-0.157$ & 0.021 & $-0.004$       & 0.010 \\
2000 & May & 15 & 3.48 & 0.19 & \phm{$-$}0.214 & 0.015 & \phm{$-$}0.012 & 0.010 & 3.15 & 0.17 & $-0.403$ & 0.028 & \phm{$-$}0.021 & 0.013 \\
2000 & Aug & 07 & 4.53 & 0.28 & \phm{$-$}0.002 & 0.019 & \phm{$-$}0.014 & 0.013 & 2.36 & 0.12 & $-0.652$ & 0.076 & \phm{$-$}0.002 & 0.017 \\
2000 & Nov & 05 & 4.21 & 0.24 & \phm{$-$}0.089 & 0.015 & \phm{$-$}0.065 & 0.012 & 1.97 & 0.10 & $-0.490$ & 0.066 & $-0.003$       & 0.016 \\
2000 & Nov & 06 & 4.63 & 0.25 & \phm{$-$}0.139 & 0.017 & \phm{$-$}0.024 & 0.015 & 1.87 & 0.10 & $-0.565$ & 0.049 & $-0.037$       & 0.013 \\
2001 & Mar & 31 & 3.54 & 0.23 & \phm{$-$}0.212 & 0.021 & \phm{$-$}0.012 & 0.010 & 3.62 & 0.21 & $-0.225$ & 0.034 & $-0.054$       & 0.016 \\
2001 & Jun & 29 & 3.82 & 0.40 & \phm{$-$}0.285 & 0.041 & \phm{$-$}0.029 & 0.013 & 3.82 & 0.26 & $-0.259$ & 0.082 & $-0.021$       & 0.019 \\
2001 & Oct & 19 & 3.52 & 0.21 & \phm{$-$}0.193 & 0.017 & \phm{$-$}0.075 & 0.011 & 3.00 & 0.15 & $-0.390$ & 0.042 & \phm{$-$}0.010 & 0.020 \\
2001 & Dec & 21 & 4.34 & 0.28 & \phm{$-$}0.176 & 0.023 & \phm{$-$}0.016 & 0.011 & 2.48 & 0.14 & $-0.507$ & 0.077 & $-0.055$       & 0.010 \\
2002 & Apr & 14 & 3.58 & 0.18 & \phm{$-$}0.165 & 0.015 & \phm{$-$}0.055 & 0.010 & 2.59 & 0.13 & $-0.385$ & 0.016 & $-0.058$       & 0.011 \\
2002 & Jul & 14 & 2.88 & 0.30 & \phm{$-$}0.315 & 0.033 & \phm{$-$}0.049 & 0.010 & 3.42 & 0.19 & $-0.191$ & 0.068 & $-0.053$       & 0.023 \\
2002 & Nov & 20 & 2.62 & 0.23 & \phm{$-$}0.439 & 0.039 & \phm{$-$}0.086 & 0.010 & 4.23 & 0.23 & $-0.132$ & 0.048 & $-0.003$       & 0.013 \\
2003 & Jan & 26 & 2.72 & 0.26 & \phm{$-$}0.448 & 0.033 & \phm{$-$}0.138 & 0.010 & 4.59 & 0.23 & $-0.145$ & 0.056 & \phm{$-$}0.012 & 0.016 \\
2003 & May & 18 & 2.81 & 0.15 & \phm{$-$}0.600 & 0.015 & \phm{$-$}0.086 & 0.011 & 3.84 & 0.19 & $-0.094$ & 0.022 & \phm{$-$}0.022 & 0.011 \\
2003 & Sep & 08 & 3.02 & 0.18 & \phm{$-$}0.648 & 0.016 & \phm{$-$}0.106 & 0.011 & 3.26 & 0.18 & $-0.066$ & 0.032 & \phm{$-$}0.019 & 0.012 \\
2003 & Dec & 05 & 3.20 & 0.16 & \phm{$-$}0.175 & 0.015 & \phm{$-$}0.009 & 0.011 & 2.84 & 0.14 & $-0.548$ & 0.019 & $-0.087$       & 0.010 \\
2004 & Mar & 06 & 3.13 & 0.16 & \phm{$-$}0.189 & 0.015 & \phm{$-$}0.069 & 0.011 & 2.51 & 0.13 & $-0.504$ & 0.023 & $-0.033$       & 0.014 \\
2004 & May & 18 & 2.58 & 0.15 & \phm{$-$}0.041 & 0.016 & $-0.122$       & 0.010 & 2.51 & 0.13 & $-0.640$ & 0.039 & $-0.264$       & 0.018 \\
2004 & Jun & 26 & 2.55 & 0.25 & \phm{$-$}0.376 & 0.027 & \phm{$-$}0.064 & 0.011 & 2.41 & 0.12 & $-0.324$ & 0.109 & $-0.078$       & 0.021 \\
2004 & Dec & 11 & 2.62 & 0.14 & \phm{$-$}0.113 & 0.018 & \phm{$-$}0.029 & 0.010 & 2.90 & 0.15 & $-0.565$ & 0.032 & $-0.041$       & 0.013 \\
2005 & Jan & 15 & 2.64 & 0.14 & \phm{$-$}0.033 & 0.018 & \phm{$-$}0.080 & 0.010 & 2.92 & 0.15 & $-0.651$ & 0.030 & \phm{$-$}0.028 & 0.014 \\
2005 & May & 28 & 2.90 & 0.15 & $-0.005$       & 0.015 & \phm{$-$}0.060 & 0.010 & 2.79 & 0.14 & $-0.701$ & 0.021 & \phm{$-$}0.020 & 0.011 \\
2005 & Jul & 16 & 2.74 & 0.14 & $-0.008$       & 0.015 & \phm{$-$}0.011 & 0.010 & 2.71 & 0.14 & $-0.732$ & 0.023 & $-0.022$       & 0.011 \\
\enddata
\tablecomments{[1,2] Flux density and flux-density standard error of
component C1; [3--6] Position ($\alpha$ and $\delta$), relative to
image phase center (see definition in text), and position standard
error of component C1; [7,8] Flux density and flux-density standard
error of component C2; [9--12] Position ($\alpha$ and $\delta$),
relative to image phase center, and position standard error of
component C2.  The tabulated flux-density standard errors do not
include the estimated 10\% standard error in the VLBI flux-density
scale at each epoch.}
\end{deluxetable}

\begin{figure}
\plotone{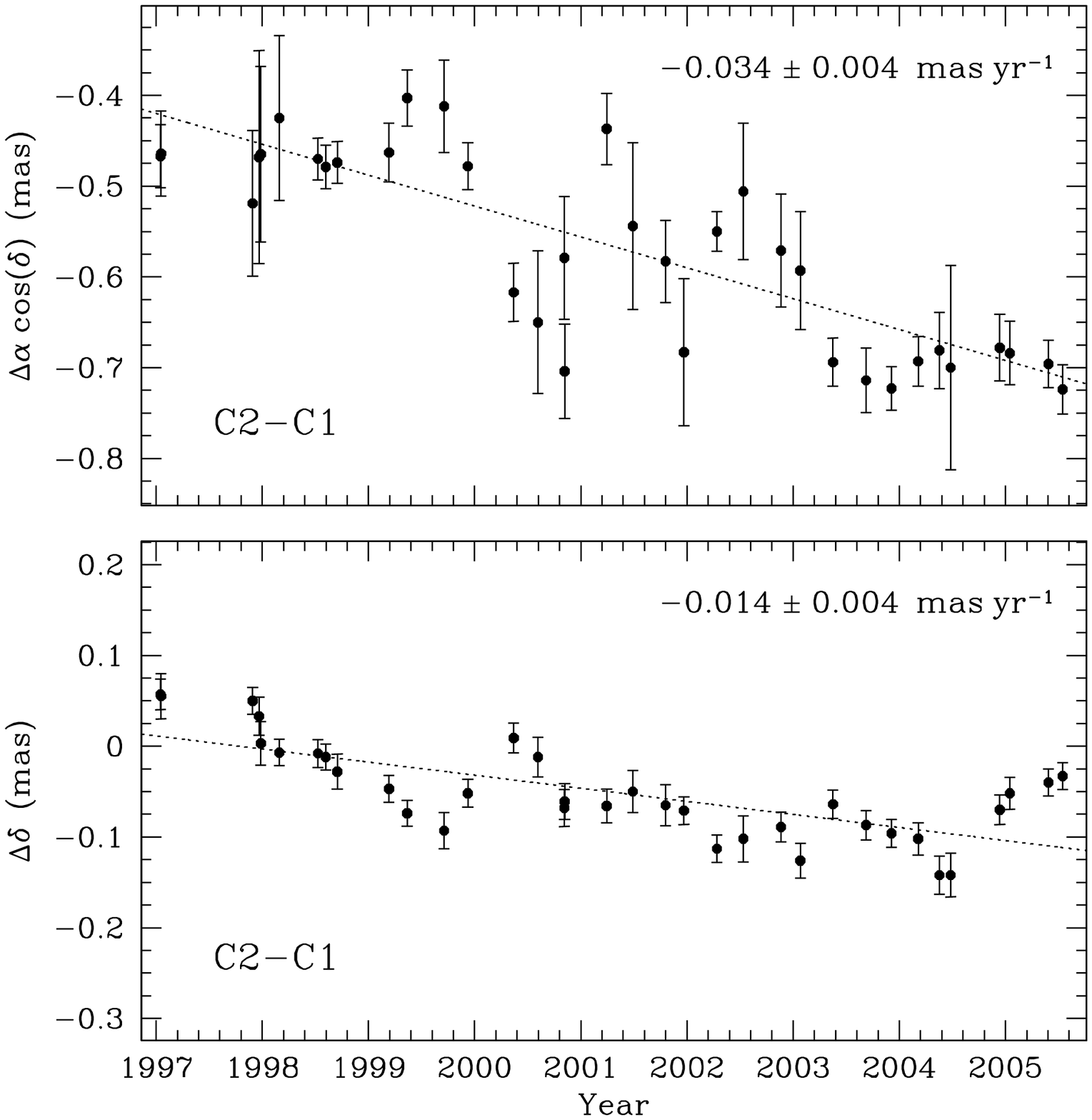}
\figcaption{Position of 3C~454.3 core-region model component C2
relative to core-region model component C1.  Error bars are the
root-sum-square (rss) of the standard errors in the position estimates
of C1 and C2.  Our estimated proper motion of C2 relative to C1 is
indicated by the dotted lines.  Note that the residuals are not
distributed randomly about the straight lines.  The apparent
``oscillations'' may represent real epoch-to-epoch changes in the
separation of C2 from C1, due to changes in the unresolved structure
in the core region (see \S~\ref{coreregion}).
\label{c1c2relpos}}
\end{figure}

In Figure~\ref{c1c2relpos}, we plot at each epoch the separation of C2
from C1.  We see that this separation increased by $\sim$0.25~mas over
the $\sim$8.5~yr period of our observing program, largely in the
$\alpha$ coordinate.  We determined, via a weighted-least-squares fit
of a straight line to the separation at all 35 epochs, a relative
proper motion for C2 with respect to C1 of $-0.034 \pm 0.003$~\masyr\
in $\alpha$ and $-0.014 \pm 0.003$~\masyr\ in $\delta$.  (We use
weighted least-squares for all proper-motion determinations.)  The
quoted uncertainties are the statistical standard errors of the fit,
scaled by a common factor so that the $\chi^2$ per degree of freedom
is unity.  We performed the same weighted fit to the separations
determined for the common-resolution images, and found that the
relative proper motion was 0.5$\sigma$ larger in $\alpha$ and
0.2$\sigma$ smaller in $\delta$, where $\sigma$ is the statistical
standard error of the nominal (differing-resolution) case.  Taking the
difference as an estimate of the systematic error in the determination
of the relative proper motion in the nominal case, and adding in
quadrature the statistical standard error for the nominal case, we get
a relative proper motion (nominal value and standard error) for C2
with respect to C1 of $-0.034 \pm 0.004$~\masyr\ in $\alpha$ and
$-0.014 \pm 0.004$~\masyr\ in $\delta$.  (We compute the standard
errors in the relative proper motions of components in the same way
throughout \S~\ref{implanefits}.)

The two-component model described above obviously does not account for
all the structure seen in, or very near, the core region at each
epoch.  In particular, the two-component model fails to accurately
describe low-level emission extending west-southwest from C2 with
surface brightness 5--25\% of the peak brightness.  At many epochs,
this emission appears to be associated with the emergence of a new
component into the inner-jet region (see, e.g.,
Figure~\ref{windowandcompid}).  Unfortunately, our resolution is not
sufficient to adequately constrain a more elaborate model for the core
region.  In an attempt to simultaneously describe the bright
components of the core region and what we call a transition component
that moves between the core and inner-jet regions, we extended at each
epoch the core-region fitting window $\sim$1~mas west into the
inner-jet region and fit a model consisting of three point sources.
For 13 epochs, the fit positions and flux densities of each of C1 and
C2 in the three-component and two-component models differ by no more
than two times the standard error in the two-component model (see
Table~\ref{3Ccorecomp}).  For these epochs, we (somewhat arbitrarily)
consider the three-component model to give an acceptable description
of the two core-region components and the transition component.  The
fit results for the transition component for these epochs and a brief
description of the results for the remaining epochs are presented in
\S~\ref{innerjetregion}.  In addition to the extension due to the
transition component, there appears to also be core-region emission
north of C2, especially at our later epochs.  This emission is at a
very low level and likely does not affect the results from the
two-component model for either C1 or C2.

How well does our simple two-component model represent the underlying
structure of the core region of 3C~454.3?  VLBI images at 43~GHz and
86~GHz reveal a great complexity and variability in the easternmost
$\sim$1~mas of 3C~454.3.  At the extreme east is a very compact
component \citep[$\lesssim$0.03~mas FWHM at
86~GHz;][]{Krichbaum+2006b}, considered by various authors to be the
VLBI ``core.''  New components are occasionally ejected from
approximately the position of this core and then move west-northwest
(along $\rm{p.a.} \approx -70\arcdeg$) at superluminal apparent
velocities in the range 3--13\,$c$ \citep{Jorstad+2001b,Jorstad+2005}.
At a distance of $\sim$0.65~mas from the core, the moving components
are thought to encounter a bend or a recollimation shock in the jet
\citep[see][]{GomezMA1999} and brighten to produce at 43~GHz a
component which is approximately stationary relative to the core at
that frequency.  The separation of the core and the ``stationary''
component at 43~GHz is similar to the separation of the two
easternmost components observed at 10.7~GHz by
\citet{Pauliny-Toth+1987} and also to the mean separation of C1 and C2
in our 8.4~GHz images.  Diffuse emission appears in the 43~GHz images
to the north and northwest of the ``stationary'' component at various
epochs, but the moving components emerge primarily to the west or
southwest of the ``stationary'' component
\citep[e.g.,][]{Jorstad+2005}.  The ejection from the core of new jet
components, and the motions of these components toward and beyond the
``stationary'' component, may be responsible for the apparent
``oscillations'' we observe in the 8.4~GHz separation of C2 relative
to C1 (see Figure~\ref{c1c2relpos}).  The proper motion of C2 relative
to C1 could be a consequence of the westward motion of a particularly
bright jet component through the position of the ``stationary''
component in the latter part of our observing period (see
\S~\ref{corediscuss}).

\begin{figure}
\plotone{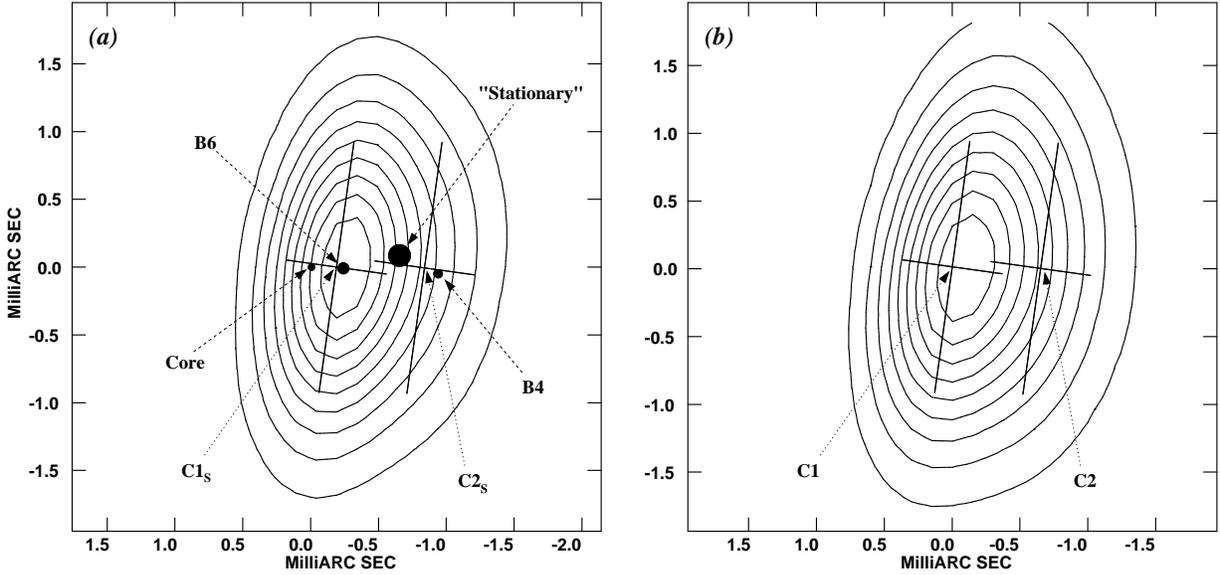}
\figcaption{$(a)$ Simulated 8.4~GHz image for 2000 July 17 and $(b)$
actual 8.4~GHz image (``zoomed-in'' on the core region) for 2000
August 7.  Contour levels are $10$, $20$, $30$, $40$, $50$, $60$,
$70$, $80$, and $90$\% of the peak brightness in each image.  The
43~GHz model for 2000 July 17 \citep[see][]{Jorstad+2005}, upon which
the simulated image is based, consists of four components (filled
black circles): Core, B6, ``Stationary,'' and B4.  With the exception
of the Core, the size of the circle represents the diameter (FWHM) of
the component.  The Core is modeled as a point source.  The positions
of 8.4~GHz model components $\rm{C1}_s$ and $\rm{C2}_s$ in $(a)$ and
C1 and C2 in $(b)$ are indicated by crosses which reflect the size and
orientation of the restoring beam.  The restoring beam is the same for
both images.  The images are centered on the center of $\rm{C1}_s$ in
$(a)$ and C1 in $(b)$.
\label{c1c2corestat}}
\end{figure}

To better quantify the spatial relationship of component C1 in our
8.4~GHz images to the core as imaged at 43~GHz, and likewise that of
component C2 to the ``stationary'' component, we used the results from
the studies of \citet{Jorstad+2001b,Jorstad+2005} to simulate 8.4~GHz
images of the core region of 3C~454.3.  In particular, we used model
data for the core, ``stationary'' component, and moving jet components
at nine epochs, each within 30 days of one of our observations (see
Table~\ref{3Csim}).  We retained the nominal position of each 43~GHz
component at each epoch, but modified the flux densities of the
components in approximate accordance with spectra presented in the
literature (e.g., \citealt{Pagels+2004}; \citealt{Chen+2005};
\citealt{Krichbaum+2006a}).  We generated the simulated 8.4~GHz image
for each of the relevant epochs by (1) Fourier transforming the
flux-density-adjusted 43~GHz model onto a spatial frequency ``grid''
corresponding to our 8.4~GHz \uv\ coverage, (2) imaging (i.e.,
Fourier-inverting and deconvolving with the CLEAN algorithm) the
``gridded'' data using the same restoring beam as for the actual
8.4~GHz image, (3) fitting the two-point-source model (described
previously) to the image, and finally (4) repeating steps 1--3,
tweaking the flux densities of the 43~GHz model components, until the
flux densities and separation of the two point sources in the
simulated 8.4~GHz image agreed with those in the actual 8.4~GHz image
to approximately within the errors given in Table~\ref{3Ccorecomp}.
We present for illustrative purposes in Figure~\ref{c1c2corestat} the
simulated image and corresponding actual image for one of the nine
epoch pairs.  We give the results of the two-component model for the
simulated image at each of the nine epochs in Table~\ref{3Csim}.  We
refer to the simulated easternmost component as $\rm{C1}_s$ and the
simulated westernmost component as $\rm{C2}_s$.  All positions in the
table are given with respect to the 43~GHz core position.  The fit
position of $\rm{C1}_s$ (columns 1 and 2 in Table~\ref{3Csim})
therefore gives explicitly the offset (in each coordinate) of this
component from the 43~GHz core.  The mean $\rm{C1}_s - \rm{core}$
displacement is $0.18 \pm 0.06$~mas towards the west, where the
uncertainty represents the rms variation in the position of
$\rm{C1}_s$.  We note that the largest displacements occur shortly
after the ejection of new jet components, when the flux density of the
core is relatively low compared to that of the jet component (see
Figure~\ref{c1c2corestat}).  Columns 5 and 6 in Table~\ref{3Csim} give
the offset in each coordinate of $\rm{C2}_s$ from the 43~GHz
``stationary'' component.  The mean $\rm{C2}_s -
\rm{stationary}$-component displacement is $0.06 \pm 0.08$~mas towards
the west-southwest.  The smaller value of this displacement, compared
to the $\rm{C1}_s - \rm{core}$ displacement, is reasonable, since jet
components move both into and out of the position defined by the
``stationary'' component, whereas jet components move only westward
away from the core.  The larger peak-to-peak variations in the
$\rm{C2}_s - \rm{stationary}$-component displacement, compared to the
$\rm{C1}_s - \rm{core}$ displacement (see Table~\ref{3Csim}), reflect
the complexity of the structure seen at 43~GHz at the western end of
the core region.

\begin{deluxetable}{l@{~}l@{~}l@{~~}l@{~}l@{~}l@{~~}c@{}c@{}c@{}c@{}c@{}c}
\tabletypesize{\scriptsize}
\tablecaption{3C~454.3 Simulated-Image Component Positions \label{3Csim}}
\tablewidth{0pt}
\tablehead{
  \multicolumn{3}{c}{43-GHz} &
  \multicolumn{3}{c}{8.4-GHz} &
  \multicolumn{2}{c}{$\rm{C1}_s$--Core} &
  \multicolumn{2}{c}{$\rm{C2}_s$--Core} &
  \multicolumn{2}{c}{$\rm{C2}_s$--Stat} \\
  \multicolumn{3}{c}{Epoch} &
  \multicolumn{3}{c}{Epoch} &
  \colhead{$\Delta$$\alpha$} &
  \colhead{$\Delta$$\delta$} &
  \colhead{$\Delta$$\alpha$} &
  \colhead{$\Delta$$\delta$} &
  \colhead{$\Delta$$\alpha$} &
  \colhead{$\Delta$$\delta$} \\
  \multicolumn{3}{c}{} &
  \multicolumn{3}{c}{} &
  \colhead{(mas)} &
  \colhead{(mas)} &
  \colhead{(mas)} &
  \colhead{(mas)} &
  \colhead{(mas)} &
  \colhead{(mas)} \\
  \multicolumn{3}{c}{} &
  \multicolumn{3}{c}{} &
  \colhead{[1]} &
  \colhead{[2]} &
  \colhead{[3]} &
  \colhead{[4]} &
  \colhead{[5]} &
  \colhead{[6]}
}
\startdata
1996 & Dec & 23 & 1997 & Jan & 16 & $-0.12$ & 0.03 & $-0.60$ & \phm{$-$}0.14 & \phm{$-$}0.05 & $-0.06$ \\
1998 & Mar & 25 & 1998 & Mar & 01 & $-0.07$ & 0.00 & $-0.54$ & \phm{$-$}0.07 & $-0.05$       & \phm{$-$}0.00 \\
1998 & Aug & 01 & 1998 & Aug & 08 & $-0.21$ & 0.11 & $-0.72$ & \phm{$-$}0.16 & $-0.11$       & $-0.02$ \\
1998 & Oct & 05 & 1998 & Sep & 16 & $-0.23$ & 0.08 & $-0.75$ & \phm{$-$}0.05 & $-0.03$       & \phm{$-$}0.01 \\
1999 & Apr & 29 & 1999 & May & 15 & $-0.25$ & 0.00 & $-0.65$ & \phm{$-$}0.10 & $-0.01$       & $-0.01$ \\
1999 & Oct & 06 & 1999 & Sep & 18 & $-0.22$ & 0.05 & $-0.73$ & \phm{$-$}0.06 & $-0.08$       & $-0.01$ \\
1999 & Dec & 05 & 1999 & Dec & 09 & $-0.16$ & 0.01 & $-0.68$ & \phm{$-$}0.04 & $-0.01$       & \phm{$-$}0.00 \\
2000 & Jul & 17 & 2000 & Aug & 07 & $-0.19$ & 0.00 & $-0.84$ & $-0.01$       & $-0.23$       & $-0.09$ \\
2001 & Apr & 14 & 2001 & Mar & 31 & $-0.20$ & 0.00 & $-0.67$ & $-0.09$       & $-0.03$       & \phm{$-$}0.01 \\
\enddata
\tablecomments{[1,2] Position ($\alpha$ and $\delta$) of simulated
8.4~GHz component $\rm{C1}_s$ relative to the 43~GHz core; [3,4]
Position ($\alpha$ and $\delta$) of simulated 8.4~GHz component
$\rm{C2}_s$ relative to the 43~GHz core; [5,6] Position ($\alpha$ and
$\delta$) of simulated 8.4~GHz component $\rm{C2}_s$ relative to the
43~GHz ``stationary'' component.}
\end{deluxetable}

\subsubsection{Inner-jet Region \label{innerjetregion}}

Our images of 3C~454.3 show that new components occasionally emerge to
the west or southwest of the position approximately defined by
core-region component C2, and move through the inner-jet region toward
the extended jet.  At any given epoch, the inner-jet region itself may
contain zero, one, or multiple components in transit between the core
region and extended jet.  Unfortunately, given the general complexity
of the radio structure in the inner-jet region, the relative weakness
of the components, and the resolution of our VLBI arrays, we were able
to confidently ``isolate'' inner-jet-region components for only a
limited number of epochs (see below).

\begin{figure}
\plotone{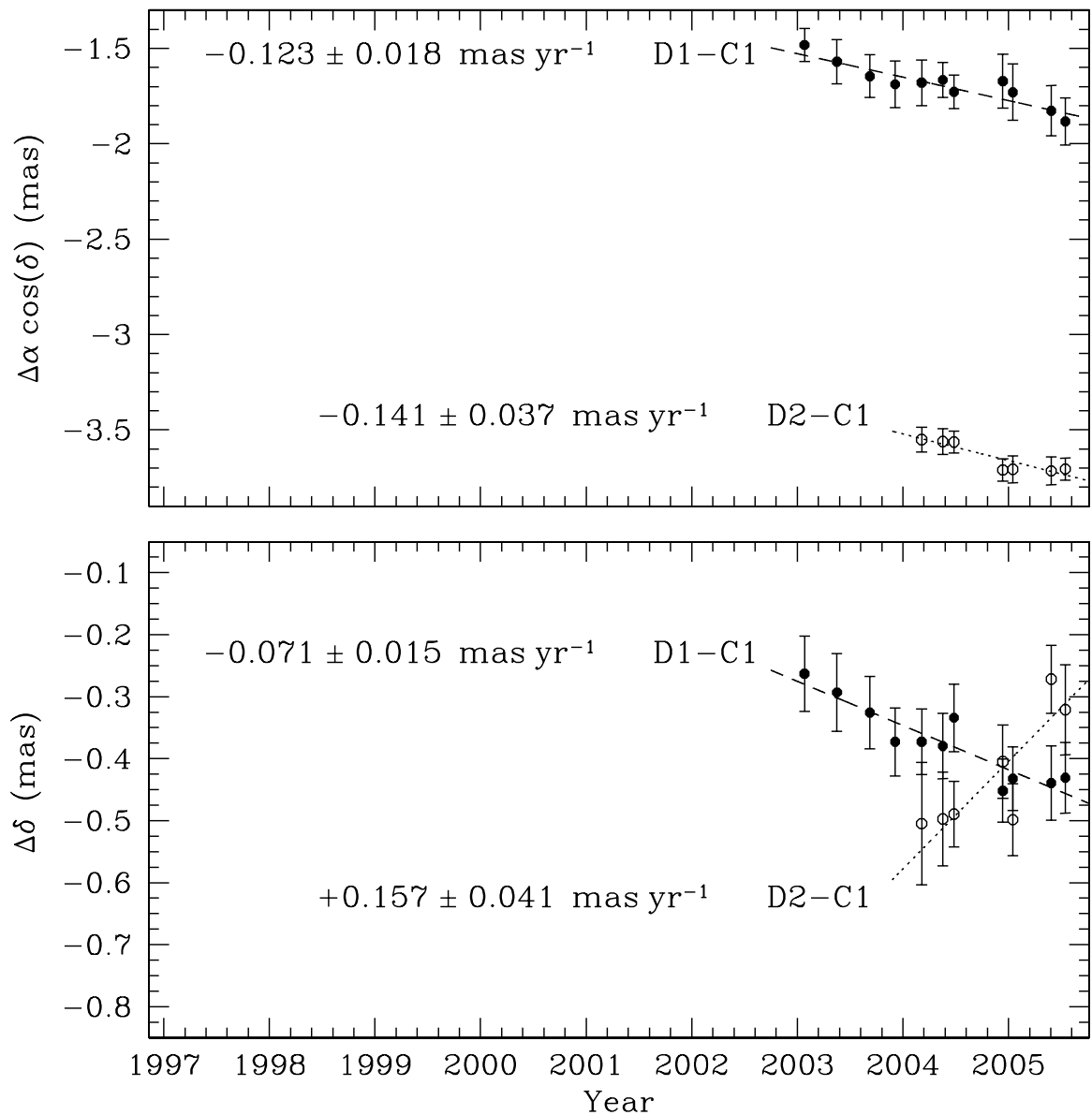} 
\figcaption{Positions of 3C~454.3 inner-jet model components D1
(filled circles) and D2 (open circles) relative to core-region model
component C1. Error bars are the rss of the standard errors in the
position estimates of C1 and D1 or D2, as appropriate.  The proper
motions of D1 and D2 relative to C1 are indicated by the dashed and
dotted lines.  Note that different scales are used for the $\alpha$
and $\delta$ coordinates.
\label{d1andd2c1relpos}}
\end{figure}

To estimate the proper motion (relative to C1) of components
transitioning between the core region and inner-jet region, we
attempted, as described in \S~\ref{coreregion}, to simultaneously
describe the core-region components and a transition component by
fitting a three point-source model to the easternmost $\sim$2~mas of
the 3C~454.3 radio structure.  For 22 of the 35 epochs, we found the
fit positions and flux densities of core-region components C1$\arcmin$
and C2$\arcmin$ in the three-component model to differ significantly
(i.e., by more than two times the respective standard errors given in
Table~\ref{3Ccorecomp}) from those of C1 and C2 in the two-component
model.  Moreover, we found that the position of C1$\arcmin$ for these
22 epochs did not align with the features seen in the images;
specifically, C1$\arcmin$ was positioned too far east to produce the
sharp cutoff seen at the eastern edge of the source.  We therefore
consider the three-component model unacceptable for these epochs.  Of
the 13 epochs with acceptably fit models, 11 fall within the
$\sim$2.5~yr interval from 2003 January to 2005 July.  For these 11
epochs, we repeated the three-component fit but with the position of
C1$\arcmin$ fixed to the position of C1 found in the two-component
model.  We plot in Figure~\ref{d1andd2c1relpos} the separation from C1
of the $0.6 \pm 0.1$~Jy transition component, D1.  We determined for
the 11 epochs a relative proper motion for D1 with respect to C1 of
$-0.123 \pm 0.018$~\masyr\ in $\alpha$ and $-0.071 \pm
0.015$~\masyr\ in $\delta$.

We were also able to track the motion of a second component, D2, for
seven consecutive epochs from 2004 March to 2005 July, by fitting a
model consisting of a single point source (D2) to the $\sim$2-mas-wide
portion of the inner-jet region west of D1.  We plot in
Figure~\ref{d1andd2c1relpos} the separation from C1 of the $0.05 \pm
0.01$~Jy D2.  We determined for the seven epochs a relative proper
motion for D2 with respect to C1 of $-0.141 \pm 0.037$~\masyr\ in
$\alpha$ and $+0.157 \pm 0.041$~\masyr\ in $\delta$.

While the proper motion in $\alpha$ is similar for components D1 and
D2, the difference in the proper motion in $\delta$ from a negative
value for D1 to a positive value for D2 clearly demonstrates a bend in
the jet between these components.  A similar bend in the inner-jet
region was noted by \citet{Pauliny-Toth1998} for a component at 5 and
8.4~GHz that moved from a position (relative to their ``core'') near
that of D1 (relative to C1) to a position beyond that of D2.  The bend
demonstrated by component motion is also clearly observed in the
low-surface-brightness emission in VLBI images between 5 and 15~GHz
(\citealt{CawthorneG1996}; \citealt{Pauliny-Toth1998};
\citealt{Kellermann+1998}; \citealt{Chen+2005};
\citealt{Lister+2009}).  The magnitudes of the proper motions of D1
and D2 are a factor 2--4 lower than that seen at 5 and 8.4~GHz for a
component traversing a region 1.5--5.0~mas west of the core between
1983 and 1991 \citep{Pauliny-Toth1998}, but consistent with that seen
at 15~GHz for a component traversing a region 2.5--3.5~mas west of the
core between 2006 and 2009 \citep[see][]{Vercellone+2010}.

\subsubsection{Extended-jet Region \label{outerjetregion}}

Beyond the brightness peak $\sim$5.5~mas west of component C1, the
extended jet of 3C~454.3 has no clearly discernible sub-components, at
least none that can be traced from epoch to epoch.  We fit to the
extended-jet region at each epoch a model consisting of a point source
and an elliptical Gaussian, representing, respectively, the compact
emission component near the (jet) brightness peak and the
$\sim$6.5~mas (mean major-axis length) extended jet itself.  The
results of the fit are given in Table~\ref{3Cextjetcomp}.  We refer to
the compact component as J1 and the extended component as Jext.  The
uncertainties in the major axis, minor axis, and p.a.\ of Jext are
statistical standard errors.

\begin{deluxetable}{l@{~}l@{~}l@{~~}c@{}c@{}c@{}c@{}c@{}c@{}c@{}c@{}c@{}c@{}c@{}c@{}c@{}c@{}c@{}c@{}c@{}c}
\tabletypesize{\tiny}
\rotate
\tablecaption{3C~454.3 Image-Plane 8.4~GHz Model Parameters: Extended-Jet Region \label{3Cextjetcomp}}
\tablewidth{0pt}
\tablehead{
  \multicolumn{3}{c}{Start Date} &
  \multicolumn{6}{c}{----------------------Component J1----------------------} &
  \multicolumn{12}{c}{---------------------------------------------------------Component Jext---------------------------------------------------------} \\
  \multicolumn{3}{c}{} &
  \colhead{$S_{8.4}$} &
  \colhead{$\sigma_{S}$} &
  \colhead{$\Delta$$\alpha$} &
  \colhead{$\sigma_{\alpha}$} &
  \colhead{$\Delta$$\delta$} &
  \colhead{$\sigma_{\delta}$} &
  \colhead{$S_{8.4}$} &
  \colhead{$\sigma_{S}$} &
  \colhead{$\Delta$$\alpha$} &
  \colhead{$\sigma_{\alpha}$} &
  \colhead{$\Delta$$\delta$} &
  \colhead{$\sigma_{\delta}$} &
  \colhead{$\rm{Maj}$} &
  \colhead{$\sigma_{\rm{Maj}}$} &
  \colhead{$\rm{Min}$} &
  \colhead{$\sigma_{\rm{Min}}$} &
  \colhead{$\rm{p.a.}$} &
  \colhead{$\sigma_{\rm{p.a.}}$} \\
  \multicolumn{3}{c}{} &
  \colhead{(Jy)} &
  \colhead{(Jy)} &
  \colhead{(mas)} &
  \colhead{(mas)} &
  \colhead{(mas)} &
  \colhead{(mas)} &
  \colhead{(Jy)} &
  \colhead{(Jy)} &
  \colhead{(mas)} &
  \colhead{(mas)} &
  \colhead{(mas)} &
  \colhead{(mas)} &
  \colhead{(mas)} &
  \colhead{(mas)} &
  \colhead{(mas)} &
  \colhead{(mas)} &
  \colhead{($\arcdeg$)} &
  \colhead{($\arcdeg$)} \\
  \multicolumn{3}{c}{} &
  \colhead{[1]} &
  \colhead{[2]} &
  \colhead{[3]} &
  \colhead{[4]} &
  \colhead{[5]} &
  \colhead{[6]} &
  \colhead{[7]} &
  \colhead{[8]} &
  \colhead{[9]} &
  \colhead{[10]} &
  \colhead{[11]} &
  \colhead{[12]} &
  \colhead{[13]} &
  \colhead{[14]} &
  \colhead{[15]} &
  \colhead{[16]} &
  \colhead{[17]} &
  \colhead{[18]}
}
\startdata
1997 & Jan & 16 & 0.267 & 0.027 & $-5.343$ & 0.055 & 0.329 & 0.077 & 1.32 & 0.07 & $-6.075$ & 0.045 & 1.650 & 0.046 & 7.28 & 0.10 & 1.77 & 0.11 & $-40.5$ & 0.7 \\
1997 & Jan & 18 & 0.263 & 0.026 & $-5.366$ & 0.058 & 0.341 & 0.079 & 1.34 & 0.07 & $-6.093$ & 0.055 & 1.674 & 0.057 & 7.43 & 0.13 & 1.78 & 0.15 & $-41.1$ & 0.9 \\
1997 & Nov & 29 & 0.357 & 0.036 & $-5.354$ & 0.054 & 0.454 & 0.073 & 1.51 & 0.08 & $-6.201$ & 0.043 & 1.894 & 0.044 & 7.53 & 0.09 & 1.87 & 0.12 & $-41.5$ & 0.7 \\
1997 & Dec & 21 & 0.332 & 0.033 & $-5.399$ & 0.065 & 0.449 & 0.073 & 1.58 & 0.09 & $-6.174$ & 0.066 & 1.854 & 0.061 & 7.29 & 0.15 & 1.83 & 0.19 & $-40.7$ & 1.1 \\
1997 & Dec & 27 & 0.308 & 0.031 & $-5.377$ & 0.059 & 0.443 & 0.073 & 1.47 & 0.08 & $-6.195$ & 0.049 & 1.840 & 0.045 & 7.02 & 0.09 & 1.85 & 0.11 & $-40.1$ & 0.7 \\
1998 & Mar & 01 & 0.284 & 0.028 & $-5.410$ & 0.062 & 0.439 & 0.073 & 1.48 & 0.08 & $-6.063$ & 0.051 & 1.695 & 0.041 & 7.09 & 0.08 & 1.74 & 0.09 & $-40.6$ & 0.6 \\
1998 & Jul & 12 & 0.276 & 0.028 & $-5.486$ & 0.055 & 0.460 & 0.086 & 1.28 & 0.07 & $-6.307$ & 0.045 & 1.911 & 0.048 & 6.78 & 0.11 & 1.82 & 0.15 & $-41.2$ & 1.0 \\
1998 & Aug & 08 & 0.275 & 0.028 & $-5.488$ & 0.054 & 0.463 & 0.090 & 1.41 & 0.07 & $-6.140$ & 0.042 & 1.726 & 0.045 & 6.49 & 0.10 & 1.83 & 0.14 & $-38.8$ & 1.0 \\
1998 & Sep & 16 & 0.298 & 0.030 & $-5.466$ & 0.053 & 0.473 & 0.084 & 1.46 & 0.07 & $-6.236$ & 0.040 & 1.841 & 0.042 & 6.99 & 0.08 & 1.89 & 0.11 & $-40.5$ & 0.7 \\
1999 & Mar & 13 & 0.238 & 0.024 & $-5.418$ & 0.053 & 0.506 & 0.077 & 1.33 & 0.07 & $-6.245$ & 0.045 & 1.885 & 0.048 & 6.45 & 0.11 & 1.77 & 0.15 & $-40.2$ & 1.0 \\
1999 & May & 15 & 0.246 & 0.025 & $-5.365$ & 0.054 & 0.491 & 0.070 & 1.40 & 0.07 & $-6.180$ & 0.047 & 1.840 & 0.049 & 6.28 & 0.12 & 1.78 & 0.15 & $-39.5$ & 1.1 \\
1999 & Sep & 18 & 0.297 & 0.030 & $-5.393$ & 0.055 & 0.507 & 0.080 & 1.54 & 0.08 & $-6.273$ & 0.049 & 1.903 & 0.051 & 6.26 & 0.12 & 1.88 & 0.17 & $-40.5$ & 1.3 \\
1999 & Dec & 09 & 0.269 & 0.027 & $-5.465$ & 0.056 & 0.548 & 0.081 & 1.34 & 0.07 & $-6.399$ & 0.042 & 2.028 & 0.045 & 6.31 & 0.09 & 1.91 & 0.14 & $-40.8$ & 1.0 \\
2000 & May & 15 & 0.226 & 0.023 & $-5.540$ & 0.069 & 0.717 & 0.058 & 1.45 & 0.07 & $-6.405$ & 0.038 & 1.893 & 0.039 & 6.29 & 0.06 & 1.86 & 0.09 & $-37.9$ & 0.6 \\
2000 & Aug & 07 & 0.240 & 0.024 & $-5.513$ & 0.074 & 0.721 & 0.073 & 1.43 & 0.07 & $-6.459$ & 0.041 & 2.076 & 0.041 & 6.48 & 0.07 & 1.94 & 0.11 & $-40.3$ & 0.8 \\
2000 & Nov & 05 & 0.217 & 0.022 & $-5.427$ & 0.068 & 0.710 & 0.068 & 1.41 & 0.07 & $-6.274$ & 0.040 & 1.926 & 0.041 & 6.07 & 0.08 & 1.84 & 0.11 & $-39.9$ & 0.8 \\
2000 & Nov & 06 & 0.310 & 0.031 & $-5.494$ & 0.057 & 0.709 & 0.080 & 1.35 & 0.07 & $-6.577$ & 0.045 & 2.372 & 0.048 & 6.13 & 0.10 & 2.08 & 0.16 & $-40.4$ & 1.2 \\
2001 & Mar & 31 & 0.243 & 0.024 & $-5.487$ & 0.061 & 0.770 & 0.061 & 1.27 & 0.07 & $-6.549$ & 0.043 & 2.147 & 0.043 & 6.18 & 0.09 & 1.78 & 0.12 & $-38.8$ & 0.9 \\
2001 & Jun & 29 & 0.207 & 0.021 & $-5.521$ & 0.076 & 0.811 & 0.069 & 1.29 & 0.07 & $-6.458$ & 0.053 & 2.084 & 0.040 & 5.78 & 0.07 & 1.87 & 0.11 & $-38.2$ & 0.9 \\
2001 & Oct & 19 & 0.222 & 0.022 & $-5.532$ & 0.068 & 0.864 & 0.081 & 1.47 & 0.08 & $-6.557$ & 0.048 & 2.275 & 0.054 & 6.34 & 0.13 & 1.98 & 0.19 & $-38.2$ & 1.4 \\
2001 & Dec & 21 & 0.244 & 0.024 & $-5.540$ & 0.063 & 0.848 & 0.077 & 1.32 & 0.07 & $-6.594$ & 0.043 & 2.340 & 0.040 & 6.08 & 0.07 & 1.96 & 0.11 & $-39.8$ & 0.8 \\
2002 & Apr & 14 & 0.241 & 0.024 & $-5.489$ & 0.071 & 0.937 & 0.092 & 1.29 & 0.07 & $-6.554$ & 0.040 & 2.287 & 0.042 & 5.92 & 0.09 & 1.98 & 0.15 & $-39.0$ & 1.2 \\
2002 & Jul & 14 & 0.194 & 0.019 & $-5.491$ & 0.068 & 0.876 & 0.066 & 1.43 & 0.07 & $-6.557$ & 0.048 & 2.237 & 0.039 & 6.02 & 0.07 & 1.85 & 0.10 & $-39.2$ & 0.8 \\
2002 & Nov & 20 & 0.199 & 0.020 & $-5.587$ & 0.072 & 0.929 & 0.075 & 1.28 & 0.07 & $-6.672$ & 0.053 & 2.379 & 0.039 & 5.95 & 0.07 & 1.84 & 0.11 & $-39.9$ & 0.8 \\
2003 & Jan & 26 & 0.183 & 0.018 & $-5.569$ & 0.079 & 0.887 & 0.077 & 1.40 & 0.07 & $-6.593$ & 0.050 & 2.216 & 0.044 & 5.79 & 0.09 & 1.82 & 0.14 & $-40.5$ & 1.2 \\
2003 & May & 18 & 0.169 & 0.017 & $-5.693$ & 0.058 & 0.983 & 0.078 & 1.30 & 0.07 & $-6.648$ & 0.039 & 2.250 & 0.041 & 5.54 & 0.07 & 1.80 & 0.12 & $-38.8$ & 1.0 \\
2003 & Sep & 08 & 0.144 & 0.015 & $-5.628$ & 0.099 & 0.976 & 0.080 & 1.30 & 0.07 & $-6.633$ & 0.043 & 2.261 & 0.045 & 5.56 & 0.09 & 1.73 & 0.14 & $-39.1$ & 1.1 \\
2003 & Dec & 05 & 0.175 & 0.018 & $-5.582$ & 0.070 & 0.927 & 0.076 & 1.42 & 0.07 & $-6.621$ & 0.039 & 2.267 & 0.040 & 5.81 & 0.07 & 1.78 & 0.11 & $-39.1$ & 0.8 \\
2004 & Mar & 06 & 0.176 & 0.018 & $-5.577$ & 0.063 & 0.951 & 0.081 & 1.49 & 0.08 & $-6.634$ & 0.040 & 2.375 & 0.042 & 6.05 & 0.08 & 1.86 & 0.12 & $-40.0$ & 0.9 \\
2004 & May & 18 & 0.194 & 0.019 & $-5.577$ & 0.088 & 0.944 & 0.092 & 1.44 & 0.07 & $-6.693$ & 0.038 & 2.441 & 0.039 & 5.86 & 0.07 & 1.91 & 0.11 & $-40.3$ & 0.9 \\
2004 & Jun & 26 & 0.166 & 0.017 & $-5.480$ & 0.084 & 0.946 & 0.072 & 1.40 & 0.07 & $-6.598$ & 0.042 & 2.346 & 0.034 & 5.80 & 0.04 & 1.82 & 0.05 & $-39.5$ & 0.4 \\
2004 & Dec & 11 & 0.176 & 0.018 & $-5.628$ & 0.074 & 0.931 & 0.077 & 1.39 & 0.07 & $-6.782$ & 0.036 & 2.440 & 0.033 & 5.99 & 0.03 & 1.85 & 0.05 & $-40.5$ & 0.4 \\
2005 & Jan & 15 & 0.181 & 0.018 & $-5.635$ & 0.067 & 0.936 & 0.075 & 1.36 & 0.07 & $-6.811$ & 0.037 & 2.469 & 0.035 & 5.90 & 0.04 & 1.86 & 0.07 & $-40.9$ & 0.5 \\
2005 & May & 28 & 0.189 & 0.019 & $-5.633$ & 0.066 & 0.978 & 0.076 & 1.38 & 0.07 & $-6.756$ & 0.037 & 2.476 & 0.037 & 5.80 & 0.06 & 1.83 & 0.08 & $-40.3$ & 0.7 \\
2005 & Jul & 16 & 0.187 & 0.019 & $-5.657$ & 0.065 & 0.955 & 0.083 & 1.37 & 0.07 & $-6.821$ & 0.034 & 2.506 & 0.033 & 5.87 & 0.03 & 1.84 & 0.05 & $-40.9$ & 0.4 \\
\enddata
\tablecomments{[1,2] Flux density and flux-density standard error of
component J1; [3--6] Position ($\alpha$ and $\delta$), relative to
component C1, and position standard error of component J1; [7,8]
Integrated flux density and flux-density standard error of component
Jext; [9--12] Position ($\alpha$ and $\delta$), relative to component
C1, and position standard error of the peak of component Jext; [13,14]
Major-axis length (FWHM) and major-axis-length statistical standard
error of component Jext; [15,16] Minor-axis length (FWHM) and
minor-axis-length statistical standard error of component Jext;
[17,18] Position angle (east of north) of major axis, and
position-angle statistical standard error, of component Jext.  The
tabulated flux-density standard errors do not include the estimated
10\% standard error in the VLBI flux-density scale at each epoch.}
\end{deluxetable}

\begin{figure}
\plotone{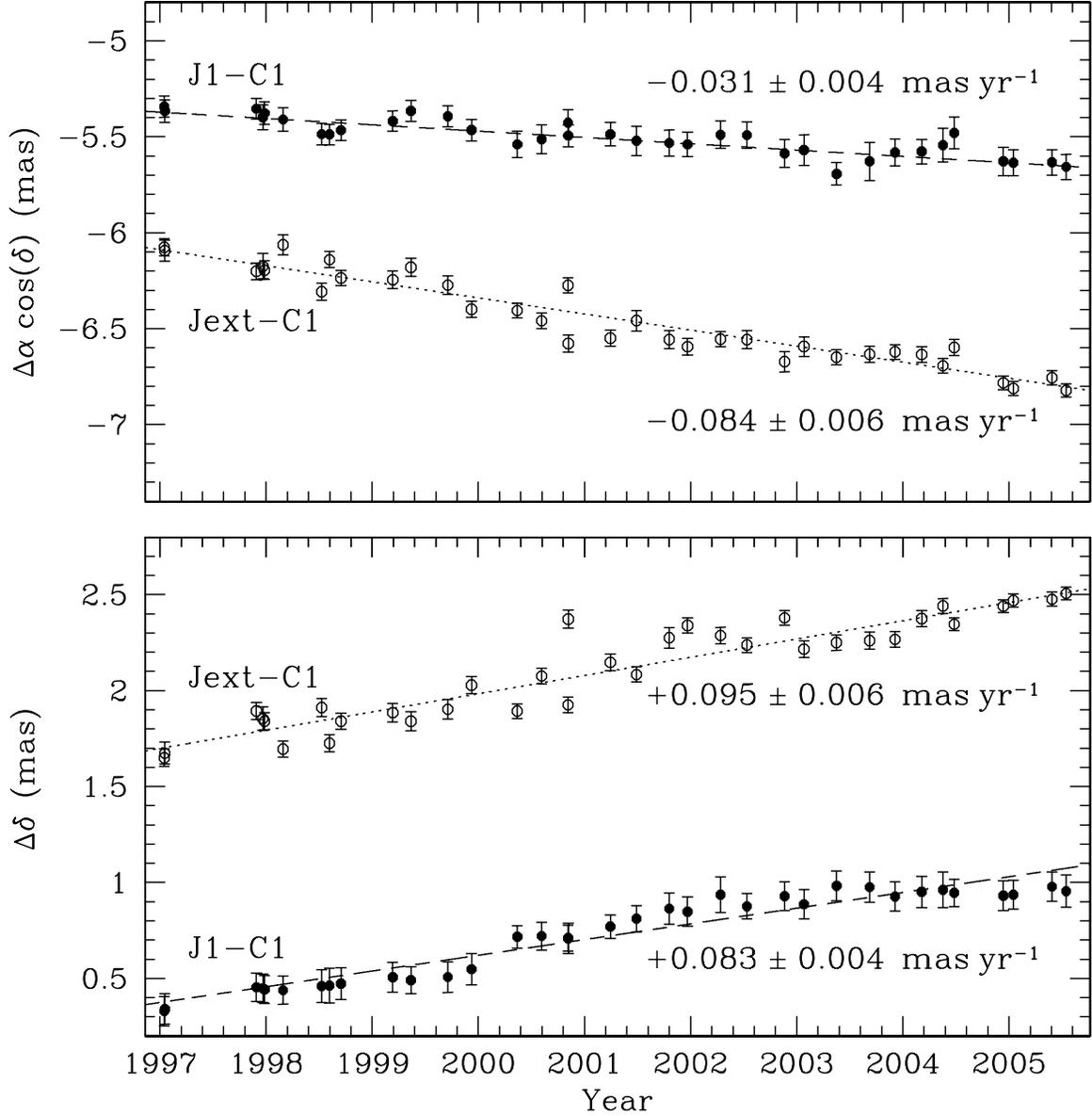}
\figcaption{Positions of 3C~454.3 extended-jet model components J1
(filled circles) and Jext (open circles) relative to core-region model
component C1.  Error bars are the rss of the standard errors in the
position estimates of C1 and J1 or Jext, as appropriate.  Note, the
larger error bars for the relative J1 positions, compared to those for
the relative Jext positions, reflect in part the $\sim$70\% larger
deconvolution errors in the former (see \S~\ref{deconv}).  The mean
proper motions of J1 and Jext relative to C1 are indicated by the
dashed and dotted lines.
\label{j1andextc1relpos}}
\end{figure}

In Figure~\ref{j1andextc1relpos}, we plot the separations at each
epoch of J1 from C1 and the peak of Jext from C1.  We see that J1
moved $\sim$0.7~mas relative to C1 over the $\sim$8.5~yr period of our
observing program, largely in the $\delta$ coordinate.  We determined
a mean relative proper motion for J1 with respect to C1 of $-0.031 \pm
0.004$~\masyr\ in $\alpha$ and $+0.083 \pm 0.004$~\masyr\ in
$\delta$.  The proper motion of J1 (see
Figure~\ref{j1andextc1relpos}), though approximately uniform in
$\alpha$, changed in $\delta$ between the first 26 epochs
($\sim$6.5~yr) and the remaining nine epochs ($\sim$2.0~yr).  Fitting
to each time period separately, we determined the relative proper
motion in $\delta$ to decrease from $+0.104 \pm 0.006$~\masyr\ for
the first 26 epochs to $+0.002 \pm 0.021$~\masyr\ for the remaining
nine.  The result for the first 26 epochs suggests that the northward
bend in the jet that starts at a position $\sim$3~mas west of C1,
i.e., at a position between transient components D1 and D2, continues
to a position $\sim$5.6~mas west of C1.  The rapid change at this
point in the direction of the proper motion may signify a second sharp
bend in the jet.  Additionally, the drastic decrease at this point in
the magnitude of the proper motion could indicate a transition away
from the highly collimated flow seen in the core and inner-jet
regions.

Figure~\ref{j1andextc1relpos} shows that the peak of Jext moved
northwest $\sim$1.1~mas relative to C1 over the $\sim$8.5~yr period of
our observing program.  We determined a relative proper motion for the
peak of Jext with respect to C1 of $-0.084 \pm 0.006$~\masyr\ in
$\alpha$ and $+0.095 \pm 0.006$~\masyr\ in $\delta$.  The proper
motion of the peak of Jext (along $\rm{p.a.} = -41\arcdeg \pm
3\arcdeg$) was precisely aligned with the component's major axis
($\rm{p.a.} = -40\arcdeg \pm 1\arcdeg$).  This result is consistent
with the 1.7~GHz space-VLBI observations of \citet{Chen+2005}, which
showed for an epoch in 1999 June that the extended jet region was
composed of 3 or 4 components.  Moreover, the decrease in the
major-axis length (FWHM) of the ``moving'' Jext from $\sim$7.3~mas to
$\sim$5.9~mas (see Table~\ref{3Cextjetcomp}) suggests that no ``new''
component transited from the inner-jet region to the extended-jet
region during our $\sim$8.5~yr observing period.  Component J1, as
discussed above, may have been close to this transition near the end
of our program.  We note that the major axis of Jext is approximately
aligned with the axis of the arcsecond-scale jet of 3C~454.3
\citep*[see][]{CheungWC2005}, suggesting that the trajectory
directions of components beyond J1 are roughly constant over
$\sim$5.2$\arcsec$ ($\sim$40~kpc).

\subsubsection{Deconvolution Errors \label{deconv}}

Deconvolution errors are known to limit the fidelity of images
produced using the CLEAN deconvolution algorithm \citep[see,
e.g.,][]{Briggs1995,Briggs+1999}.  We performed two tests to
investigate the reliability of the radio structure seen in the CLEAN
deconvolved images of 3C~454.3.  The first test was a simulation study
in which we produced \uv\ data on a ``grid'' corresponding to our
typical\footnote{We take as our typical \uv\ coverage that of epoch
2003 May~18 (see Table\ref{antab}).  We repeated the simulation study
described here for non-typical instances of our \uv\ coverage (1997
January~18, 2002 April~14, and 2002 November~20) and found
flux-density and position variations for the 3C~454.3 model components
consistent with the uncertainties derived here for the typical
coverage.}  \uv\ coverage and consisting of the Fourier transforms of
simple geometrical models similar to 3C~454.3 as described in
\S~\ref{coreregion}--\S~\ref{outerjetregion}.  Each variation of the
model used different positions and flux densities for the two bright
core-region components (C1 and C2), and in a few cases also used
different positions and flux densities for the weaker components (D1,
D2, J1, and Jext).  We Fourier-inverted the simulated data for each
variation and used the CLEAN deconvolution algorithm to produce
images.  We subsequently compared to the input models the resulting
simulated images and new models fit to the simulated images.  (The new
model, or ``output'' model, contains the same number and type of
components as the input model.)  For each variation of the input
model, we found the model structure to be reasonably well reproduced
in the simulated images (taking into account the size and orientation
of the CLEAN restoring beam), but noted that additional ``ripples''
were present in the core region with peak-to-peak amplitudes of 1--3\%
of the image peak and in the inner-jet and extended-jet regions with
amplitudes of 0.2--1\% of the image peak.  The Fourier counterparts of
these ripples occur at spatial frequencies just outside the bulk of
the \uv\ coverage.  The maximum fraction (considering several
variations of the input model) of the rms flux density in ripples near
the positions of the input-model components to the flux density in the
components themselves was $\sim$5\% for the core-region components (C1
and C2) and extended-jet component (Jext) and $\sim$10\% for the
compact-jet components (D1, D2, and J1).  When we compared the flux
densities of the output-model components to those of the input-model
components, we found corresponding differences of up to $\sim$5\% for
C1, C2, and Jext and $\sim$10\% for D1, D2, and J1.  We therefore take
these values ($\sim$5\% for each of C1, C2, and Jext and $\sim$10\%
for each of D1, D2, and J1) as a conservative estimate of the
fractional flux-density uncertainty due to deconvolution errors.  To
estimate the amount by which the ripples affect the positions of the
model components, we compared the positions of the output and input
model components for each variation.  We found the maximum difference
between the output and input component positions to be $\sim$0.020~mas
for C1 and C2 ($\sim$0.015~mas in $\alpha$ and $\sim$0.010~mas in
$\delta$), $\sim$0.100~mas for D1 ($\sim$0.080~mas in $\alpha$ and
$\sim$0.050~mas in $\delta$), $\sim$0.070~mas for D2 and J1
($\sim$0.050~mas in each coordinate), and $\sim$0.040~mas for Jext
($\sim$0.030~mas in each coordinate).  We take the maximum difference
between the positions of the output and input model components quoted
above as conservative estimates of the various position uncertainties
due to deconvolution errors.

As a second reliability test we used another deconvolution algorithm,
namely the maximum entropy method \citep*[MEM;
e.g.,][]{Cornwell+1999}, to produce images of 3C~454.3 at select
epochs.  We then performed the same fits in the image plane as
described in \S~\ref{coreregion}--\S~\ref{outerjetregion}.  We found
that the differences in the position determinations of the relevant
components in the MEM deconvolved images compared to the CLEAN
deconvolved images were well within the limits given above.

\subsection{\uv-Plane Model Fitting at 8.4~GHz \label{uvplanefits}}

\begin{figure}
\plotone{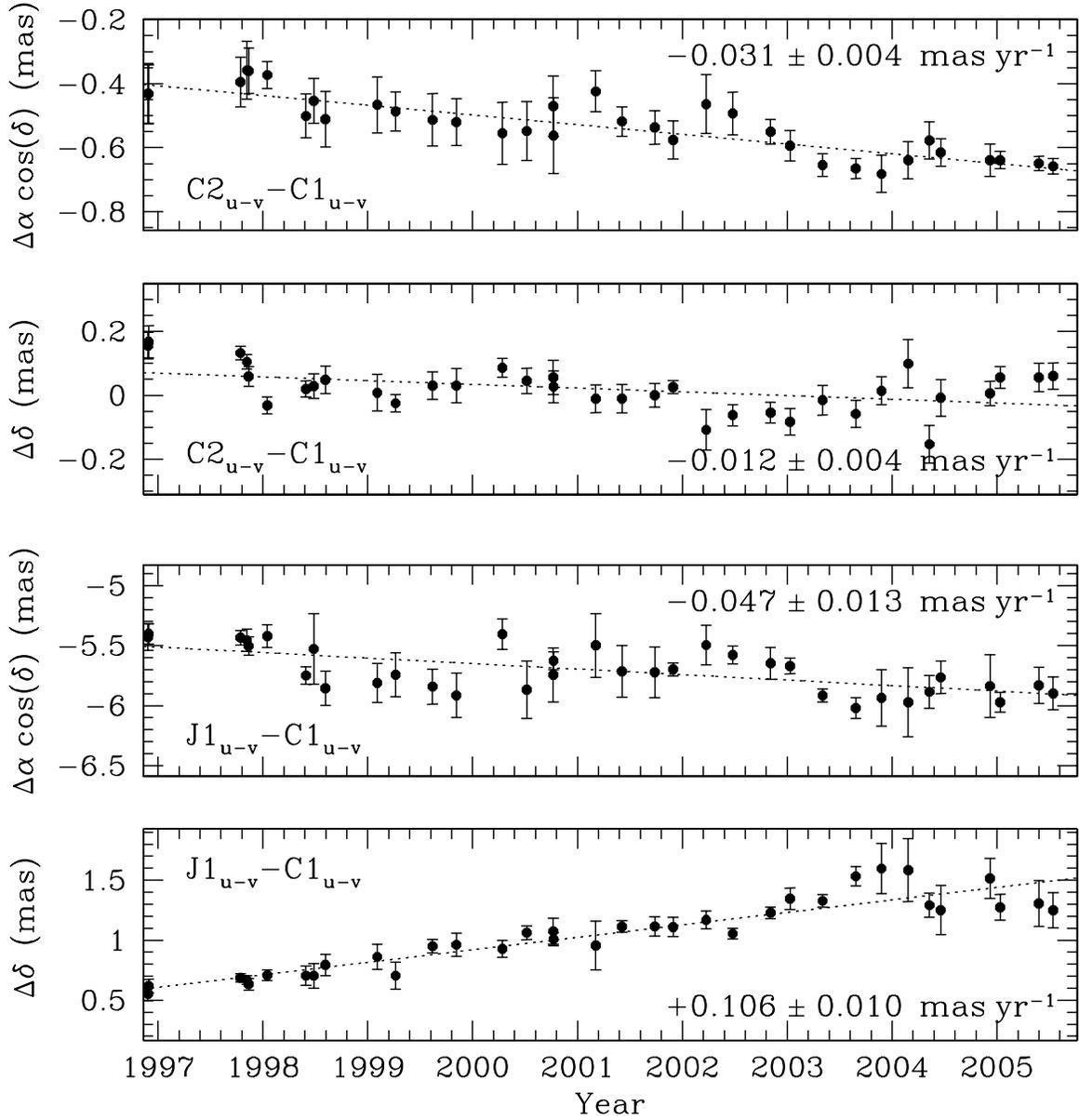}
\figcaption{Positions of 3C~454.3 \uv-plane-based model components
$\rm{C2}_{\uv}$ (top panels) and $\rm{J1}_{\uv}$ (bottom panels)
relative to $\rm{C1}_{\uv}$.  Error bars are the rss of the
statistical errors in the position estimates of each pair of
components.  The proper motions of $\rm{C2}_{\uv}$ and $\rm{J1}_{\uv}$
relative to $\rm{C1}_{\uv}$ are indicated by dotted lines.  The
uncertainties quoted in the proper motion values are statistical
errors.  Note that different scales are used for the
$\rm{C2}_{\uv}$$-$$\rm{C1}_{\uv}$ and
$\rm{J1}_{\uv}$$-$$\rm{C1}_{\uv}$ positions.
\label{uvrelpos}}
\end{figure}

We also performed an analysis of the changing 8.4~GHz radio structure
of 3C~454.3 via model fitting in the \uv\ plane.  Fitting in the \uv\
plane is not affected by deconvolution errors (as described in
\S~\ref{deconv}) and is not as sensitive to differences from epoch to
epoch in the precise \uv\ coverage.  On the other hand, we cannot,
given our \uv\ coverage, simultaneously constrain the
high-flux-density components in the core region and the
low-flux-density or low-surface-brightness components in the inner-jet
and extended-jet regions.  We fit to the \uv\ data at each epoch a
model consisting of three point sources, which describe,
approximately, the core-region components C1 and C2 as well as the
compact jet component J1 (see \S~\ref{implanefits}).  For comparison
with the image-plane models, we plot in Figure~\ref{uvrelpos} the
separations in the \uv-plane model of $\rm{C2}_{\uv}$ and
$\rm{C1}_{\uv}$ (top panels) and $\rm{J1}_{\uv}$ and $\rm{C1}_{\uv}$
(bottom panels).  The proper motion of $\rm{C2}_{\uv}$ with respect to
$\rm{C1}_{\uv}$ is consistent with that found in the image-plane
analysis.  The proper motion of $\rm{J1}_{\uv}$ with respect to
$\rm{C1}_{\uv}$ differs at the 1.5\,$\sigma$ level from that found in
the image-plane analysis.  This difference may result from structural
changes in the unmodeled extended jet over the $\sim$8.5~yr period of
our observing program.  Given the limitations of the \uv-plane-based
analysis for 3C~454.3, we believe our image-plane models (see
\S~\ref{implanefits}) provide the better overall description of the
evolving structure of this source.

\subsection{VLBI Images at 5.0 and 15.4~GHz}

\begin{figure}
\plotone{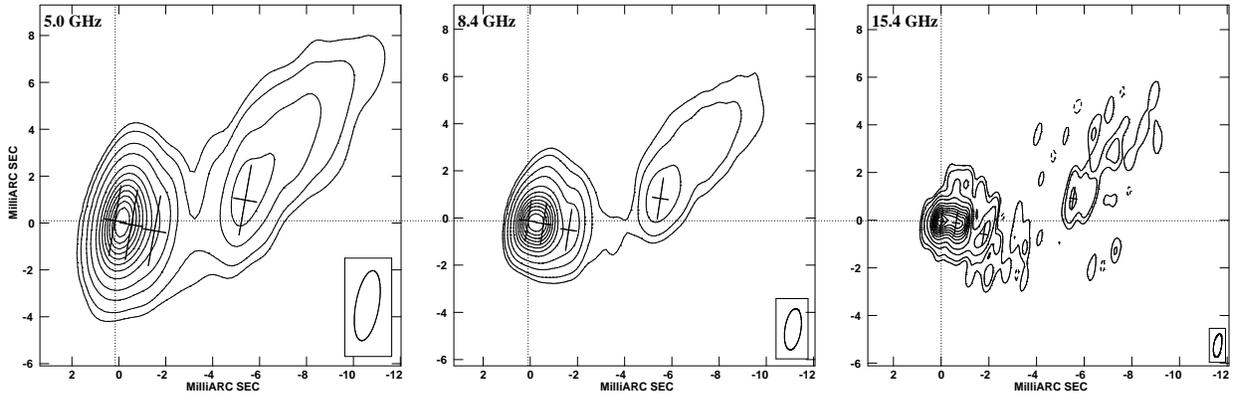}
\figcaption{5.0, 8.4, and 15.4~GHz VLBI images of 3C~454.3 for epoch
2004 May~18.  In each image the contour levels are ($-1$), $1$, $2$,
$5$, $10$, $20$, $30$, $40$, $50$, $60$, $70$, $80$, and $90$\% of the
peak brightness.  (Note, parentheses around a contour level indicates
a level that does not appear in every image.) Image characteristics
are summarized in Table~\ref{3Cimstat}.  The restoring beam is
indicated in the bottom-right-hand corner of each image.  The images
are aligned, as indicated by the dotted lines, on the center of
component C1.  The positions (from east to west) of components C1, C2,
D1, and J1 (see Table~\ref{3CcompCXU}) are indicated on each image by
crosses which reflect the size and orientation of the restoring beam.
\label{3CimagesCXU}}
\end{figure}

We present in Figure~\ref{3CimagesCXU} the 5.0, 8.4, and 15.4~GHz
images of 3C~454.3 generated from our nearly simultaneous VLBI
observations on 2004 May~18 (see \S~\ref{obs}).  The 8.4~GHz image is
the same as that presented in Figure~\ref{3Cimages} except for the
choice of contour levels.  The images are aligned on the center of the
easternmost component of the radio structure (see below).  The image
characteristics are given in Table~\ref{3Cimstat}.  The 5.0 and
15.4~GHz images are slightly noisier than the 8.4~GHz image, mainly
because the total time on source was $\sim$4~hr at each of 5.0 and
15.4~GHz compared to $\sim$7~hr at 8.4~GHz.  Nevertheless, the images
at all three frequencies show a core region extended nearly east-west
and a jet bending toward the north.  To quantitatively compare the
positions and flux densities of the core- and inner-jet-region
components, we fit to the combined core/inner-jet region in each of
the 5.0 and 15.4~GHz images the three-component model (C1, C2, and D1)
described in \S~\ref{innerjetregion}.  Similarly, to compare the
positions and flux densities of the extended-jet-region component(s),
we fit to the extended-jet region in the 5.0~GHz image the
two-component model (J1 and Jext) described in
\S~\ref{outerjetregion}.  However, at 15.4~GHz we fit only a single
point source (J1) to the region within $\sim$2~mas of the jet peak,
because the surface brightness of the extended jet in this image is
too low to allow us to adequately constrain an elliptical Gaussian
component.  The results of the fits are given in Table~\ref{3CcompCXU}
and illustrated in Figure~\ref{3CimagesCXU}.  The uncertainties in the
flux densities and positions include contributions, added in
quadrature, of the statistical standard error of the fit and an
estimate (using an analysis similar to that described in
\S~\ref{deconv}) of the systematic error associated with deconvolution
errors in the corresponding image.

\begin{deluxetable}{c@{}c@{}c@{}c@{}c@{}c@{}c@{}c@{}c@{}c@{}c@{}c@{}c@{}c@{}c@{}c@{}c@{}c@{}c@{}c@{}c@{}c}
\tabletypesize{\tiny}
\rotate
\tablecaption{3C~454.3 Image-Plane Model Parameters for 2004 May~18 \label{3CcompCXU}}
\tablewidth{0pt}
\tablehead{
  \colhead{} &
  \multicolumn{2}{c}{Component C1} &
  \multicolumn{6}{c}{-----------------------Component C2-----------------------} &
  \multicolumn{6}{c}{-----------------------Component D1-----------------------} &
  \multicolumn{6}{c}{-----------------------Component J1-----------------------} \\
  \colhead{$\nu$} &
  \colhead{$S_{\nu}$} &
  \colhead{$\sigma_{S}$} &
  \colhead{$S_{\nu}$} &
  \colhead{$\sigma_{S}$} &
  \colhead{$\Delta$$\alpha$} &
  \colhead{$\sigma_{\alpha}$} &
  \colhead{$\Delta$$\delta$} &
  \colhead{$\sigma_{\delta}$} &
  \colhead{$S_{\nu}$} &
  \colhead{$\sigma_{S}$} &
  \colhead{$\Delta$$\alpha$} &
  \colhead{$\sigma_{\alpha}$} &
  \colhead{$\Delta$$\delta$} &
  \colhead{$\sigma_{\delta}$} &
  \colhead{$S_{\nu}$} &
  \colhead{$\sigma_{S}$} &
  \colhead{$\Delta$$\alpha$} &
  \colhead{$\sigma_{\alpha}$} &
  \colhead{$\Delta$$\delta$} &
  \colhead{$\sigma_{\delta}$} \\
  \colhead{(GHz)} &
  \colhead{(Jy)} &
  \colhead{(Jy)} &
  \colhead{(Jy)} &
  \colhead{(Jy)} &
  \colhead{(mas)} &
  \colhead{(mas)} &
  \colhead{(mas)} &
  \colhead{(mas)} &
  \colhead{(Jy)} &
  \colhead{(Jy)} &
  \colhead{(mas)} &
  \colhead{(mas)} &
  \colhead{(mas)} &
  \colhead{(mas)} &
  \colhead{(Jy)} &
  \colhead{(Jy)} &
  \colhead{(mas)} &
  \colhead{(mas)} &
  \colhead{(mas)} &
  \colhead{(mas)} \\
  \colhead{} &
  \colhead{[1]} &
  \colhead{[2]} &
  \colhead{[3]} &
  \colhead{[4]} &
  \colhead{[5]} &
  \colhead{[6]} &
  \colhead{[7]} &
  \colhead{[8]} &
  \colhead{[9]} &
  \colhead{[10]} &
  \colhead{[11]} &
  \colhead{[12]} &
  \colhead{[13]} &
  \colhead{[14]} &
  \colhead{[15]} &
  \colhead{[16]} &
  \colhead{[17]} &
  \colhead{[18]} &
  \colhead{[19]} &
  \colhead{[20]}
}
\startdata
\phm{0}5.0 & \phm{000}2.94\phm{000} & \phm{000}0.20\phm{000} & 2.73 & 0.14 & $-0.701$ & 0.022 & $-0.178$ & 0.021 & 0.946 & 0.096 & $-1.685$ & 0.081 & $-0.436$ & 0.055 & 0.328 & 0.033 & $-5.553$ & 0.052 & 0.878 & 0.061 \\
\phm{0}8.4 & \phm{000}2.42\phm{000} & \phm{000}0.15\phm{000} & 2.45 & 0.13 & $-0.650$ & 0.021 & $-0.120$ & 0.016 & 0.621 & 0.078 & $-1.665$ & 0.088 & $-0.380$ & 0.053 & 0.194 & 0.019 & $-5.577$ & 0.051 & 0.944 & 0.054 \\
15.4       & \phm{000}1.41\phm{000} & \phm{000}0.08\phm{000} & 1.22 & 0.07 & $-0.626$ & 0.018 & $-0.077$ & 0.017 & 0.271 & 0.028 & $-1.782$ & 0.052 & $-0.551$ & 0.050 & 0.096 & 0.010 & $-5.544$ & 0.036 & 0.950 & 0.045 \\
\enddata
\tablecomments{[1,2] Flux density and flux-density standard error of
component C1; [3,4] Flux density and flux-density standard error of
component C2; [5--8] Position ($\alpha$ and $\delta$), relative to
component C1, and position standard error of component C2; [9,10] Flux
density and flux-density standard error of component D1; [11--14]
Position ($\alpha$ and $\delta$), relative to component C1, and
position standard error of component D1; [15,16] Flux density and
flux-density standard error of component J1; [17--20] Position
($\alpha$ and $\delta$), relative to component C1, and position
standard error of component J1.  The tabulated flux-density standard
errors do not include the estimated 5\% standard error in the VLBI
flux-density scale for 2004 May~18 (see text).}
\end{deluxetable}

We choose in Figure~\ref{3CimagesCXU} to align the images at each
frequency on the (fit-estimated) center of component C1.  This is a
reasonable choice, since our comparison of the structure of the core
region at 8.4 and 43~GHz in \S~\ref{coreregion} shows that the 8.4~GHz
C1 is located within $\sim$0.2~mas of the 43-GHz-identified core.  But
how well do the images at 5.0, 8.4, and 15.4~GHz actually line-up?  To
answer this question we look at the separation of J1 and C1 at each
frequency.  The jet peak is clearly identified at each frequency, and
thus serves as a useful marker in the radio structure.  The largest
difference between the three images in the J1--C1 separation (see the
last four columns of Table~\ref{3CcompCXU}) is $0.033 \pm 0.063$~mas
in $\alpha$ and $0.072 \pm 0.076$~mas in $\delta$.  We therefore
believe that the images at 5.0, 8.4, and 15.4~GHz are aligned
(relative to these internal reference points) to within 0.10~mas in
$\alpha$ and 0.15~mas in $\delta$; i.e., to better than $\sim$15\% of
the resolution of the 5.0~GHz image.

The separations of C2 and D1 from C1 are more variable with frequency
than the separation of J1 from C1.  This difference is not surprising
considering the complexity of the underlying structure in the core and
inner-jet regions.  The $\sim$1.8 times higher angular resolution of
the 15.4~GHz image, compared to that of the 8.4~GHz image, reveals
that the core and inner-jet regions for 2004 May~18 contain more than
three components.  Specifically, there appears to be a component
located between the positions of C1 and C2, and another component
located west-southwest of the position of C2.  Can we understand the
differences in the fit-estimated C2--C1 and D1--C1 separations at each
frequency in terms of the frequency-dependent flux densities of these
unmodeled components?  Generally speaking, the spectral index ,
$\alpha$ ($S \sim \nu^{\alpha}$), of a component is steeper (i.e.,
more negative) the farther west it is from the core
\citep[e.g.,][]{Pagels+2004,Chen+2005}.  Thus, relative to C1 and C2,
a component lying between them will likely have a larger fractional
flux density at 15.4~GHz than at 8.4 or 5.0~GHz, while a component
west of C2 will have a larger fractional flux density at 5.0~GHz than
at either 8.4 or 15.4~GHz.  These flux-density differences would
create at 15.4~GHz a ``plateau'' between C1 and C2 (as observed) and
at 5.0~GHz a west-broadened C2.  For a core-region modeled only by C1
and C2, we might then expect a C2--C1 separation which is smaller at
15.4~GHz than at 5.0~GHz.  This is indeed what we find in our model
fits (see Table~\ref{3CcompCXU}).  In the inner-jet region, the
situation is reversed.  Here, the component west-southwest of C2 might
be expected to create at 5.0~GHz a plateau between C2 and D1, and
yield a D1--C1 separation which is smaller at this frequency than at
15.4~GHz.  These expectations, too, are confirmed by our model fits.

We give in Table~\ref{3CspecCXU} the flux-density ratio
($\frac{S_{\nu_{high}}}{S_{\nu_{low}}}$) and resulting point-to-point
spectral index ($\alpha = \log{\frac{S_{\nu_{high}}}{S_{\nu_{low}}}} /
\log{\frac{\nu_{high}}{\nu_{low}}}$) for each model component for each
pair of frequencies.  We also give for comparison the total (CLEAN)
flux-density ratio (using the values in Table~\ref{3Cimstat}) and
resulting point-to-point spectral index for each pair of frequencies.
Note that to compute the standard errors in the flux densities, we
added in quadrature to the other contributions a 5\% uncertainty
representative of the standard error in the VLBI flux-density scale.
We adopt a 5\% uncertainty for this epoch rather than the nominal 10\%
(see \S~\ref{VLBIamp}), because the VLBI- and VLA-determined flux
densities at each frequency differ by only 2--7\% (see
Table~\ref{3Cimstat}).  The slightly lower ratio of VLBI-determined to
VLA-determined flux density at 5.0~GHz compared to those at 8.4 and
15.4~GHz is likely due to the insensitivity of our VLBI observations
to low-surface-brightness (and steep-spectral-index) emission in the
extended jet at distances $\gtrsim$15~mas northwest of the core
\citep[see][]{Chen+2005}.

\begin{deluxetable}{c@{~~}c@{~~}c@{~~}c@{~~}c@{~~}c@{~~}c@{~~}c@{~~}c@{~~}c@{~~}c}
\tabletypesize{\scriptsize}
\rotate
\tablecaption{3C~454.3 Spectral Indices for 2004 May~18 \label{3CspecCXU}}
\tablewidth{0pt}
\tablehead{
  \colhead{$\nu_{high}$/$\nu_{low}$} &
  \colhead{C1} &
  \colhead{$\alpha_{C1}$} &
  \colhead{C2} &
  \colhead{$\alpha_{C2}$} &
  \colhead{D1} &
  \colhead{$\alpha_{D1}$} &
  \colhead{J1} &
  \colhead{$\alpha_{J1}$} &
  \colhead{Total} &
  \colhead{$\alpha_{\rm{Tot}}$} \\
  \colhead{[1]} &
  \colhead{[2]} &
  \colhead{[3]} &
  \colhead{[4]} &
  \colhead{[5]} &
  \colhead{[6]} &
  \colhead{[7]} &
  \colhead{[8]} &
  \colhead{[9]} &
  \colhead{[10]} &
  \colhead{[11]}
}
\startdata
\phm{0}8.4/5.0 & $0.823 \pm 0.096$ & $-0.38 \pm 0.22$ & $0.897 \pm 0.097$ & $-0.21 \pm 0.20$ & $0.656 \pm 0.081$ & $-0.81 \pm 0.25$ & $0.592 \pm 0.093$ & $-1.01 \pm 0.30$ & $0.864 \pm 0.061$ & $-0.28 \pm 0.14$ \\
15.4/8.4       & $0.583 \pm 0.064$ & $-0.89 \pm 0.19$ & $0.498 \pm 0.052$ & $-1.14 \pm 0.17$ & $0.392 \pm 0.061$ & $-1.37 \pm 0.28$ & $0.495 \pm 0.079$ & $-1.16 \pm 0.26$ & $0.692 \pm 0.049$ & $-0.61 \pm 0.12$ \\
15.4/5.0       & $0.480 \pm 0.055$ & $-0.65 \pm 0.11$ & $0.447 \pm 0.046$ & $-0.72 \pm 0.09$ & $0.287 \pm 0.043$ & $-1.11 \pm 0.14$ & $0.293 \pm 0.047$ & $-1.09 \pm 0.14$ & $0.598 \pm 0.042$ & $-0.46 \pm 0.06$ \\
\enddata
\tablecomments{[1] Frequency pair for which flux-density ratios and
spectral indices are computed; [2,3] Flux-density ratio and
point-to-point spectral index (see text) for component C1; [4,5]
Flux-density ratio and point-to-point spectral index for component C2;
[6,7] Flux-density ratio and point-to-point spectral index for
component D1; [8,9] Flux-density ratio and point-to-point spectral
index for component J1; [10,11] CLEAN flux-density ratio and
point-to-point spectral index.}
\end{deluxetable}

The spectral indices for C1, C2, and D1 in Table~\ref{3CspecCXU}
should be viewed with some caution.  As noted above, the structure in
the core and inner-jet regions is more complicated than that described
by our three-component model.  Moreover, the flux density in each
unmodeled component is probably allocated differently to C1, C2, and
D1 at each frequency due to differences in resolution and opacity.
Nonetheless, we can still make some general comments and compare our
results with previous multi-frequency studies: (1) Estimates of the
spectral index of the 43- and 86-GHz-identified core range between
$\alpha \approx +0.6$ and $\alpha \approx -0.4$
(\citealt{GomezMA1999}; \citealt{Pagels+2004}; \citealt{Chen+2005};
\citealt{Krichbaum+2006a}).  Our estimate of the spectral index of C1
between 5.0 and 8.4~GHz ($\alpha = -0.4 \pm 0.2$) is at the low end of
this range, and consistent with an isolated core several months
removed from its most recent ejection \citep{Pagels+2004}.  Our
spectral-index estimate for C1 between 8.4 and 15.4~GHz ($\alpha =
-0.9 \pm 0.2$) suggests that the component between C1 and C2 is
contributing fractionally more flux density to C1 at 8.4~GHz than at
15.4~GHz (where it is partially resolved). (2) The ``stationary''
component $\sim$0.65~mas from the core is modeled by \citet{Chen+2005}
for their 1999 May epoch as a synchrotron self-absorbed source, with a
turnover frequency between 5.0 and 8.4~GHz.  The flat spectral index
we estimate for C2 between 5.0 and 8.4~GHz ($\alpha = -0.2 \pm 0.2$)
is consistent with this interpretation.  Our spectral-index estimate
for C2 between 8.4 and 15.4~GHz ($\alpha = -1.1 \pm 0.2$) suggests
that the component west-southwest of C2 is contributing fractionally
more flux density to C2 at 8.4~GHz than at 15.4~GHz. (3)
\citet{Chen+2005} find for a component located $\sim$1.5~mas
west-southwest of the core a spectral index of $\alpha \approx -0.9$.
Our spectral-index estimates for D1 ($\alpha$ $\approx$ $-1.4$ to
$-0.8$) are roughly consistent with this value, but again suggest that
components between C2 and D1 are contributing fractionally more flux
density to D1 at 5.0 and 8.4~GHz than at 15.4~GHz.

In contrast to the core and inner-jet regions, the extended-jet region
appears to be reasonably well represented by our two-component model
at 5.0 and 8.4~GHz, and likewise the (immediate) jet-peak region by
our one-component model at 15.4~GHz.  Our estimates of the spectral
index of J1 ($\alpha$ $\approx$ $-1.2$ to $-1.0$) are self-consistent
and also consistent with the spectral index ($\alpha \approx -0.9$)
estimated by \citet{Chen+2005} for a component approximately
$\sim$5.5~mas west of the core.  We note that the lower-level
structure extending west of J1 in the 15.4~GHz image is likely real,
and not a noise spike or sidelobe of one of the bright core-region
components.  Its appearance is robust against changes (e.g., to the
positioning of CLEAN windows) in the imaging process.  The east-west
extension at J1 at all three frequencies is, perhaps, a consequence of
a near superposition of jet components undergoing the change in
trajectory described in \S~\ref{outerjetregion}.

Can we determine the ejection epoch for the component seen between C1
and C2 in our 15.4~GHz image for 2004 May~18, and perhaps also the
time of transit of the component from C1 to C2?  Our 8.4~GHz image for
2003 September~8 shows a ``sudden'' brightening of C1 relative to C2
not seen 2003 May~18 or earlier.  On 2003 December~5, the 8.4~GHz
brightness peak is very nearly aligned with the center of C1.  Over
the next several epochs, the brightness peak moves westward, and by
2005 January~15 is very nearly aligned with the center of C2.  If we
assume that the brightening in 2003 September~8 temporally coincides
(within $\sim$1--2 months) with the ejection of a new component from
the ``core,'' then we may interpret the subsequent shifts in the
position of the brightness peak as roughly tracking the motion of the
new component between the positions defined by C1 and C2.  The total
transit time, using 2003 September~8 as the epoch of ejection and 2005
January~15 as the epoch of arrival, is $\sim$1.4~yr.  The 2004 May~18
epoch represents the approximate mid-point of this interval.

\section{Quasar B2250+194 \label{2250}}

\subsection{VLBI Images at 8.4~GHz \label{2250imagestext}}

\begin{figure}
\plotone{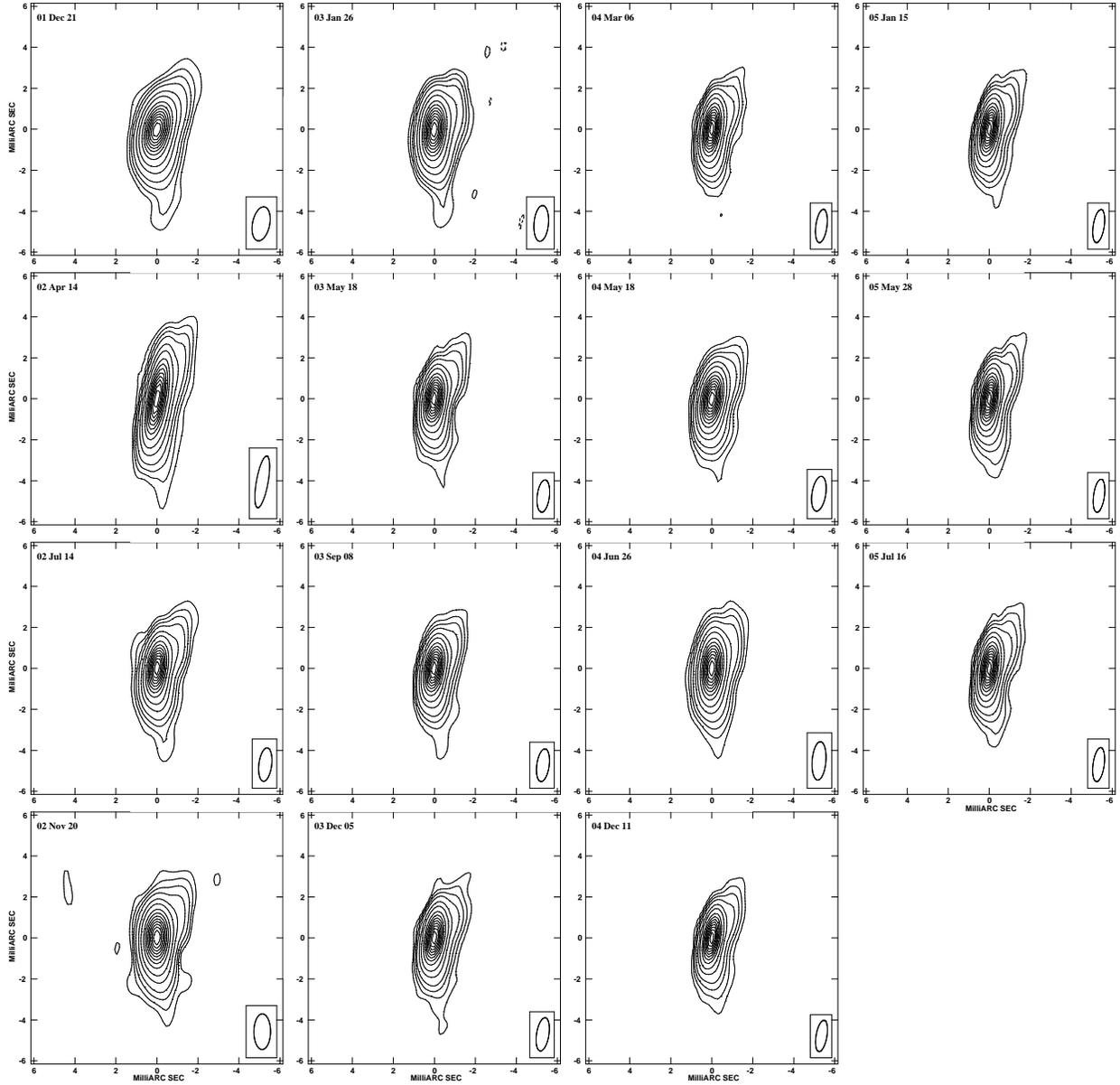}
\figcaption{8.4~GHz VLBI images of B2250+194 for each of our 35
observing sessions between 1997 January and 2005 July.  In each image
the contour levels are ($-0.2$), $0.2$, $0.5$, $1$, $2$, $5$, $10$,
$20$, $30$, $40$, $50$, $60$, $70$, $80$, and $90$\% of the peak
brightness.  Image characteristics are summarized in
Table~\ref{2250imstat}.  The restoring beam is indicated in the
bottom-right-hand corner of each image.  The image for each epoch is
centered on the brightness peak, which is offset from the phase center
(i.e., origin of coordinates) by $<$0.02~mas in each of $\alpha$ and
$\delta$.
\label{2250imagesplot}}
\end{figure}

We present in Figure~\ref{2250imagesplot} the 8.4~GHz images of
B2250+194 generated from each of the 35 sessions of VLBI observations
made between 1997~January and 2005~July.  The image characteristics
are summarized in Table~\ref{2250imstat}.  The images reveal a
relatively compact source.  Near the brightness peak, the source shows
no deviation from the restoring beam, down to the $\sim$10\% level.
Below this level, i.e., at the brightness level of the outermost five
or six contours, the source shows a distinct ``kidney-bean'' shape:
The northern portion of the source bends $\sim$30\arcdeg\ to the
west, while the southern portion of the source bends $\sim$10\arcdeg\@
to the west.  Other than the $\sim$25\% increase in the peak
brightness over the $\sim$8.5~yr period of our observing program,
there is no significant indication of radio structure evolution in the
images.  The low-level kidney-bean structure may trace (1) continuous
emission along a single, bent jet, (2) emission from well-separated
regions in a single, bent jet viewed nearly straight on, or (3)
emission from a pair of bent (or at least not quite counter-aligned)
jets.  If scenario 1 were correct, then the core of the quasar could
be located at either the north or south end of the source, and the
centrally-located brightness peak would be a bright jet component.
If, on the other hand, scenario 2 or 3 were correct, then the
brightness peak could be nearly aligned with the core.  We elaborate
more on these options in \S~\ref{2250uvplanefits} and
\S~\ref{2250canduimages}.

\begin{figure}
\plotone{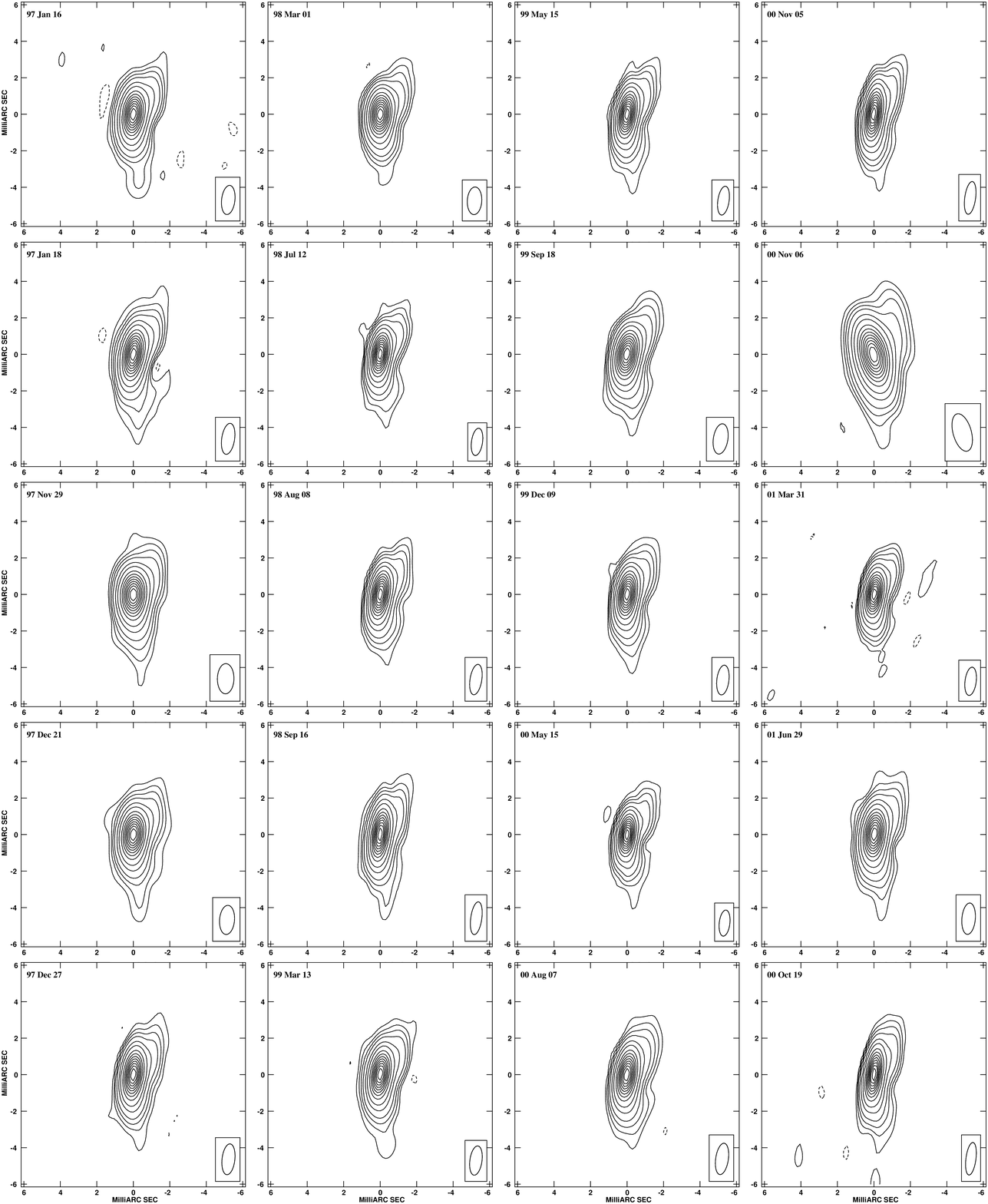}
\end{figure}

\begin{deluxetable}{l@{~}l@{~}l@{\,}r c c c c c c c c}
\tabletypesize{\scriptsize}
\tablecaption{B2250+194 Image Characteristics \label{2250imstat}}
\tablewidth{0pt}
\tablehead{
  \multicolumn{3}{c}{Start Date} &
  \colhead{} &
  \colhead{Peak} &
  \colhead{Min.} &
  \colhead{rms} &
  \colhead{$\Theta_{\rm{maj}}$} &
  \colhead{$\Theta_{\rm{min}}$} &
  \colhead{p.a.} &
  \colhead{CLEAN} &
  \colhead{$[\frac{\rm{CLEAN}}{\rm{VLA}}]$} \\
  \multicolumn{3}{c}{} &
  \colhead{} &
  \colhead{($\rm{mJy}\,\Omega_b^{-1}$)} &
  \colhead{($\rm{mJy}\,\Omega_b^{-1}$)} &
  \colhead{($\rm{mJy}\,\Omega_b^{-1}$)} &
  \colhead{(mas)} &
  \colhead{(mas)} &
  \colhead{($\arcdeg$)} &
  \colhead{(mJy)} &
  \colhead{} \\
  \multicolumn{3}{c}{} &
  \colhead{} &
  \colhead{[1]} &
  \colhead{[2]} &
  \colhead{[3]} &
  \colhead{[4]} &
  \colhead{[5]} &
  \colhead{[6]} &
  \colhead{[7]} &
  \colhead{[8]}
}
\startdata
1997 & Jan & 16 &                      & 318 & $-0.85$ & 0.18 & 1.60 & 0.73 & \phn$-6.39$    & 407 & 0.93  \\
1997 & Jan & 18 &                      & 314 & $-0.80$ & 0.14 & 1.71 & 0.70 & \phn$-8.38$    & 396 & 0.92  \\
1997 & Nov & 29 &                      & 356 & $-0.65$ & 0.18 & 1.65 & 0.91 & \phn$-0.96$    & 415 & 0.89  \\
1997 & Dec & 21 &                      & 383 & $-0.63$ & 0.14 & 1.60 & 0.82 & \phn$-4.45$    & 457 & 1.01  \\
1997 & Dec & 27 &                      & 367 & $-0.75$ & 0.16 & 1.70 & 0.70 & \phn$-6.38$    & 429 & 0.99  \\
1998 & Mar & 01 &                      & 401 & $-0.83$ & 0.15 & 1.51 & 0.79 & \phn$-3.12$    & 468 & 0.96  \\
1998 & Jul & 12 &                      & 360 & $-0.70$ & 0.15 & 1.51 & 0.61 & \phn$-8.03$    & 435 & 0.97  \\
1998 & Aug & 08 &                      & 373 & $-0.64$ & 0.12 & 1.68 & 0.64 & \phn$-8.94$    & 438 & 1.05  \\
1998 & Sep & 16 &                      & 360 & $-0.60$ & 0.14 & 1.81 & 0.61 & \phn$-7.84$    & 440 & 0.97  \\
1999 & Mar & 13 &                      & 313 & $-0.66$ & 0.12 & 1.61 & 0.66 & \phn$-7.87$    & 401 & 0.93  \\
1999 & May & 15 &                      & 347 & $-0.37$ & 0.09 & 1.59 & 0.62 & \phn$-8.41$    & 428 & 0.94  \\
1999 & Sep & 18 &                      & 382 & $-0.76$ & 0.14 & 1.67 & 0.80 & \phn$-8.65$    & 445 & 1.06  \\
1999 & Dec & 09 &                      & 360 & $-0.57$ & 0.12 & 1.64 & 0.66 & \phn$-6.67$    & 433 & 0.93  \\
2000 & May & 15 &                      & 357 & $-0.64$ & 0.15 & 1.43 & 0.58 & \phn$-6.52$    & 441 & 1.02  \\
2000 & Aug & 07 &                      & 346 & $-0.72$ & 0.12 & 1.73 & 0.69 & \phn$-8.13$    & 412 & 0.98  \\
2000 & Nov & 05 &                      & 354 & $-0.41$ & 0.10 & 1.81 & 0.59 & \phn$-7.77$    & 429 & 0.98  \\
2000 & Nov & 06 &                      & 341 & $-0.67$ & 0.17 & 2.12 & 1.00 & \phm{$-$}15.75 & 406 & 0.92  \\
2001 & Mar & 31 &                      & 331 & $-0.79$ & 0.14 & 1.56 & 0.61 & \phn$-7.17$    & 403 & 0.91  \\
2001 & Jun & 29 &                      & 320 & $-0.59$ & 0.08 & 1.72 & 0.74 & \phn$-4.87$    & 401 & 1.02  \\
2001 & Oct & 19 &                      & 269 & $-0.83$ & 0.19 & 1.79 & 0.55 & \phn$-7.32$    & 347 & 0.91  \\
2001 & Dec & 21 &                      & 321 & $-0.53$ & 0.11 & 1.69 & 0.84 & $-11.40$       & 406 &  ---  \\
2002 & Apr & 14 &                      & 276 & $-0.49$ & 0.13 & 2.56 & 0.57 & $-10.13$       & 339 & 0.94  \\
2002 & Jul & 14 &                      & 264 & $-0.46$ & 0.11 & 1.64 & 0.62 & \phn$-6.85$    & 354 &  ---  \\
2002 & Nov & 20 &                      & 314 & $-0.62$ & 0.16 & 1.74 & 0.81 & \phn$-0.34$    & 396 &  ---  \\
2003 & Jan & 26 &                      & 312 & $-0.71$ & 0.17 & 1.75 & 0.70 & \phn$-4.99$    & 401 & 0.95  \\
2003 & May & 18 &                      & 330 & $-0.52$ & 0.10 & 1.58 & 0.59 & \phn$-6.79$    & 409 & 0.96  \\
2003 & Sep & 08 &                      & 390 & $-0.61$ & 0.12 & 1.60 & 0.60 & \phn$-7.74$    & 468 & 0.98  \\
2003 & Dec & 05 &                      & 408 & $-0.80$ & 0.15 & 1.64 & 0.60 & \phn$-8.95$    & 488 &  ---  \\
2004 & Mar & 06 &                      & 437 & $-0.55$ & 0.13 & 1.63 & 0.54 & \phn$-7.54$    & 516 & 1.01  \\
2004 & May & 18 &                      & 456 & $-0.63$ & 0.16 & 1.71 & 0.67 & \phn$-8.95$    & 529 & 0.99  \\
2004 & May & 18 & (C)\tablenotemark{a} & 290 & $-0.70$ & 0.18 & 3.04 & 0.96 & $-11.21$       & 367 & 0.96  \\
2004 & May & 18 & (U)\tablenotemark{a} & 502 & $-2.89$ & 0.61 & 0.98 & 0.35 & \phn$-7.72$    & 567 & 1.01  \\
2004 & Jun & 26 &                      & 423 & $-0.55$ & 0.10 & 1.88 & 0.67 & \phn$-4.42$    & 511 & 0.97  \\
2004 & Dec & 11 &                      & 418 & $-0.48$ & 0.10 & 1.55 & 0.53 & \phn$-9.74$    & 494 & 0.97  \\
2005 & Jan & 15 &                      & 431 & $-0.54$ & 0.08 & 1.63 & 0.54 & \phn$-8.81$    & 507 & 0.96  \\
2005 & May & 28 &                      & 421 & $-0.38$ & 0.09 & 1.62 & 0.53 & \phn$-7.78$    & 495 & 0.99  \\
2005 & Jul & 16 &                      & 425 & $-0.45$ & 0.09 & 1.67 & 0.55 & \phn$-7.67$    & 494 & 0.95  \\
\enddata
\tablenotetext{a}{C indicates 5.0~GHz; U indicates 15.4~GHz; all other
observations were at 8.4~GHz.}
\tablecomments{[1,2] Positive and negative extrema ($\Omega_b$
$\equiv$ CLEAN restoring beam area); [3] rms background level; [4--6]
CLEAN restoring beam major axis (FWHM), minor axis (FWHM), and
position angle (east of north); [7] Total CLEAN flux density; [8]
Ratio of VLBI-determined (column 7) to VLA-determined
(Table~\ref{fluxdensities}) flux density for all epochs at which the
VLA observed.  The estimated standard error in the VLBI-determined
flux densities is 10\%.}
\end{deluxetable}

\subsection{\uv-Plane Model Fitting at 8.4~GHz \label{2250uvplanefits}}

\begin{figure}
\plotone{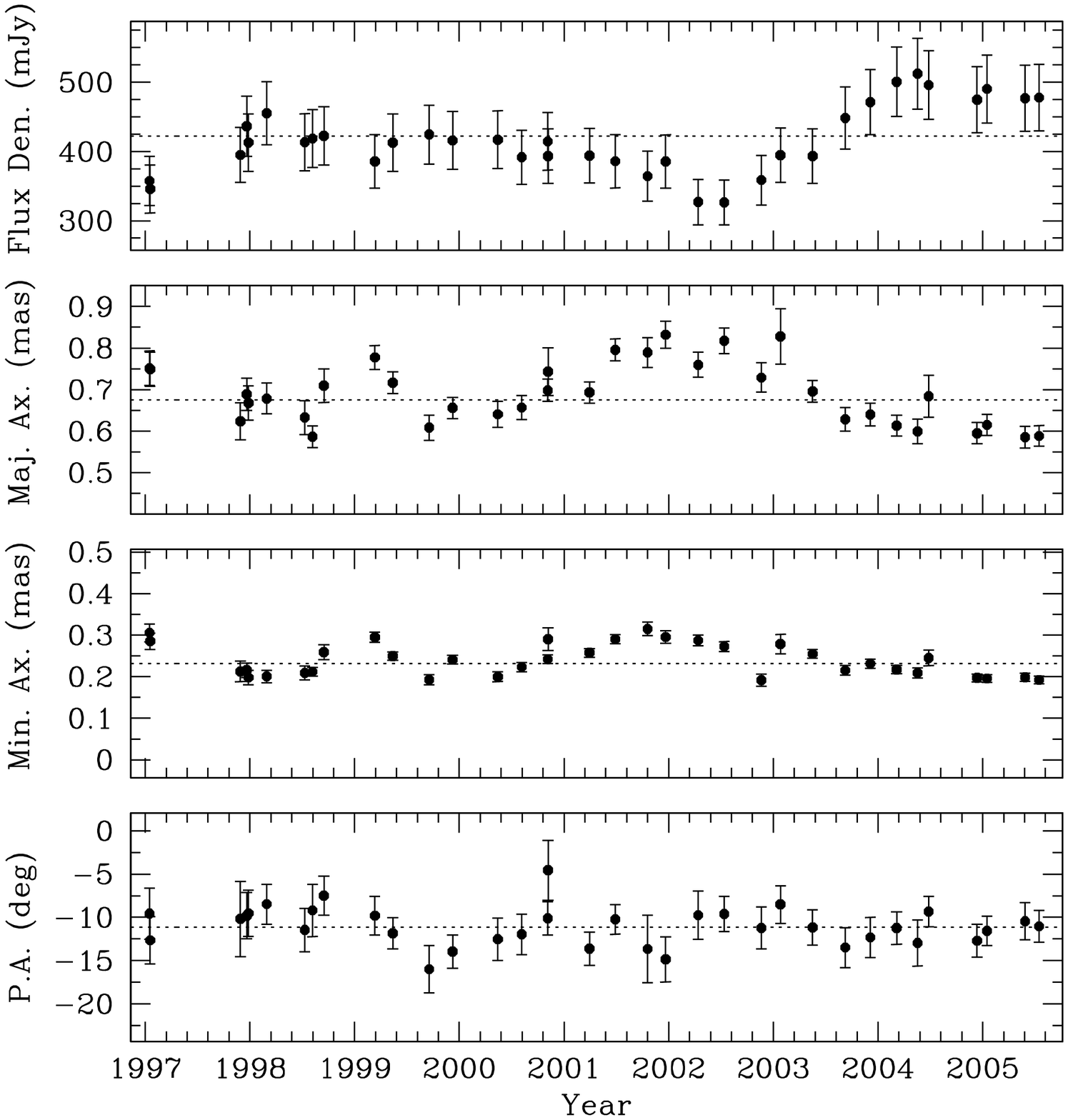} 
\figcaption{B2250+194 model parameters (see also
Table~\ref{2250model}).  Panels, top to bottom, are integrated flux
density, major-axis length (FWHM), minor-axis length (FWHM), and
position angle (east of north) of the major axis.  Error bars
represent the standard errors in the estimates of the corresponding
parameter.  The dotted line in each panel represents the weighted mean
of the parameter value over the 35 epochs.
\label{2250modelplot}}
\end{figure}

To more quantitatively investigate changes in the radio structure of
B2250+194, we fit to the \uv\ data, not including baselines to the
70-m DSN Tidbinbilla (Ti) telescope (see below), a model consisting of
a single elliptical Gaussian.  The results for the fits are given in
Table~\ref{2250model} and illustrated in Figure~\ref{2250modelplot}.
The uncertainty in each parameter at each epoch is a standard error;
it includes contributions, added in quadrature, from both the
statistical standard error of the fit and the systematic error
associated with the particular \uv\ coverage.  To estimate the
systematic error we performed the fit at each epoch an additional five
or six times, using for each additional fit only a subset (8--12
telescopes) of the full VLBI array (minus Ti).  We take the systematic
error for a given parameter at a given epoch to be the rms variation
of the subset values about the value derived from all of the available
data at that epoch.  The uncertainty in the tabulated flux density at
each epoch includes also the 10\% standard error in the VLBI
flux-density scale.

\begin{deluxetable}{l@{~}l@{~}l@{~~}c@{}c@{}c@{}c@{}c@{}c@{}c@{}c}
\tabletypesize{\scriptsize}
\tablecaption{B2250+194 \uv-Plane 8.4~GHz Model Parameters \label{2250model}}
\tablewidth{0pt}
\tablehead{
  \multicolumn{3}{c}{Start Date} &
  \colhead{$S_{8.4}$} &
  \colhead{$\sigma_{S}$} &
  \colhead{$\rm{Maj}$} &
  \colhead{$\sigma_{\rm{Maj}}$} &
  \colhead{$\rm{Min}$} &
  \colhead{$\sigma_{\rm{Min}}$} &
  \colhead{$\rm{p.a.}$} &
  \colhead{$\sigma_{\rm{p.a.}}$} \\
  \multicolumn{3}{c}{} &
  \colhead{(mJy)} &
  \colhead{(mJy)} &
  \colhead{(mas)} &
  \colhead{(mas)} &
  \colhead{(mas)} &
  \colhead{(mas)} &
  \colhead{($\arcdeg$)} &
  \colhead{($\arcdeg$)} \\
  \multicolumn{3}{c}{} &
  \colhead{[1]} &
  \colhead{[2]} &
  \colhead{[3]} &
  \colhead{[4]} &
  \colhead{[5]} &
  \colhead{[6]} &
  \colhead{[7]} &
  \colhead{[8]}
}
\startdata
1997 & Jan & 16 & 357 & 36 & 0.75 & 0.04 & 0.31 & 0.02 & $-10$    & 3 \\
1997 & Jan & 18 & 346 & 35 & 0.75 & 0.04 & 0.29 & 0.02 & $-13$    & 3 \\
1997 & Nov & 29 & 395 & 40 & 0.62 & 0.05 & 0.21 & 0.02 & $-10$    & 4 \\
1997 & Dec & 21 & 436 & 44 & 0.69 & 0.04 & 0.22 & 0.02 & $-10$    & 3 \\
1997 & Dec & 27 & 413 & 41 & 0.67 & 0.04 & 0.20 & 0.02 & $-10$    & 3 \\
1998 & Mar & 01 & 455 & 46 & 0.68 & 0.04 & 0.20 & 0.02 & \phn$-8$ & 2 \\
1998 & Jul & 12 & 413 & 41 & 0.63 & 0.04 & 0.21 & 0.02 & $-11$    & 3 \\
1998 & Aug & 08 & 419 & 42 & 0.59 & 0.03 & 0.21 & 0.01 & \phn$-9$ & 3 \\
1998 & Sep & 16 & 423 & 42 & 0.71 & 0.04 & 0.26 & 0.02 & \phn$-8$ & 2 \\
1999 & Mar & 13 & 386 & 39 & 0.78 & 0.03 & 0.29 & 0.01 & $-10$    & 2 \\
1999 & May & 15 & 413 & 41 & 0.72 & 0.03 & 0.25 & 0.01 & $-12$    & 2 \\
1999 & Sep & 18 & 425 & 42 & 0.61 & 0.03 & 0.19 & 0.01 & $-16$    & 3 \\
1999 & Dec & 09 & 416 & 42 & 0.66 & 0.03 & 0.24 & 0.01 & $-14$    & 2 \\
2000 & May & 15 & 417 & 42 & 0.64 & 0.03 & 0.20 & 0.01 & $-13$    & 2 \\
2000 & Aug & 07 & 392 & 39 & 0.66 & 0.03 & 0.22 & 0.01 & $-12$    & 2 \\
2000 & Nov & 05 & 415 & 41 & 0.70 & 0.03 & 0.24 & 0.01 & $-10$    & 2 \\
2000 & Nov & 06 & 393 & 39 & 0.74 & 0.06 & 0.29 & 0.03 & \phn$-5$ & 3 \\
2001 & Mar & 31 & 394 & 39 & 0.69 & 0.03 & 0.26 & 0.01 & $-14$    & 2 \\
2001 & Jun & 29 & 386 & 39 & 0.80 & 0.03 & 0.29 & 0.01 & $-10$    & 2 \\
2001 & Oct & 19 & 364 & 36 & 0.79 & 0.04 & 0.31 & 0.02 & $-14$    & 4 \\
2001 & Dec & 21 & 385 & 39 & 0.83 & 0.03 & 0.30 & 0.01 & $-15$    & 3 \\
2002 & Apr & 14 & 327 & 33 & 0.76 & 0.03 & 0.29 & 0.01 & $-10$    & 3 \\
2002 & Jul & 14 & 327 & 33 & 0.82 & 0.03 & 0.27 & 0.01 & $-10$    & 2 \\
2002 & Nov & 20 & 359 & 36 & 0.73 & 0.04 & 0.19 & 0.01 & $-11$    & 2 \\
2003 & Jan & 26 & 395 & 39 & 0.83 & 0.07 & 0.28 & 0.02 & \phn$-9$ & 2 \\
2003 & May & 18 & 393 & 39 & 0.70 & 0.03 & 0.25 & 0.01 & $-11$    & 2 \\
2003 & Sep & 08 & 449 & 45 & 0.63 & 0.03 & 0.22 & 0.01 & $-13$    & 2 \\
2003 & Dec & 05 & 471 & 47 & 0.64 & 0.03 & 0.23 & 0.01 & $-12$    & 2 \\
2004 & Mar & 06 & 501 & 50 & 0.61 & 0.02 & 0.22 & 0.01 & $-11$    & 2 \\
2004 & May & 18 & 512 & 51 & 0.60 & 0.03 & 0.21 & 0.01 & $-13$    & 3 \\
2004 & Jun & 26 & 496 & 50 & 0.68 & 0.05 & 0.25 & 0.02 & \phn$-9$ & 2 \\
2004 & Dec & 11 & 475 & 47 & 0.59 & 0.03 & 0.20 & 0.01 & $-13$    & 2 \\
2005 & Jan & 15 & 490 & 49 & 0.62 & 0.03 & 0.20 & 0.01 & $-12$    & 2 \\
2005 & May & 28 & 477 & 48 & 0.59 & 0.03 & 0.20 & 0.01 & $-10$    & 2 \\
2005 & Jul & 16 & 478 & 48 & 0.59 & 0.02 & 0.19 & 0.01 & $-11$    & 2 \\
\enddata
\tablecomments{[1,2] Integrated flux density and its standard error;
[3,4] Major-axis length (FWHM) and its standard error; [5,6]
Minor-axis length (FWHM) and its standard error; [7,8] Position angle
(east of north) of major axis, and its standard error.  The tabulated
flux-density standard errors include the estimated 10\% standard error
in the VLBI flux-density scale at each epoch.}
\end{deluxetable}

Figure~\ref{2250modelplot} shows that the size of the Gaussian
(parametrized by major and minor axis lengths) is variable from epoch
to epoch, and that there is an anti-correlation between the size and
flux density of the Gaussian.  In contrast, the orientation of the
Gaussian is approximately constant (p.a.\ of the major axis $=
-11\arcdeg \pm 2\arcdeg$).  While the apparently random epoch-to-epoch
fluctuations in the size of the Gaussian may well be largely due to
limitations of the one-component model, the anti-correlation between
size and flux density suggests that activity is at least mainly
confined to a relatively compact region, namely the core or a shocked
region of the jet.

We did not include data from baselines to Ti in our above analysis
since Ti was not used consistently over the $\sim$8.5~yr period of our
observing program, and because including these data led to systematic
variations in the results for the size and orientation of the
elliptical Gaussian.  (Because their radio structures are extended
largely east-west, the results presented for sources 3C~454.3 and
B2252+172 are far less sensitive to the inclusion of the long,
north-south oriented, Ti baselines.)  The major-axis length (FWHM) of
the Gaussian is significantly smaller and less variable when Ti
baselines are included than when they are left out: $0.47 \pm
0.03$~mas instead of $0.66 \pm 0.06$~mas (for the same epochs).  This
comparison suggests that the Ti baselines further constrain the
compact central region.  Additionally, the major axis of the Gaussian
rotates when Ti baselines are included: $-11\arcdeg \pm 2\arcdeg$
p.a.\ becomes $-26\arcdeg \pm 3\arcdeg$.  The $-26\arcdeg$ estimate
plausibly better represents the alignment of the jet axis near the
core (see \S~\ref{2250canduimages}).

We attempted to separate the changes in the compact central region
from changes in the lower-level structure to the northwest and
south-southwest by also fitting to the \uv\ data a model consisting
of three point sources.  Unfortunately, we were unable to adequately
separate the three components: We found the same anti-correlation
between overall source size and total flux density with the
three-component model as with the one-component model; i.e., we found
that during periods of increased flux density (1) the flux densities
of all three components increased, and (2) the northwest and southwest
components moved closer to the central component (which we fixed to
the position of the brightness peak at each epoch).  We also found no
significant change with time of the angles of the position vectors
drawn from the center component to the northwest component or from the
center component to the southwest component.

\subsection{VLBI Images at 5.0 and 15.4~GHz \label{2250canduimages}}

\begin{figure}
\plotone{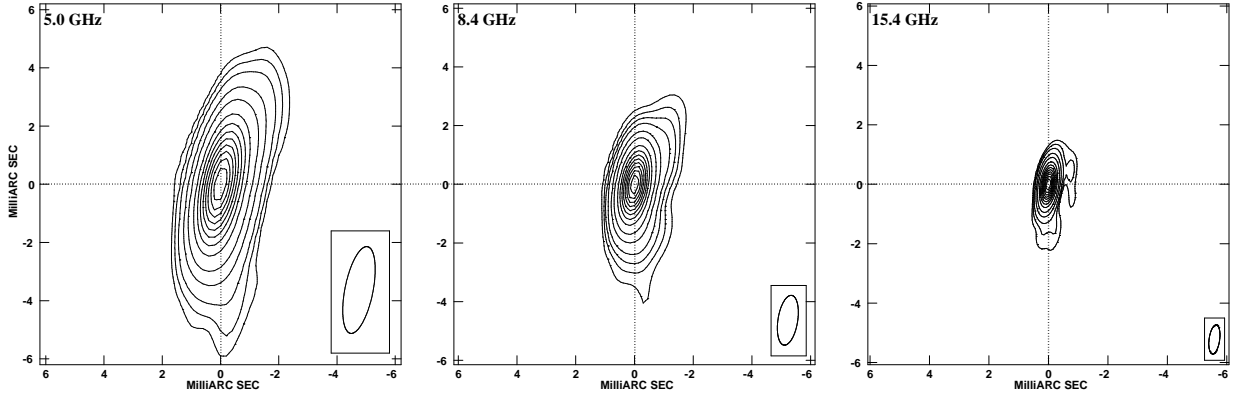}
\figcaption{5.0, 8.4, and 15.4~GHz VLBI images of B2250+194 for epoch
2004 May~18.  Contour levels are $0.5$, $1$, $2$, $5$, $10$, $20$,
$30$, $40$, $50$, $60$, $70$, $80$, and $90$\% of the peak brightness
in each image.  Image characteristics are summarized in
Table~\ref{2250imstat}.  The restoring beam is indicated in the
bottom-right-hand corner of each image.  The images are centered, as
indicated by the dotted lines, on the brightness peak in each image.
\label{2250imagesCXU}}
\end{figure}

We present in Figure~\ref{2250imagesCXU} the 5.0, 8.4, and 15.4~GHz
images of B2250+194 generated for the VLBI observations on 2004
May~18.  The 8.4~GHz image is the same as that presented in
Figure~\ref{2250imagesplot} except for the omission of the lowest
contour level.  The image characteristics are given in
Table~\ref{2250imstat}.

The images at all three frequencies show a compact emission region
near the brightness peak and low-level emission to the northwest and
south or south-southwest.  The upper limit on the angular size of the
compact emission region is set approximately by the angular resolution
of our VLBI array at 15.4~GHz (see below).  Between 5.0 and 15.4~GHz,
the peak flux density in the images increases $\sim$73\%, and the
total CLEAN flux density increases $\sim$55\%.  When taken together
with our observation that the flux density at 8.4~GHz is at or near a
maximum on 2004 May~18 (see top panel in Figure~\ref{2250modelplot}),
the small size and inverted spectrum (between 5.0 and 15.4~GHz) of the
compact central region suggest that the brightness peak in the image
at each frequency nearly coincides with the core.  Scenario 1 above
(see \S~\ref{2250imagestext}), which places the core at either the
north or south end of the source, thus seems unlikely.

We fit to the 5.0 and 15.4~GHz \uv\ data the same one-component model
described in \S~\ref{2250uvplanefits}.  The results are given in
Table~\ref{2250compCXU}.  The uncertainties are standard errors (see
\S~\ref{2250uvplanefits}), with the flux-density uncertainties
including also a 5\% uncertainty in the VLBI flux-density scale for
2004 May~18.  We adopt a (conservative) 5\% uncertainty for this epoch
rather than the nominal 10\% (see \S~\ref{VLBIamp}), because the VLBI-
and VLA-determined flux densities at each frequency differ by only
1--4\% (see Table~\ref{2250imstat}).  The results at each frequency
for the flux density, size, and orientation of the Gaussian are
consistent with a compact emission region near the brightness peak
that is progressively more dominant, relative to the surrounding
low-level emission, as the frequency increases from 5.0 to 15.4~GHz.
We estimate, using the $3\sigma$ upper limit on the major-axis length
(FWHM) of the Gaussian at 15.4~GHz, a maximum diameter for the compact
emission region of $0.40$~mas.  The orientation of the Gaussian at
15.4~GHz ($\rm{p.a.}  = -25\arcdeg \pm 4\arcdeg$) is consistent with
that at 8.4~GHz when Ti baselines are included in the fit.  This
consistency supports the accuracy of the results obtained by including
Ti baselines at 8.4~GHz, and provides additional evidence for the
underlying orientation of the jet axis near the core.  (Note that
telescope Ti did not observe at either 5.0 or 15.4~GHz.)  The decrease
in the spectral index from $\alpha = 0.79 \pm 0.15$ between 5.0 and
8.4~GHz to $\alpha = 0.12 \pm 0.13$ between 8.4 and 15.4~GHz is
consistent with the spectrum for the compact emission region peaking
at a frequency near 15.4~GHz.

\begin{deluxetable}{c@{}c@{}c@{}c@{}c@{}c@{}c@{}c@{}c}
\tabletypesize{\scriptsize}
\tablecaption{B2250+194 \uv-Plane Model Parameters for 2004 May~18 \label{2250compCXU}}
\tablewidth{0pt}
\tablehead{
  \colhead{$\nu$} &
  \colhead{$S_{\nu}$} &
  \colhead{$\sigma_{S}$} &
  \colhead{$\rm{Maj}$} &
  \colhead{$\sigma_{\rm{Maj}}$} &
  \colhead{$\rm{Min}$} &
  \colhead{$\sigma_{\rm{Min}}$} &
  \colhead{$\rm{p.a.}$} &
  \colhead{$\sigma_{\rm{p.a.}}$} \\
  \colhead{(GHz)} &
  \colhead{(mJy)} &
  \colhead{(mJy)} &
  \colhead{(mas)} &
  \colhead{(mas)} &
  \colhead{(mas)} &
  \colhead{(mas)} &
  \colhead{($\arcdeg$)} &
  \colhead{($\arcdeg$)} \\
  \colhead{} &
  \colhead{[1]} &
  \colhead{[2]} &
  \colhead{[3]} &
  \colhead{[4]} &
  \colhead{[5]} &
  \colhead{[6]} &
  \colhead{[7]} &
  \colhead{[8]}
}
\startdata
\phm{0}5.0 & 342 & 19 & 0.81 & 0.13 & 0.51 & 0.02 & $-15$ & 5 \\
\phm{0}8.4 & 512 & 27 & 0.60 & 0.03 & 0.21 & 0.01 & $-13$ & 3 \\
15.4       & 549 & 32 & 0.31 & 0.03 & 0.12 & 0.02 & $-25$ & 4 \\
\enddata
\tablecomments{[1,2] Integrated flux density and its standard error;
[3,4] Major-axis length (FWHM) and its standard error; [5,6]
Minor-axis length (FWHM) and its standard error; [7,8] Position angle
(east of north) of major axis, and its standard error.  The tabulated
flux-density standard errors include the estimated 5\% standard error
in the VLBI flux-density scale for 2004 May~18.}
\end{deluxetable}

\section{Radio Source B2252+172 \label{2252}}

\subsection{VLBI Images at 8.4~GHz \label{2252imagestext}}

\begin{figure}
\centerline{\includegraphics[totalheight=0.69\textheight,clip]{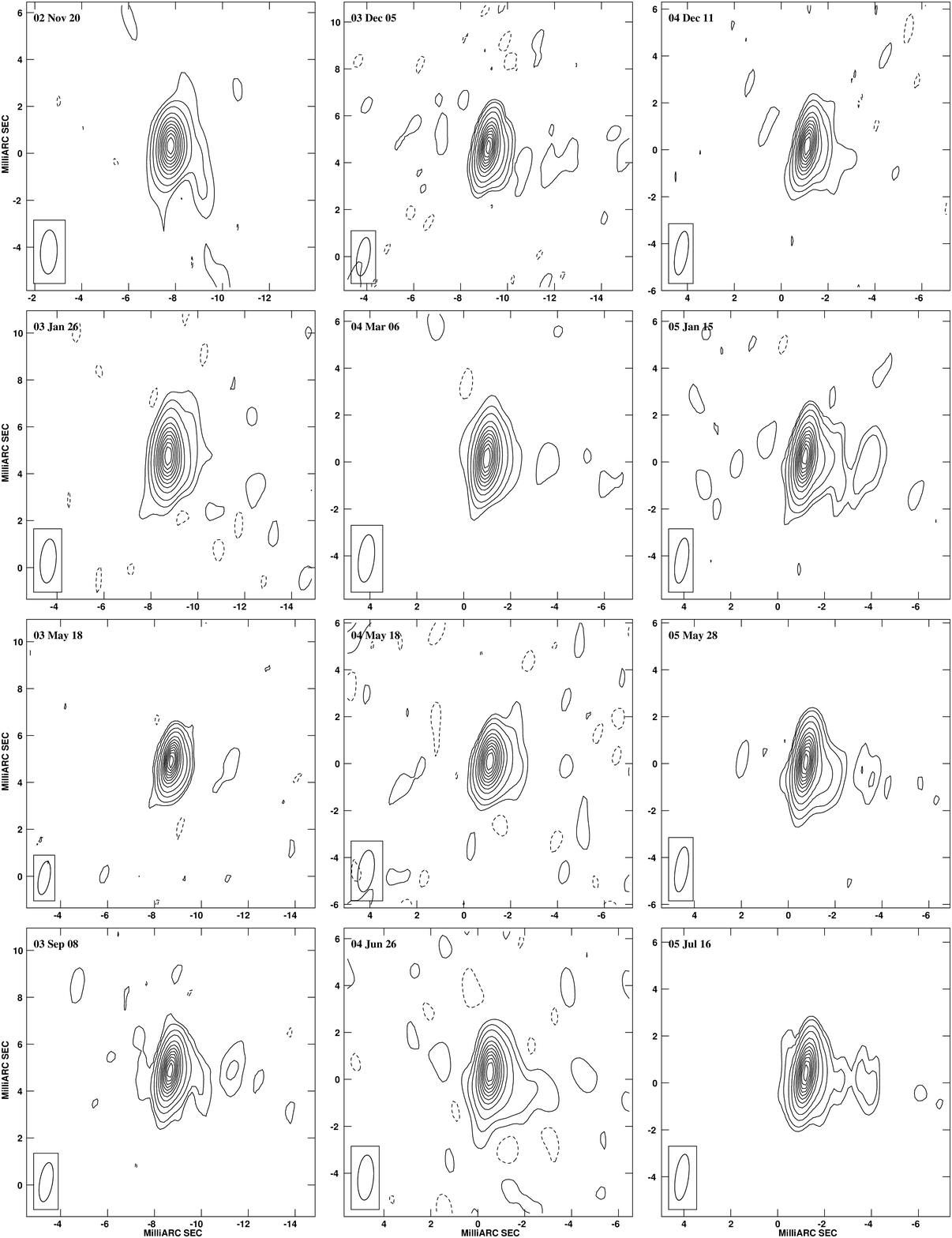}}
\figcaption{8.4~GHz VLBI images of B2252+172 for each of the 12
observing sessions between 2002 November and 2005 July.  The lowest
contour level in each image is chosen to display some noise, and is
either $\pm 1$\% or $\pm 2$\% of the peak brightness in the image.  In
images where the lowest contour is $\pm 1$\%, the $2$\% contour is
also shown.  The remaining contour levels in each image are $5$, $10$,
$20$, $30$, $40$, $50$, $60$, $70$, $80$, and $90$\% of the peak
brightness in the image.  Image characteristics are summarized in
Table~\ref{2252imstat}.  The restoring beam is indicated in the
bottom-left-hand corner of each image.  The image for each epoch is
centered on the brightness peak.  The up to $\sim$10~mas offset in
each image between the position of the brightness peak and the phase
center (i.e., origin of coordinates) largely reflects inaccuracies in
the assumed positions of B2252+172 and (phase reference source)
3C~454.3 used at the correlator.
\label{2252imagesplot}}
\end{figure}

We present in Figure~\ref{2252imagesplot} the 8.4~GHz images of
B2252+172 generated from each of the 12 sessions of VLBI observations
made between 2002~November and 2005~July.  The image characteristics
are summarized in Table~\ref{2252imstat}.  The images reveal a source
consisting of a relatively bright, compact emission region and
extended emission at or just above the noise level.  Near the
brightness peak, the source shows no deviation from the restoring
beam, down to the $\sim$10\% level.  Below this level, i.e., at the
brightness level of the outermost three or four contours, the source
extends west of the brightness peak.  For our last three epochs, a
very weak emission component, itself extended east-west, appears
(above the new peaks in the background) $\sim$2.6~mas west-southwest
of the brightness peak.  This component is likely real, since its
appearance is robust for these epochs against changes (e.g., to the
positioning of CLEAN windows) in the imaging process.  The flux
density of this component is $\sim$0.7~mJy.  Variations in the quality
of the images from epoch to epoch\footnote{The quality of the image
for B2252+172 for a given epoch is a function of the accuracy of the
phase referencing, as well as the brightness distribution and flux
density of the source (see \S~\ref{2252phaseref}).} make it difficult
to trace any motion of this component relative to the brightness peak.
However, there is evidence for the evolution of the low-level emission
$\sim$0.5~mas to $\sim$1.5~mas west of the brightness peak.  The peak
flux density in the image for 2003 May is at or near the maximum we
observed during the $\sim$2.7~yr span of our observations of this
source.  The low-level emission at this 2003 May epoch shows minimal
separation from the brightness peak.  In subsequent epochs, the
low-level emission extends increasingly westward toward the
$\sim$0.7~mJy component.  This evolution is consistent with the
ejection and subsequent motion of a new jet component.  The brightness
peak in each image near the easternmost edge of the source and the
apparent westward motion of the low-level emission suggests that the
brightness peak nearly coincides with the core.

\begin{deluxetable}{l@{~}l@{~}l@{\,}r c c c c c c c c}
\tabletypesize{\scriptsize}
\tablecaption{B2252+172 Image Characteristics \label{2252imstat}}
\tablewidth{0pt}
\tablehead{
  \multicolumn{3}{c}{Start Date} &
  \colhead{} &
  \colhead{Peak} &
  \colhead{Min.} &
  \colhead{rms} &
  \colhead{$\Theta_{\rm{maj}}$} &
  \colhead{$\Theta_{\rm{min}}$} &
  \colhead{p.a.} &
  \colhead{CLEAN} &
  \colhead{$[\frac{\rm{CLEAN}}{\rm{VLA}}]$} \\
  \multicolumn{3}{c}{} &
  \colhead{} &
  \colhead{($\rm{mJy}\,\Omega_b^{-1}$)} &
  \colhead{($\rm{mJy}\,\Omega_b^{-1}$)} &
  \colhead{($\rm{mJy}\,\Omega_b^{-1}$)} &
  \colhead{(mas)} &
  \colhead{(mas)} &
  \colhead{($\arcdeg$)} &
  \colhead{(mJy)} &
  \colhead{} \\
  \multicolumn{3}{c}{} &
  \colhead{} &
  \colhead{[1]} &
  \colhead{[2]} &
  \colhead{[3]} &
  \colhead{[4]} &
  \colhead{[5]} &
  \colhead{[6]} &
  \colhead{[7]} &
  \colhead{[8]}
}
\startdata
2002 & Nov & 20 &                      & 14.5    & $-0.41$ & 0.10 & 1.88 & 0.71 & \phn$-2.14$       & 18.0 &  ---  \\
2003 & Jan & 26 &                      & 14.3    & $-0.65$ & 0.12 & 1.88 & 0.67 & \phn$-4.93$       & 20.6 &  0.82 \\
2003 & May & 18 &                      & 18.7    & $-0.34$ & 0.08 & 1.44 & 0.48 & $-10.68$          & 22.9 &  0.85 \\
2003 & Sep & 08 &                      & 15.9    & $-0.28$ & 0.06 & 1.69 & 0.55 & \phn$-9.40$       & 18.7 &  0.87 \\
2003 & Dec & 05 &                      & 16.4    & $-0.34$ & 0.08 & 1.66 & 0.53 & \phn$-9.84$       & 19.1 &  ---  \\
2004 & Mar & 06 &                      & 12.4    & $-0.37$ & 0.08 & 2.03 & 0.67 & \phn$-6.62$       & 15.2 &  0.89 \\
2004 & May & 18 &                      & \phn9.7 & $-0.51$ & 0.15 & 1.79 & 0.67 & \phn$-8.85$       & 12.5 &  0.78 \\
2004 & May & 18 & (C)\tablenotemark{a} & 10.2    & $-0.56$ & 0.17 & 3.20 & 0.95 & $-10.96$          & 14.3 &  0.83 \\
2004 & May & 18 & (U)\tablenotemark{a} & \phn9.3 & $-2.31$ & 0.60 & 1.16 & 0.59 & \phn\phm{$-$}9.62 & 10.2 &  0.89 \\
2004 & Jun & 26 &                      & \phn8.9 & $-0.36$ & 0.10 & 1.92 & 0.67 & \phn$-4.22$       & 13.0 &  0.82 \\
2004 & Dec & 11 &                      & 11.6    & $-0.39$ & 0.09 & 1.87 & 0.53 & \phn$-9.14$       & 15.5 &  0.85 \\
2005 & Jan & 15 &                      & 11.6    & $-0.16$ & 0.05 & 1.94 & 0.55 & \phn$-8.26$       & 15.4 &  0.88 \\
2005 & May & 28 &                      & 16.8    & $-0.19$ & 0.05 & 1.93 & 0.54 & \phn$-8.30$       & 20.8 &  0.90 \\
2005 & Jul & 16 &                      & 16.9    & $-0.18$ & 0.04 & 1.98 & 0.56 & \phn$-7.98$       & 21.0 &  0.80 \\
\enddata
\tablenotetext{a}{C indicates 5.0~GHz; U indicates 15.4~GHz; all other
observations were at 8.4~GHz.}  \tablecomments{[1,2] Positive and
negative extrema ($\Omega_b$ $\equiv$ CLEAN restoring beam area); [3]
rms background level; [4--6] CLEAN restoring beam major axis (FWHM),
minor axis (FWHM), and position angle (east of north); [7] Total CLEAN
flux density; [8] Ratio of VLBI-determined (column 7) to
VLA-determined (Table~\ref{fluxdensities}) flux density for all epochs
at which the VLA observed.  The flux-density ratios (column 8)
indicate that the CLEAN components do not contain all the flux density
from the low-level emission.  The estimated standard error in the
VLBI-determined peak flux densities is 10\%.}
\end{deluxetable}

\subsection{\uv-Plane Model Fitting at 8.4~GHz \label{2252uvplanefits}}

To more quantitatively investigate changes in the radio structure of
B2252+172 near its brightness peak, we fit to the \uv\ data a model
consisting of a single elliptical Gaussian.  Our results (not
tabulated) show that the emission region near the brightness peak is
elongated along $\rm{p.a.} = 73\arcdeg \pm 8\arcdeg$ and has
major-axis length (FWHM) $0.35 \pm 0.07$~mas.  The quoted parameter
values and uncertainties are the mean and rms variation about the mean
over our 12 observing sessions.  The orientation of the Gaussian
suggests that the jet axis may bend slightly from $\rm{p.a.} \sim
73\arcdeg$ near the position of the brightness peak to $\rm{p.a.} \sim
90\arcdeg$ at distances $\gtrsim$1~mas from the brightness peak (see
also Figure~\ref{2252imagesplot}).  No systematic variations in either
the orientation or the size of the Gaussian are evident over our 12
epochs.

We also fit to the data a two-point-source model in an attempt to
characterize the changes observed in the low-level emission
immediately west of the brightness peak.  Unfortunately, the SNR of
the phase-referenced data was not sufficiently high to adequately
constrain both the relatively bright component near the brightness
peak and the weak secondary component.

\subsection{VLBI Images at 5.0 and 15.4~GHz \label{2252canduimages}}

\begin{figure}[ht]
\plotone{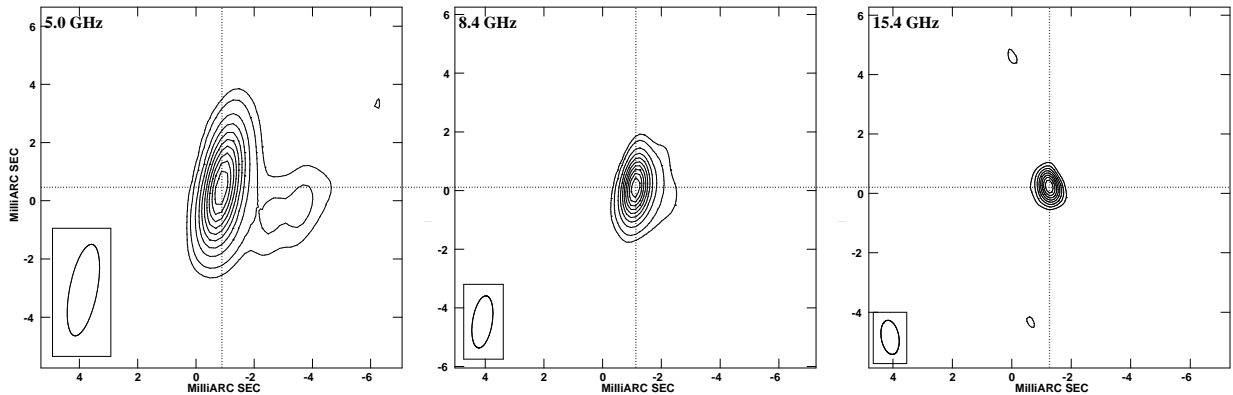}
\figcaption{5.0, 8.4, and 15.4~GHz VLBI images of B2252+172 for epoch
2004 May~18.  Contour levels in the 5.0 and 8.4~GHz images are $0.5$,
$1$, $2$, $5$, $10$, $20$, $30$, $40$, $50$, $60$, $70$, $80$, and
$90$\% of the peak brightness in each image.  Contour levels in the
15.4~GHz image are $20$, $30$, $40$, $50$, $60$, $70$, $80$, and
$90$\% of the peak brightness in the image.  Image characteristics are
summarized in Table~\ref{2252imstat}.  The restoring beam is indicated
in the bottom-left-hand corner of each image.  The images are
centered, as indicated by the dotted lines, on the brightness peak in
each image.
\label{2252imagesCXU}}
\end{figure}

We present in Figure~\ref{2252imagesCXU} the 5.0, 8.4, and 15.4~GHz
images of B2252+172 generated for the VLBI observations on 2004
May~18.  The 8.4~GHz image is the same as that presented in
Figure~\ref{2252imagesplot} except for the omission of the lowest
contour level.  The image characteristics are given in
Table~\ref{2252imstat}.

The images at 5.0 and 8.4~GHz each show a compact emission region near
the brightness peak and low-level emission to its west.  The image at
15.4~GHz shows that the compact emission region is marginally extended
northeast-southwest.  The component $\sim$2.6~mas west-southwest of
the brightness peak, identified in the 8.4~GHz images only at later
epochs, already appears prominently in the 5.0~GHz image for 2004
May~18.  The greater prominence at 5.0~GHz of this component is
consistent with steep-spectral-index emission from one or more jet
components perhaps several years after ejection.  The evolving
low-level emission observed at 8.4~GHz closer to the brightness peak
(see \S~\ref{2252imagestext}) may derive from a more recent ejection
event.

We fit to the 5.0 and 15.4~GHz \uv\ data the same one-component model
described in \S~\ref{2252uvplanefits}.  At 5.0~GHz, the best-fit
Gaussian is elongated along $\rm{p.a.} = 67\arcdeg \pm 11\arcdeg$ and
has major-axis length (FWHM) $0.57 \pm 0.09$~mas.  At 15.4~GHz, the
Gaussian is elongated along $\rm{p.a.} = 70\arcdeg \pm 6\arcdeg$ and
has major-axis length (FWHM) $0.33 \pm 0.04$~mas.  The uncertainties
are standard errors, for which the systematic contributions were
estimated using a procedure similar to that described in
\S~\ref{2250uvplanefits}.  The results for the orientation are
consistent with the mean value ($\rm{p.a.} = 73\arcdeg \pm 8\arcdeg$)
found at 8.4~GHz for the 12 epochs we observed this source.  Using the
$3\sigma$ upper limit on the major-axis length (FWHM) of the Gaussian
at 15.4~GHz, we estimate a maximum diameter of $0.45$~mas for the
compact emission region about the brightness peak.  The flat spectrum
of the emission between 5.0 and 15.4~GHz ($\alpha = -0.17 \pm 0.07$)
suggests that the brightness peak in the image at each frequency
nearly coincides with the core.

\section{Discussion \label{discuss}}

\subsection{8.4~GHz Core of Quasar 3C~454.3 \label{corediscuss}}

Our 35 images at 8.4~GHz, obtained from data gathered between 1997
January and 2005 July, show that the core region (easternmost
$\sim$1~mas) of 3C~454.3 is quite variable.  For each epoch, we fit to
the core region in the image a model consisting of two point sources.
We refer to the eastern component as C1 and the western component as
C2.  Our 8.4~GHz two-component model only approximately represents
the complex underlying structure seen in VLBI images at higher
frequencies (namely 43~GHz and 86~GHz).  Nevertheless, we determined
via a simulation study that C1 and C2 are closely aligned with the two
principal 43~GHz components identified by
\citet{Jorstad+2001b,Jorstad+2005}: C1 is located $0.18 \pm 0.06$~mas
west of the 43~GHz core and C2 is located $0.06 \pm 0.08$~mas
west-southwest of the ``stationary'' component located $\sim$0.65~mas
west of the 43~GHz core.  Since component C1 is the more nearly
aligned with the 43~GHz core, we consider it to be the 8.4~GHz core.
The uncertainties quoted above represent real variations in the
positions of C1 and C2, which appear to be largely caused by the
westward motion of new components along the jet axis.  The 43~GHz
images of \citet{Jorstad+2001b,Jorstad+2005} show the passage of four
new jet components between the core and ``stationary'' component
between 1997 January and 2001 March (the time period in common between
our and Jorstad et al.'s VLBI programs).  If we consider a simplified
picture for the core region consisting of the core, new jet component,
and the ``stationary'' component, then we can interpret the position
changes of C1 and C2 as follows: After its ejection, the jet component
moves west away from the core.  Since the 8.4~GHz images do not
resolve the core and jet component, C1 shifts first slightly west.  As
the jet component continues westward, separating itself further from
the core, C1 eventually ``relaxes'' back east toward the core.  The
0.06~mas value of the rms scatter suggests that the forward-and-back
``motion'' of C1 spans not much more than 0.12~mas.  A similar
scenario occurs as the jet component approaches and subsequently
passes the ``stationary'' component.  Here, C2 shifts first east, then
west, and finally back to its starting point.  The rms scatter
(0.08~mas) suggests that the back-forward-back ``motion'' of C2 spans
not much more than 0.16~mas.

\begin{figure}[ht]
\plotone{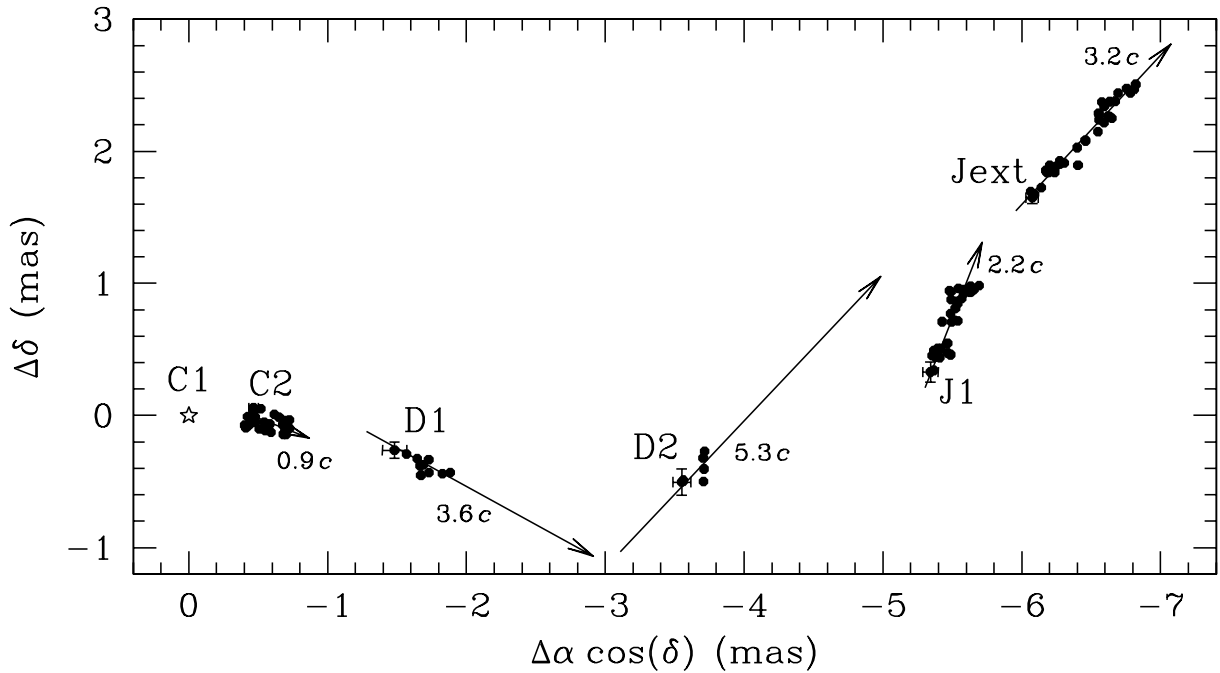}
\figcaption{Two-dimensional motions of 3C~454.3 model components C2,
D1, D2, J1, and Jext relative to the 8.4~GHz core (C1).  The black dots
are the relative position determinations at each epoch, and correspond
to those plotted separately in $\alpha$ and $\delta$ in
Figure~\ref{c1c2relpos} (C2), Figure~\ref{d1andd2c1relpos} (D1 and
D2), and Figure~\ref{j1andextc1relpos} (J1 and Jext).  The first dot in
the sequence for each component is accompanied by an error bar which
reflects the standard error in the relative position determination.
The arrows and quoted values indicate mean transverse velocity between
1997 January and 2005 July for the estimated distance of 1610~Mpc (see
Paper~III).
\label{3Ccompmotions}}
\end{figure}

The proper motion of C2 with respect to C1 between 1997 January and
2005 July was $-0.034 \pm 0.004$~\masyr\ in $\alpha$ and $-0.014 \pm
0.004$~\masyr\ in $\delta$.  This relative proper motion corresponds
to an apparent transverse velocity of $\sim$0.9\,$c$ (see
Figure~\ref{3Ccompmotions}).  This result indicates that over the
$\sim$8.5~yr period of our observing program either C1 ``moved''
systematically east-northeast and/or C2 ``moved'' systematically
west-southwest.  Given the proximity of C1 to the 43~GHz core, the
latter is more likely.  Moreover, VLBI images at 43 and 86~GHz
produced from observations between 2005 July and November
\citep{Krichbaum+2006a,Jorstad+2010} show the presence of a bright,
extended jet component with peak $\sim$0.4~mas west of the
``stationary'' component.  A plausible westward motion of this
component at earlier epochs, first toward, and then beyond, the
nominal position of the ``stationary'' component, would have caused
our C2 to ``move'' approximately westward away from C1 as we observed.
The various component models of 3C~454.3 for the time period of our
VLBI program appear to be able to accommodate all of the features and
changes we observed at 8.4~GHz in the core region of this source.

Based on its position near the easternmost edge of the 8.4~GHz radio
structure, close spatial association with the 43~GHz core, and
relatively flat spectrum, we believe C1 is the best choice for the
ultimate reference point for the \GPB\ guide star.  The general
westward motion away from C1 of the other 8.4~GHz components (see
Figure~\ref{3Ccompmotions}) further supports this choice.  In
Paper~III, we estimate the proper motion of C1 relative to the updated
CRF.

\subsection{Motion Along the Jet Axis of Quasar 3C~454.3}

We can trace in our 8.4~GHz images the motions of several jet
components at varying distances from C1.  We determined the proper
motions (relative to C1) of D1, D2, and J1, respectively located
$\sim$1.7, $\sim$3.7, and $\sim$5.5~mas west of the core.  The proper
motion of D1 was $0.14 \pm 0.02$~\masyr\ ($\sim$4\,$c$) directed
west-southwest, and that of D2 was $0.21 \pm 0.06$~\masyr\@
($\sim$5\,$c$) directed northwest (see Figure~\ref{3Ccompmotions}).
The difference in direction is consistent with the jet having a bend
at a distance from the core of $\sim$3~mas.

The proper motion of J1 (see Figure \ref{3Ccompmotions}) changed from
a northwestward $0.13 \pm 0.01$ mas\,yr$^{-1}$ ($\sim$3\,$c$) between
1997 January and 2003 May to a westward $0.03 \pm 0.01$~\masyr\@
($\sim$0.8\,$c$) between 2003 September and 2005 July.  The change in
direction and magnitude of the proper motion of J1 may signify a
second bend in the jet axis, and possibly a transition away from the
highly collimated flow seen in the easternmost $\sim$5.5~mas.  We
observed ``bulk motion'' ($0.13 \pm 0.01$~\masyr\ directed northwest,
see Figure~\ref{3Ccompmotions}) in the low-surface-brightness emission
beyond J1, but did not identify individual components in this extended
region.

\subsection{Proper Motion of the Brightness Peak of Quasar B2250+194?}

Our 35 images at 8.4~GHz of B2250+194 reveal a source consisting of a
bright, compact emission region and low-level emission which extends
both northwest and south-southwest of the brightness peak.  We were
unable using a three-component model to discern any relative motions
in the radio structure.  However, we determined using a model
consisting of a single elliptical Gaussian that the size and flux
density of this source were anti-correlated over the course of our
observing program.  This result indicates that the compact central
region was likely more dominant during periods of increased flux
density.  By also fitting the elliptical Gaussian model to our 5.0 and
15.4~GHz data separately for 2004 May~18, we determined that near a
flux-density maximum the compact central region (1) had an inverted
spectrum, (2) had a diameter (at 15.4~GHz) $<$0.40~mas, and (3) was
elongated (at 15.4~GHz) along $\rm{p.a} \sim -25\arcdeg$.  The
brightness peak in the image for 2004 May~18 for each frequency likely
nearly coincides with the core.  If periodic flux density increases
were accompanied by the ejection and subsequent motion of new
components, then we may expect the 8.4~GHz brightness peak to have
shifted approximately north-northwest or south-southeast on the sky
between each flux-density maximum and the subsequent minimum.  Since
the overall 8.4~GHz structure of B2250+194 appears approximately
static over the course of our observing program, it is likely that any
shifts in the position of the brightness peak were small (peak-to-peak
variations $\lesssim$0.2~mas).  In Paper~III, we bound the proper
motion of the 8.4~GHz brightness peak of B2250+194 relative to the
updated CRF.

\subsection{Proper Motion of the Brightness Peak of Quasar B2252+172?}

Our 12 images at 8.4~GHz of B2252+172 reveal a source consisting of a
relatively bright, compact emission region and low-level emission
which extends west-southwest at least $\sim$2.6~mas from the
brightness peak.  Using a model consisting of a single elliptical
Gaussian, we determined that the compact emission region was elongated
along $\rm{p.a.} = 73\arcdeg \pm 8\arcdeg$ and had a diameter $0.35
\pm 0.7$~mas (FWHM).  By also fitting the elliptical Gaussian model to
our 5.0 and 15.4~GHz data separately for 2004 May~18, we determined
that (at this epoch) the spectrum of the compact emission region
between these frequencies was nearly flat.  The brightness peak in the
image for 2004 May~18 for each frequency likely nearly coincides with
the core.  Given the observed evolution of the low-level emission
immediately west of the compact emission region, we may expect the
8.4~GHz brightness peak to have shifted slightly westward between 2002
November and 2005 July, as a possible new component was first ejected
(on or near 2003 May) and then moved away from the core.  Any such
shift was likely small (peak-to-peak variations $\lesssim$0.2~mas),
since the low-level emission is $\lesssim$10\% of the brightness peak
in the 8.4~GHz images.  In Paper~III, we bound the proper motion of
the 8.4~GHz brightness peak of B2252+172 relative to the updated CRF.

\section{Conclusions \label{conclus}}

Here we give a summary of our results and conclusions (all quoted
uncertainties are estimated standard errors, including allowance for
systematic errors):

1. Each of our 8.4 GHz images of quasar 3C~454.3, obtained from data
at 35 epochs spread between 1997 January and 2005 July, shows a
bright, relatively compact core region and a bent milliarcsecond-scale
jet.  The images reveal significant structural changes in the core
region (easternmost $\sim$1~mas), and three compact, moving jet
components at various distances from the core.

2. We modeled the structure of the 3C~454.3 core region with two point
sources oriented approximately east-west.  We refer to the eastern
component as C1 and the western component as C2.  We determined a
relative proper motion for C2 with respect to C1 of $-0.034 \pm
0.004$~\masyr\ in $\alpha$ and $-0.014 \pm 0.004$~\masyr\ in
$\delta$, an apparent transverse velocity of $\sim$0.9\,$c$.

3. C1 and C2 are closely aligned with the two principal core-region
components identified in VLBI images at 43~GHz by
\citet{Jorstad+2001b,Jorstad+2005}: C1 is located $0.18 \pm 0.06$~mas
west of the 43~GHz core and C2 is located $0.06 \pm 0.08$~mas
west-southwest of a component observed at 43~GHz to be located
$\sim$0.65~mas west of the core.  Since component C1 is most nearly
aligned with the 43~GHz core, we consider it to be the 8.4~GHz core.
The ``oscillations'' we observed in the 8.4~GHz separation of C2
relative to C1 probably reflect the motions of new jet components
periodically ejected from near the position of the 43~GHz core.  The
proper motion of C2 relative to C1 could be a consequence of the
westward motion of a bright jet component, which was visible at the
westernmost edge of the core region in 43 and 86~GHz images made 1--4
months after our last observing session.

4. We determined the proper motions relative to C1 of the 3C~454.3 jet
components D1, D2, and J1, respectively located $\sim$1.7, $\sim$3.7,
and $\sim$5.5~mas west of the core.  The proper motion of D1 was $0.14
\pm 0.02$~\masyr\ ($\sim$4\,$c$) towards the west-southwest, and that
of D2 was $0.21 \pm 0.06$~\masyr\ ($\sim$5\,$c$) towards the
northwest.  The change in the direction presumably demonstrates a bend
in the jet axis $\sim$3~mas from the core.  The proper motion of J1
changed from a northwestward $0.13 \pm 0.01$~\masyr\ ($\sim$3\,$c$)
between 1997 January and 2003 May to a westward $0.08 \pm
0.01$~\masyr\ ($\sim$0.8\,$c$) between 2003 September and 2005 July.
The change in direction and magnitude of the proper motion of J1 may
signify a second bend in the jet axis $\sim$5.5~mas from the core.  We
observed ``bulk motion'' ($0.13 \pm 0.01$~\masyr\ directed northwest)
in the low-surface-brightness emission beyond J1.

5. Our 15.4~GHz image of 3C~454.3 for 2004 May~18 shows a ``new'' jet
component between C1 and C2.  Our 8.4~GHz images suggest that this
component was ejected from near the position of C1 on or about 2003
September and arrived near the position of C2 $\sim$1.4~yr later.

6. Each of our 35 8.4-GHz images of quasar B2250+194 shows a bright,
compact emission region and low-level emission which extends slightly,
both northwest and south-southwest of the brightness peak.

7. We fit an elliptical Gaussian to the radio structure of B2250+194.
We found the size and flux density of the Gaussian to be
anti-correlated, suggesting that the variability of the source was
greatest near the brightness peak.

8. The 8.4~GHz flux density of B2250+194 reached a maximum in our
observations on or about 2004 May.  Our 5.0, 8.4, and 15.4~GHz
elliptical-Gaussian models for 2004 May~18 show that the compact
central region (1) had an inverted spectrum, (2) had a diameter (at
15.4~GHz) $<$0.40~mas, and (3) was elongated (at 15.4~GHz) along
$\rm{p.a.} \sim -25\arcdeg$.  The brightness peak in the image for
2004 May~18 for each frequency likely nearly coincides with the core.

9. Each of our 12 8.4-GHz images of B2252+172 shows a relatively
bright, compact emission region and low-level emission which extends
west-southwest at least $\sim$2.6~mas from the brightness peak.

10. We fit an elliptical Gaussian to the radio structure of B2252+172
and found the Gaussian to be elongated along $\rm{p.a.} = 73\arcdeg
\pm 8\arcdeg$ and to have a diameter $0.35 \pm 0.7$~mas (FWHM).  Our
5.0, 8.4, and 15.4~GHz elliptical-Gaussian models for 2004 May~18 show
that the compact emission region had a flat spectrum.  The brightness
peak in the image for 2004 May~18 for each frequency likely nearly
coincides with the core.

11. Based primarily on its close spatial association with the 43~GHz
core, we believe 3C~454.3 component C1 to be the best choice for the
ultimate reference point for the \GPB\ guide star.  The inverted to
flat spectra and small bounds for any changes near their brightness
peaks make B2250+194 and B2252+172 useful secondary reference sources.

\acknowledgements

ACKNOWLEDGMENTS.  We thank the anonymous referee for a constructive
review of the paper and for comments helpful in the preparation of the
final manuscript.  This research was primarily supported by NASA,
through a contract with Stanford University to the Smithsonian
Astrophysical Observatory (SAO), and a subcontract from SAO to York
University.  The National Radio Astronomy Observatory (NRAO) is a
facility of the National Science Foundation operated under cooperative
agreement by Associated Universities, Inc.  The DSN is operated by
JPL/Caltech, under contract with NASA.  We thank S.~G.\ Jorstad for
providing the 43~GHz models used in the simulation study of 3C~454.3.
We have made use of the United States Naval Observatory (USNO) Radio
Reference Frame Image Database (RRFID) and NASA's Astrophysics Data
System Abstract Service, developed and maintained by SAO.

\end{document}